\documentclass[preprint,twocolumn]{aastex63}
\usepackage{float, bm, graphicx, amsmath, morefloats, soul}
\usepackage[caption=false]{subfig}
\turnoffeditone
\turnoffedittwo

\shorttitle{ZTF FBOTs}
\shortauthors{Ho et al.}

\newcommand{\ncand}{\mbox{$\rm 38$}}

\newcommand{\ncandvis}{\mbox{$\rm 38$}}
\newcommand{\nfbot}{\mbox{$\rm 28$}}

\newcommand{\swift}{\textit{Swift}}
\newcommand{\galex}{\textit{GALEX}}
\newcommand{\wise}{\textit{WISE}}
\newcommand{\cigale}{\textsc{cigale}}

\newcommand{\erg}{\mbox{$\rm erg$}}

\newcommand{\km}{\mbox{$\rm km$}}

\newcommand{\pmpc}{\mbox{$\rm Mpc^{-1}$}}

\newcommand{\pcmsq}{\mbox{$\rm cm^{-2}$}}

\newcommand{\degsq}{\mbox{$\rm deg^{2}$}}

\newcommand{\pgpccub}{\mbox{$\rm Gpc^{-3}$}}

\newcommand{\days}{\mbox{$\rm d$}}
\newcommand{\psec}{\mbox{$\rm s^{-1}$}}
\newcommand{\pyr}{\mbox{$\rm yr^{-1}$}}

\newcommand{\phz}{\mbox{$\rm Hz^{-1}$}}

\begin{document}

\title{
A Search for Extragalactic Fast Blue Optical Transients in ZTF and \\ the Rate of AT2018cow-like Transients
}

\author[0000-0002-9017-3567]{Anna Y. Q.~Ho}
\affiliation{Miller Institute for Basic Research in Science, 468 Donner Lab, Berkeley, CA 94720, USA}
\affiliation{Department of Astronomy, University of California, Berkeley, 501 Campbell Hall, Berkeley, CA, 94720, USA}
\affiliation{Lawrence Berkeley National Laboratory, 1 Cyclotron Road, MS 50B-4206, Berkeley, CA 94720, USA}
\affiliation{Department of Astronomy, Cornell University, Ithaca, NY 14853, USA}

\author[0000-0001-8472-1996]{Daniel A.~Perley}
\affiliation{Astrophysics Research Institute, Liverpool John Moores University, IC2, Liverpool Science Park, 146 Brownlow Hill, Liverpool L3 5RF, UK}

\author{Avishay Gal-Yam}
\affiliation{Department of Particle Physics and Astrophysics, Weizmann Institute of Science, 234 Herzl St, 76100 Rehovot, Israel}

\author[0000-0001-9454-4639]{Ragnhild Lunnan}
\affiliation{The Oskar Klein Centre, Department of Astronomy, Stockholm University, AlbaNova, SE-10691 Stockholm, Sweden}

\author[0000-0003-1546-6615]{Jesper Sollerman}
\affiliation{The Oskar Klein Centre, Department of Astronomy, Stockholm University, AlbaNova, SE-10691 Stockholm, Sweden}

\author[0000-0001-6797-1889]{Steve Schulze}
\affiliation{The Oskar Klein Centre, Department of Astronomy, Stockholm University, AlbaNova, SE-10691 Stockholm, Sweden}

\author[0000-0001-8372-997X]{Kaustav K.~Das}
\affiliation{Cahill Center for Astrophysics, 
California Institute of Technology, MC 249-17, 
1200 E California Boulevard, Pasadena, CA, 91125, USA}

\author[0000-0003-0699-7019]{Dougal Dobie}
\affiliation{Centre for Astrophysics and Supercomputing, Swinburne University of Technology, Hawthorn, Victoria, Australia}
\affiliation{ARC Centre of Excellence for Gravitational Wave Discovery (OzGrav), Hawthorn, Victoria, Australia}

\author[0000-0001-6747-8509]{Yuhan Yao}
\affiliation{Cahill Center for Astrophysics, 
California Institute of Technology, MC 249-17, 
1200 E California Boulevard, Pasadena, CA, 91125, USA}

\author[0000-0002-4223-103X]{Christoffer Fremling}
\affiliation{Cahill Center for Astrophysics, 
California Institute of Technology, MC 249-17, 
1200 E California Boulevard, Pasadena, CA, 91125, USA}

\author{Scott Adams}
\affiliation{Cahill Center for Astrophysics, 
California Institute of Technology, MC 249-17, 
1200 E California Boulevard, Pasadena, CA, 91125, USA}

\author{Shreya Anand}
\affiliation{Division of Physics, Mathematics and Astronomy, California Institute of Technology, Pasadena, CA 91125, USA}

\author{Igor Andreoni}
\affiliation{Cahill Center for Astrophysics, 
California Institute of Technology, MC 249-17, 
1200 E California Boulevard, Pasadena, CA, 91125, USA}

\author[0000-0001-8018-5348]{Eric C. Bellm}
\affiliation{DIRAC Institute, Department of Astronomy, University of Washington, 3910 15th Avenue NE, Seattle, WA 98195, USA}

\author[0000-0002-0786-7307]{Rachel J. Bruch}
\affil{Department of Particle Physics and Astrophysics
Weizmann Institute of Science
234 Herzl St.
76100 Rehovot, Israel}

\author[0000-0002-7226-836X]{Kevin B. Burdge}
\affiliation{Division of Physics, Mathematics and Astronomy, California Institute of Technology, Pasadena, CA 91125, USA}

\author[0000-0003-2999-3563]{Alberto J. Castro-Tirado}
\affiliation{Instituto de Astrof´ısica de Andaluc´ıa (IAA-CSIC) Glorieta
de la Astronomia E18008, Granada, Spain}

\author{Aishwarya Dahiwale}
\affiliation{Cahill Center for Astrophysics, 
California Institute of Technology, MC 249-17, 
1200 E California Boulevard, Pasadena, CA, 91125, USA}

\author{Kishalay De}
\affiliation{Cahill Center for Astrophysics, 
California Institute of Technology, MC 249-17, 
1200 E California Boulevard, Pasadena, CA, 91125, USA}

\author[0000-0002-5884-7867]{Richard Dekany}
\affiliation{Caltech Optical Observatories, California Institute of Technology, Pasadena, CA  91125}

\author{Andrew J. Drake}
\affiliation{Cahill Center for Astrophysics, 
California Institute of Technology, MC 249-17, 
1200 E California Boulevard, Pasadena, CA, 91125, USA}

\author[0000-0001-5060-8733]{Dmitry~A.~Duev}
\affiliation{Division of Physics, Mathematics, and Astronomy, California Institute of Technology, Pasadena, CA 91125, USA}

\author{Matthew~J.~Graham}
\affiliation{Cahill Center for Astrophysics, 
California Institute of Technology, MC 249-17, 
1200 E California Boulevard, Pasadena, CA, 91125, USA}

\author[0000-0003-3367-3415]{George~Helou}
\affiliation{IPAC, California Institute of Technology, 1200 E. California Blvd, Pasadena, CA 91125, USA}

\author[0000-0001-6295-2881]{David~L.~Kaplan}
\affiliation{Center for Gravitation, Cosmology, and Astrophysics, Department of Physics, University of Wisconsin-Milwaukee, P.O. Box 413, Milwaukee, WI 53201, USA}

\author{Viraj~Karambelkar}
\affiliation{Cahill Center for Astrophysics, 
California Institute of Technology, MC 249-17, 
1200 E California Boulevard, Pasadena, CA, 91125, USA}

\author[0000-0002-5619-4938]{Mansi~M.~Kasliwal}
\affiliation{Cahill Center for Astrophysics, 
California Institute of Technology, MC 249-17, 
1200 E California Boulevard, Pasadena, CA, 91125, USA}

\author[0000-0002-7252-3877]{Erik~C.~Kool}
\affiliation{The Oskar Klein Centre, Department of Astronomy, Stockholm University, AlbaNova, SE-10691 Stockholm, Sweden}

\author[0000-0001-5390-8563]{S.~R.~Kulkarni}
\affiliation{Cahill Center for Astrophysics, 
California Institute of Technology, MC 249-17, 
1200 E California Boulevard, Pasadena, CA, 91125, USA}

\author[0000-0003-2242-0244]{Ashish~A.~Mahabal}
\affiliation{Division of Physics, Mathematics and Astronomy, California Institute of Technology, Pasadena, CA 91125, USA}
\affiliation{Center for Data Driven Discovery, California Institute of Technology, Pasadena, CA 91125, USA}

\author[0000-0002-7226-0659]{Michael~S.~Medford}
\affiliation{Department of Astronomy, University of California, Berkeley, Berkeley, CA 94720}
\affiliation{Lawrence Berkeley National Laboratory, 1 Cyclotron Rd., Berkeley, CA 94720}

\author[0000-0001-9515-478X]{A.~A.~Miller}
\affiliation{Center for Interdisciplinary Exploration and Research in
             Astrophysics (CIERA) and Department of Physics and Astronomy,
             Northwestern University,
             1800 Sherman Road, Evanston, IL 60201, USA}
\affiliation{The Adler Planetarium, Chicago, IL 60605, USA}

\author{Jakob Nordin}
\affiliation{Institute of Physics, Humboldt-Universität zu Berlin, Newtonstr. 15, 12489 Berlin, Germany}

\author[0000-0002-6786-8774]{Eran Ofek}
\affiliation{Department of Particle Physics and Astrophysics, Weizmann Institute of Science, 234 Herzl St, 76100 Rehovot, Israel}

\author{Glen Petitpas}
\affiliation{Harvard-Smithsonian Center for Astrophysics, 60 Garden Street, Cambridge, MA 02138, USA}

\author{Reed Riddle}
\affiliation{Caltech Optical Observatories, California Institute of Technology, Pasadena, CA  91125}

\author{Yashvi Sharma}
\affiliation{Cahill Center for Astrophysics, 
California Institute of Technology, MC 249-17, 
1200 E California Boulevard, Pasadena, CA, 91125, USA}

\author[0000-0001-7062-9726]{Roger Smith}
\affiliation{Caltech Optical Observatories, California Institute of Technology, Pasadena, CA  91125}

\author[0000-0001-8026-5903]{Adam J. Stewart}
\affiliation{Sydney Institute for Astronomy, School of Physics, The University of Sydney, NSW 2006, Australia}

\author{Kirsty Taggart}
\affiliation{Department of Astronomy and Astrophysics, University of California, Santa Cruz, CA 95064, USA}

\author[0000-0003-3433-1492]{Leonardo Tartaglia}
\affiliation{The Oskar Klein Centre, Department of Astronomy, AlbaNova, SE-106 91 Stockholm, Sweden}
\affiliation{INAF - Osservatorio Astronomico di Padova, Vicolo dell’Osservatorio 5, I-35122 Padova, Italy}

\author{Anastasios Tzanidakis}
\affiliation{Cahill Center for Astrophysics, 
California Institute of Technology, MC 249-17, 
1200 E California Boulevard, Pasadena, CA, 91125, USA}

\author[0000-0001-6114-9173]{Jan Martin Winters}
\affiliation{Institut de Radioastronomie Millim{\'e}trique (IRAM), 300 rue de la Piscine, F-38406 St. Martin d'H{\'e}res, France}

\begin{abstract}

We present a search for extragalactic fast blue optical transients (FBOTs) during Phase I of the Zwicky Transient Facility (ZTF). We identify \ncand\ candidates with durations above half-maximum light $1\,\days< t_{1/2}<12\,$\days, of which \nfbot\ have blue ($g-r\lesssim-0.2\,$mag) colors at peak light.
Of the \ncand\ transients (\nfbot\ FBOTs), 19 (13) can be spectroscopically classified as core-collapse supernovae (SNe): 11 (8) H- or He-rich (Type~II/IIb/Ib) SNe, 6 (4) interacting (Type~IIn/Ibn) SNe, and 2 (1) H\&He-poor (Type~Ic/Ic-BL) SNe.
Two FBOTs (published previously) had high-S/N predominantly featureless spectra and luminous radio emission: AT2018lug and AT2020xnd.
Seven (five) did not have a definitive classification: AT\,2020bdh showed tentative broad H$\alpha$ in emission, and AT\,2020bot showed unidentified broad features and was 10\,kpc offset from the center of an early-type galaxy.
Ten (six) have no spectroscopic observations or redshift measurements.
We present multiwavelength (radio, millimeter, and/or X-ray) observations for five FBOTs (three Type~Ibn, one Type~IIn/Ibn, one Type~IIb).
Additionally, we search radio-survey (VLA and ASKAP) data to set limits on the presence of radio emission for 22 of the transients.
All X-ray and radio observations resulted in non-detections;
we rule out AT2018cow-like X-ray and radio behavior for five FBOTs and more luminous emission (such as that seen in the Camel) for four additional FBOTs.
We conclude that exotic transients similar to AT2018cow, the Koala, and the Camel represent a rare subset of FBOTs,
and use ZTF's SN classification experiments to measure the rate to be at most $0.1\%$ of the local core-collapse SN rate.

\end{abstract}

\section{Introduction}

In the past decade, high-cadence optical surveys have uncovered a variety of extragalactic transients with light curves that evolve faster than those of established supernova (SN) classes.
As reviewed in \citet{Inserra2019}, rapid transients have diverse origins, including massive-star explosions with low ejecta masses, thermonuclear explosions, and interaction-powered supernovae.
In recent years, a subset of rapid transients dubbed ``fast blue optical transients'' (FBOTs; \citealt{Drout2014,Pursiainen2018,Margutti2019}) have attracted significant attention due to the discovery of luminous X-ray \citep{RiveraSandoval2018,Margutti2019,Ho2019cow}, radio \citep{Margutti2019}, and submillimeter \citep{Ho2019cow} emission accompanying the nearby ($d=60$\,Mpc) FBOT AT2018cow \citep{Prentice2018,Perley2019cow}.

\edit2{A commonly used definition of ``FBOT'' is blue colors ($g-r\lesssim-0.2\,$mag) at peak light and a short duration above half-maximum light ($t_{1/2}\lesssim 12\,$d) \citep{Inserra2019}.
Approximately 100 FBOTs have been discovered in archival searches, the vast majority too late for spectroscopic follow-up observations \citep{Drout2014,Pursiainen2018}.
Single-object studies suggest that some FBOTs arise from shock-interaction with a dense wind \citep{Ofek2010} or shell \citep{Rest2018,Ho2019gep}, with spectral types ranging from hydrogen-rich \citep{Ofek2010} to hydrogen-poor \citep{Ho2019gep,Pritchard2021}.
}

\edit2{In the last few years, the improved grasp of optical surveys \citep{Bellm2016_grasp,Ofek2020} has made the \edit1{discovery of rapid transients} routine.
In this paper we present the first sample of FBOTs with spectroscopic classifications, using data from the Zwicky Transient Facility (ZTF; \citealt{Bellm2019_ztf,Graham2019}) high-cadence surveys \citep{Bellm2019_surveys}.}
In Section~\ref{sec:sample-overview} we present our selection criteria and \edit2{the ZTF FBOT sample. %We identify a comparison sample by applying similar criteria to literature objects.
In Section~\ref{sec:analysis} we analyze the photometric and spectroscopic evolution of the ZTF FBOTs, set limits on accompaying X-ray and radio emission, %perform a combined analysis of the ZTF and literature FBOTs, 
and identify several subtypes. We conclude that AT2018cow-like FBOTs are rare, and estimate their rate in} Section~\ref{sec:rates}.
We discuss the implications of our work for the progenitors in Section~\ref{sec:discussion},
and summarize in Section~\ref{sec:summary}.

Throughout the paper we assume a flat $\Lambda$CDM cosmology with $H_0=67.7\,\km\,\psec\,\pmpc$ and $\Omega_M=0.307$ \citep{Planck2016}.
Times are presented in UTC, and magnitudes are given in AB.
The optical photometry and spectroscopy will be made public through WISeREP, the Weizmann Interactive Supernova Data Repository \citep{Yaron2012}.

\section{Observations and Selection Criteria}
\label{sec:sample-overview}

\subsection{ZTF}

The ZTF custom mosaic camera \citep{Dekany2020} is mounted on the 48-inch Samuel Oschin Telescope (P48) at Palomar Observatory.
As summarized in \citet{Bellm2019_surveys},
observing time for ZTF Phase I was divided between public (40\%), partnership (40\%), and Caltech surveys (20\%).
Three custom filters are used ($g_{\mathrm{ZTF}}$, $r_{\mathrm{ZTF}}$, and $i_{\mathrm{ZTF}}$; hereafter $g$, $r$, and $i$; \citealt{Dekany2020})
and images reach a typical dark-time limiting magnitude of $r\sim20.5\,$mag.

Images are processed and reference-subtracted
by the IPAC ZTF pipeline \citep{Masci2019} using the \citet{Zackay2016} image-subtraction algorithm.
Every 5-$\sigma$ point-source detection is saved as an ``alert.''
Alerts are distributed in Avro format \citep{Patterson2019} and can be filtered based on a machine learning real-bogus metric \citep{Mahabal2019,Duev2019}; host-galaxy characteristics, including a star-galaxy classifier \citep{Tachibana2018}; and light-curve properties.
During the time period relevant for this paper (ZTF Phase I) the collaboration used a web-based system called the GROWTH marshal \citep{Kasliwal2019} to
identify, monitor, and coordinate follow-up observations for transients of interest.

Although we use observations from all programs,
the most effective surveys for discovering \edit2{FBOTs} are the high-cadence partnership survey (HC), which covered 2500\,deg$^{2}$ with six visits per night (three in $r$ and three in $g$);
the ZTF Uniform Depth Survey (ZUDS\footnote{\url{https://github.com/zuds-survey/zuds-pipeline}}), which covered 2500\,deg$^{2}$ with six visits per night (2$r$, 2$g$, and 2$i$);
the one-day cadence Caltech survey (1DC), which covered 3000\,deg$^{2}$ with 1$r$ and 1$g$ visit per night;
and one-day cadence observations for shadowing the Transiting Exoplanet Survey Satellite (TESS; \citealt{Ricker2014}) fields.

\subsection{ZTF Sample Selection}
\label{sec:ztf-sample-selection}

We searched data from ZTF Phase I, i.e., obtained from March 2018 through October 2020.
We used \texttt{ztfquery} \citep{Rigault2018} to identify fields in the primary grid with $E(B-V)<0.3\,$mag at the central field coordinate,
and only searched field-nights that had at least one observation in the same field within the preceding and subsequent five nights.
This left a total of 127,487 field nights.

For each of the 127,487 field-nights, we searched for transients fulfilling the criteria laid out in Table~\ref{tab:search}.
We performed the search with the following steps:

\begin{deluxetable}{lrr}[!ht] 
\tablecaption{Steps for selecting $1\,\days < t_{1/2} < 12\,\days$ transients in ZTF data.\label{tab:search}} 
\tablewidth{0pt} 
\tablehead{ \colhead{Step$^\dag$} & \colhead{Criteria} & \colhead{\# Candidates} }
\tabletypesize{\scriptsize} 
\startdata 
1 & Basic cuts on subtractions & 2.5M \\ 
2 & Candidate has $\geq3$ alerts & 651,920 \\
3 & Light curve has short duration & 19,715 \\
4 & Light curve is well-sampled & 6,059 \\
5 & Fast-rising in $g$ or $r$ & 1,779\\
6 & Manual inspection & \ncandvis
\enddata 
\tablenotetext{$\dag$}{Details on each step are provided in the text.}
\end{deluxetable} 

\begin{enumerate}
    \item We applied basic cuts to remove artifacts and stellar phenomena. We kept sources with a real-bogus score $\texttt{rb}>0.5$ \citep{Mahabal2019}
    and a deep learning score $\texttt{braai}>0.8$ \citep{Duev2019}.
    The \texttt{braai} score
    corresponds to a false positive rate of 0.7\% and a false negative rate of 3\% \citep{Duev2019}.
    We removed sources within 2$^{\prime\prime}$ of a counterpart with a star-galaxy score greater than 0.76 \citep{Tachibana2018},
    and sources within 15$^{\prime\prime}$ of a bright ($r<15\,$mag) star.
    We removed sources that arose from negative subtractions. This left $\sim2.5\,$M unique sources.
    \item We required that each source be detected in at least three alerts, leaving 651,920 sources.
    \item  To remove flaring and long-duration transients, we required that the time from the first to last detection (including the 30-day history in the alert packets, which uses a lower threshold than issued alerts) is between 1\,d and 120\,d\footnote{Intra-night ($t<1\,$d) transients are presented in separate work \citep{Ho2020d,Andreoni2020,Ho2022b}.}.
    \edit2{Following the FBOT definition commonly adopted in the literature \citep{Drout2014,Inserra2019}} we required that the duration above half-maximum of the light curve be $1\days < t_{1/2} < 12\,\days$. \edit2{We applied the cut to the $g$-band light curve.} This left 19,715 sources.
    \item \edit1{Because color is an important characteristic of FBOTs, we required that the light curve be well sampled with multi-band photometry,} i.e., that there is a P48 observation (resulting in either an upper limit or a detection) within 5.5\,d of the peak of the $g$-band light curve, before and after, in $g$ band and $r$ band. This left 6,059 sources.
    \item We further required the source to be fast-rising: that in either $g$ band or $r$ band it rose 1\,mag in the preceding 6.5\,d. This left 1,779 sources.
    \item We examined each of the 1,779 sources manually,
    and removed events (75\% of the total) that passed Step 5 only on the basis of spurious non-detections in between detections. We discarded an additional 20\% of events for having
    a point-like counterpart\footnote{Extragalactic transients in compact hosts could accidentally be removed by this step. However, we found it important for removing outbursts from cataclysmic variables.} (making particular use of the eighth data release of the Legacy Survey; \citealt{Dey2019}), repeated flaring behavior, or spectroscopic classifications indicating that they were stellar outbursts. We used forced photometry \citep{Yao2019} to confirm a short event duration of $1\,\days < t_{1/2,g} < 12\,$d.
    This left \ncandvis\ sources.
\end{enumerate}

\startlongtable
\begin{deluxetable*}{lrrrrrrrrrr}
\tablecaption{Candidate extragalactic transients from ZTF Phase I with durations $ 1\,\days < t_{1/2,g}<12$\,d and well-sampled light curves.\label{tab:sources_all}} 
\tablewidth{0pt} 
\tablehead{ \colhead{ZTF Name} & \colhead{R.A.} & \colhead{Dec.} & \colhead{IAU Name} & \colhead{Peak MJD} & \colhead{Peak Mag} & \colhead{$t_{1/2,g}$} & \colhead{$g-r$} & \colhead{$z$} & \colhead{Class$^{\ddag\ddag}$} & \colhead{Ref} \\[-0.3cm] \colhead{} & \colhead{(J2000)} & \colhead{(J2000)} & \colhead{} & \colhead{} & \colhead{} & \colhead{(d)} & \colhead{(mag)} & \colhead{} & \colhead{} & \colhead{}}
\tabletypesize{\scriptsize} 
\startdata 
18aakuewf & 16:14:22.65 & +35:55:04.4 & SN 2018bcc & 58230.38 & $17.46\pm0.04$ & $9.1 \pm 0.4$ & $-0.3$ & 0.0636 & Ibn & [1] \\ 
18abfcmjw$^m$ & 17:36:46.74 & +50:32:52.1 & SN 2019dge & 58583.16 & $18.40\pm0.02$ & $6.1 \pm 0.2$ & $-0.2$ & 0.0213 & Ib & [2] \\ 
18abianhw & 19:23:40.60 & +44:48:30.1 & AT 2018lwd & 58318.41 & $19.55\pm0.05$ & $6.8 \pm 1.0$ & $-0.2$ & -- & -- & -- \\ 
18abukavn$^m$ & 16:43:48.20 & +41:02:43.3 & SN 2018gep & 58374.22 & $15.91\pm0.01$ & $9.3 \pm 0.2$ & $-0.4$ & 0.03154 & Ic-BL & [3,4] \\ 
18abvkmgw & 00:37:26.87 & +15:00:51.2 & SN 2018ghd & 58377.35 & $18.49\pm0.03$ & $9.5 \pm 0.9$ & $-0.1$ & 0.03923 & Ib & [5,6] \\ 
18abvkwla$^m$ & 02:00:15.19 & +16:47:57.3 & AT 2018lug & 58374.41 & $19.34\pm0.05$ & $4.0 \pm 0.1$ & $-0.6$ & 0.2714 & Feat.;RL$^{\dag\dag}$ & [7] \\ 
18abwkrbl & 02:16:15.58 & +28:35:28.6 & SN 2018gjx & 58379.44 & $15.58\pm0.01$ & $7.4 \pm 0.1$ & $-0.2$ & 0.00999 & IIb & [8] \\ 
19aankdan & 11:53:47.14 & +44:44:44.8 & AT 2019dcm & 58572.27 & $19.09\pm0.04$ & $9.8 \pm 0.6$ & $-0.1$ & -- & -- & [9] \\ 
19aapfmki$^M$ & 14:05:43.56 & +09:30:56.6 & SN 2019deh & 58587.33 & $17.22\pm0.02$ & $10.7 \pm 0.7$ & $-0.2$ & 0.05469 & Ibn & [10--12] \\ 
19aapuudk & 15:10:03.55 & +38:07:11.8 & AT 2019aajt & 58585.27 & $19.49\pm0.05$ & $5.7 \pm 0.7$ & $-0.3$ & -- & -- & -- \\ 
19aasexmy & 13:31:54.39 & +25:44:05.9 & AT 2019aaju & 58599.33 & $19.41\pm0.02$ & $10.3 \pm 1.5$ & $-0.3$ & -- & -- & -- \\ 
19aatoboa & 12:25:40.57 & +44:44:48.8 & AT 2019esf & 58609.22 & $18.84\pm0.03$ & $7.2 \pm 0.6$ & $-0.4$ & 0.0758 & -- & [13] \\ 
19abeyvoi & 23:50:15.80 & +08:07:05.3 & AT 2019lbr & 58675.45 & $19.09\pm0.04$ & $9.2 \pm 0.8$ & $-0.4$ & -- & -- & [14,15] \\ 
19abfarpa & 11:07:09.56 & +57:06:03.2 & AT 2019kyw & 58676.18 & $18.28\pm0.04$ & $11.9 \pm 0.5$ & $-0.2$ & 0.074 & -- & -- \\ 
19abobxik$^M$ & 00:43:43.12 & +37:03:38.9 & SN 2019myn & 58706.45 & $18.84\pm0.02$ & $9.5 \pm 0.8$ & $-0.1$ & 0.1 & Ibn & -- \\ 
19abrpfps & 18:36:27.30 & +45:05:32.0 & AT 2019aajv & 58720.22 & $19.48\pm0.03$ & $3.4 \pm 0.5$ & $-0.4$ & -- & -- & -- \\ 
19abuvqgw & 19:50:06.37 & +66:04:56.5 & SN 2019php & 58730.30 & $18.68\pm0.06$ & $8.4 \pm 0.4$ & $-0.2$ & 0.087 & Ibn & [16] \\ 
19abyjzvd$^M$ & 16:48:12.90 & +48:04:50.0 & SN 2019qav & 58739.13 & $18.99\pm0.06$ & $10.8 \pm 0.5$ & $-0.3$ & 0.1353 & IIn/Ibn &[17]\\ 
19acaxbjt & 23:12:35.94 & +09:02:07.9 & AT 2019qwx & 58754.20 & $19.03\pm0.04$ & $9.9 \pm 0.7$ & $-0.3$ & -- & -- & [18] \\ 
19acayojs & 21:22:41.87 & +22:52:54.8 & SN 2019rii & 58757.18 & $18.75\pm0.02$ & $10.0 \pm 0.4$ & $-0.1$ & 0.1234 & Ibn & [19] \\ 
19accjfgv & 08:28:49.30 & +75:19:41.0 & SN 2019rta & 58759.43 & $17.88\pm0.02$ & $6.8 \pm 0.4$ & $-0.1$ & 0.027 & IIb & [20] \\ 
19accxzsc & 03:26:14.73 & +04:47:26.7 & AT 2019scr & 58763.42 & $18.91\pm0.05$ & $3.5 \pm 1.5$ & $-0.7\dag$ & -- & -- & [21] \\ 
19acsakuv & 06:21:15.36 & +53:16:39.5 & AT 2019van & 58800.55 & $18.54\pm0.11$ & $6.4 \pm 1.7$ & $-0.4$ & -- & -- & [22] \\ 
20aaelulu$^m$ & 12:22:54.92 & +15:49:25.0 & SN 2020oi & 58862.48 & $14.06\pm0.12$ & $11.0 \pm 0.6$ & $0.1$ & 0.0052 & Ic & [23--25] \\ 
20aahfqpm & 13:06:25.19 & +53:28:45.5 & SN 2020ano & 58871.45 & $19.06\pm0.03$ & $3.4 \pm 2.0$ & $-0.5$ & 0.03113 & IIb & -- \\ 
20aaivtof & 02:48:18.49 & -09:26:52.8 & AT 2020bdh & 58875.16 & $18.60\pm0.03$ & $8.9 \pm 1.4$ & $-0.1$ & 0.04106 & IIn? & [26,27] \\ 
20aakypiu & 11:31:13.75 & +34:30:00.7 & AT 2020bot & 58880.45 & $19.46\pm0.04$ & $3.7 \pm 0.4$ & $-0.1$ & 0.197 & UB$^\ddag$ & [28] \\ 
20aaxhzhc & 13:36:05.01 & +28:59:00.1 & SN 2020ikq & 58971.30 & $18.27\pm0.03$ & $11.8 \pm 1.8$ & $-0.2$ & 0.042 & IIb & [29,30] \\ 
20aayrobw & 09:31:13.19 & +38:15:14.4 & SN 2020jmb & 58981.17 & $18.51\pm0.03$ & $10.0 \pm 0.4$ & $-0.5$ & 0.061 & II & [31] \\ 
20aazchcq & 14:41:40.57 & +19:20:56.9 & SN 2020jji & 58979.25 & $19.50\pm0.09$ & $10.8 \pm 1.0$ & $0.4$ & 0.03788 & II & [32] \\ 
20aazrcbp & 11:02:20.89 & +30:51:52.1 & AT 2020mlq & 58986.21 & $19.71\pm0.06$ & $11.0 \pm 0.8$ & $0.0$ & -- & -- & [33] \\ 
20ababxjv & 16:28:39.48 & +56:13:40.6 & AT 2020kfw & 58991.33 & $19.05\pm0.03$ & $8.4 \pm 0.3$ & $-0.2$ & 0.059 & -- & [34] \\ 
%%20abjbgjj & 23:50:14.27 & +10:07:41.3 & SN 2020ntt & 59033.45 & $18.61\pm0.09$ & $13.2 \pm 1.3$ & $0.3$ & 0.074 & II & [13] \\
20abmocba & 16:34:38.89 & +50:59:26.5 & AT 2020aexw & 59051.26 & $19.39\pm0.03$ & $10.5 \pm 0.4$ & $-0.2$ & 0.0734 & -- & -- \\ 
20abummyz & 16:50:45.92 & +30:45:14.9 & AT 2020yqt & 59080.21 & $19.17\pm0.11$ & $4.0 \pm 1.0$ & $-0.6$ & 0.0986 & Feat. & [35] \\ 
20aburywx$^M$ & 01:19:56.51 & +38:11:09.5 & SN 2020rsc & 59081.47 & $19.36\pm0.07$ & $3.3 \pm 0.3$ & $-0.2$ & 0.0313 & IIb & [36] \\ 
20acigmel$^m$ & 22:20:02.02 & -02:50:25.3 & AT 2020xnd & 59136.21 & $19.24\pm0.04$ & $5.6 \pm 1.6$ & $-0.4$ & 0.2442 & UB;RL & [37] \\ 
20acigusw & 22:50:25.37 & +08:50:41.8 & SN 2020vyv & 59134.23 & $18.68\pm0.03$ & $5.6 \pm 0.3$ & $-0.3$ & 0.062 & II? & [38] \\ 
20aclfmwn & 08:17:11.29 & +64:31:34.7 & SN 2020xlt & 59141.45 & $19.59\pm0.04$ & $4.0 \pm 0.5$ & $-0.2$ & 0.0384 & IIb & -- 
\enddata 
\tablerefs{Published or classified by 
[1] \citet{Karamehmetoglu2021}, 
%[2] \cite{Ho2021_ZTF18aakuewf},
[2] \citet{Yao2020},
%[4] \citet{Costantin2018}, 
%[5] \citet{Burke2018_ZTF18abukavn},
[3] \citet{Ho2019gep},
[4] \citet{Pritchard2021},
[5] \citet{Tonry_ZTF18abvkmgw},
[6] \citet{Fremling_ZTF18abvkmgw},
[7] \citet{Ho2020b},
[8] \citet{Prentice2020_SN2018gjx},
[9] \citet{Nordin2019_ZTF19aankdan},
[10] \citet{Nordin2019_ZTF19aapfmki},
[11] \citet{Prentice2019_ZTF19aapfmki},
[12] \citet{Pellegrino2022},
[13] \citet{Nordin2019_ZTF19aatoboa},
[14] \citet{Tonry_ZTF19abeyvoi},
[15] \citet{Wiseman2019},
%[9] \citet{Gromadzki2018_ZTF18abwkrbl},
[16] \citet{Tonry2019_ZTF19abuvqgw},
[17] \citet{Chambers2019_ZTF19abyjzvd},
[18] \citet{Forster2019},
[19] \citet{Forster2021},
%[11] \citet{Ho2021_Ibn_classifications},
%[12] \citet{Ho2021_ZTF19abyjzvd},
[20] \citet{Dahiwale2019_ZTF19accjfgv},
[21] \citet{Tonry_ZTF19accxzsc},
[22] \citet{Nordin2019_ZTF19acsakuv}
%[14] \citet{Siebert2020_ZTF20aaelulu},
[23] \citet{Horesh2020},
[24] \citet{Rho2021_ZTF20aaelulu},
[25] \citet{Gagliano2022},
%[17] \citet{Ho2021_ZTF20aahfqpm},
[26] \citet{Smith2020_ZTF20aaivtof},
[27] \citet{Forster2020_ZTF20aaivtof},
[28] \citet{Nordin2020_AT2020bot},
[29] \citet{Tonry2020_ZTF20aaxhzhc},
[30] \citet{Angus2020},
[31] \citet{Dahiwale2020_ZTF20aayrobw},
[32] \citet{De2020_ZTF20aayrobw_ZTF20aazchcq},
[33] \citet{Marques-Chaves2020},
[34] \citet{Forster2020_ZTF20ababxjv},
[35] \citet{Chambers2020_ZTF20abummyz},
[36] \citet{Forster2020_ZTF20aburywx},
[37] \citet{Perley2021},
[38] \citet{Siebert2020}
% [20] \citet{Ho2021_IIb},
% [21] \citet{Ho2022_ZTF20aburywx},
% [24] \citet{Ho2022_ZTF20aclfmwn},
% [36] \citet{Ho2021_ZTF20abmocba},
% [39] \citet{Fremling2020_ZTF20acigusw},
% [40] \citet{Forster2020_ZTF20aclfmwn},
% [41] \citet{Tonry2018_ZTF18aakuewf},
% [42] \citet{Nordin2019_SN2019dge}
}
%[11] \citet{Perley2019cow}, 
\tablenotetext{\dag}{The closest $r$-band detection was 2\,d from the peak $g$-band measurement; for all other sources in the table, the $g$ and $r$ measurements were within 1\,d of each other.}
\tablenotetext{\dag\dag}{Feat.: high-S/N featureless spectrum at peak light. RL: radio loud.}
\tablenotetext{\ddag}{UB: unidentified broad features in spectrum.}
\tablenotetext{\ddag\ddag}{For a detailed review of the SN classifications in this table, see \citet{GalYam2017} and \citet{Smith2017}. Briefly, Type~II SNe have hydrogen P-Cygni features and Type~Ib and Ic SNe lack hydrogen and helium, respectively. Type~Ic-BL SNe have particularly broad P-Cygni features indicative of large ejecta velocities ($v>20,000\,$km\,s$^{-1}$). Type~Ibn and Type~IIn SNe have spectra dominated by emission lines of helium and hydrogen, respectively.}
\tablenotetext{m}{Previously published multiwavelength (UV, X-ray, and/or radio) observations available.}
\tablenotetext{M}{Multiwavelength observations published as part of this paper.}
\end{deluxetable*}  

\edit2{The \ncand\ ZTF transients with $1\,\days < t_{1/2,g} < 12\,\days$ and well-sampled light curves are listed in Table~\ref{tab:sources_all}. Most were identified in the HC and 1DC surveys in real time by filters explicitly designed to find rapidly evolving transients \citep{Ho2020d,Perley2021},
and the details of their discovery and follow-up are provided in Section~\ref{sec:discovery-details} in the Appendix.
Several of the objects in Table~\ref{tab:sources_all} have been previously published: the Type~Ibn SN\,2018bcc \citep{Karamehmetoglu2021},
the ultra-stripped Type~Ib candidate SN\,2019dge \citep{Yao2020},
the Type~Ic-BL SN\,2018gep \citep{Ho2019gep},
and the radio-loud transients AT\,2018lug (the ``Koala;'' \citealt{Ho2020b}) and AT\,2020xnd (the ``Camel;'' \citealt{Perley2021}).
}

\edit1{Our requirement of multi-band photometry (Step 4) excludes some known rapidly evolving transients, including AT2018cow itself (which was only observed by ZTF in the $r$-band filter for the first month), the Type~IIb SN\,2018jak \citep{Perley2020_BTS}, and the Type~Ibn SN\,2019aajs (Kool et al. in prep). In addition,
two rapidly evolving Type~Icn SNe, SN\,2019hgp \citep{Bruch2019_Icn,GalYam2021_Icn} and SN\,2021csp \citep{Perley2021_Icn_disc,Perley2021_Icn_class},
do not pass our cuts: SN\,2019hgp had too long a duration, and SN\,2021csp occurred outside the date range we considered. Our goal is to provide a systematically selected sample of well-observed objects, not a fully complete sample of rapidly evolving transients in ZTF.}

\begin{figure*}[p!]
    \centering
    \includegraphics[width=0.9\textwidth]{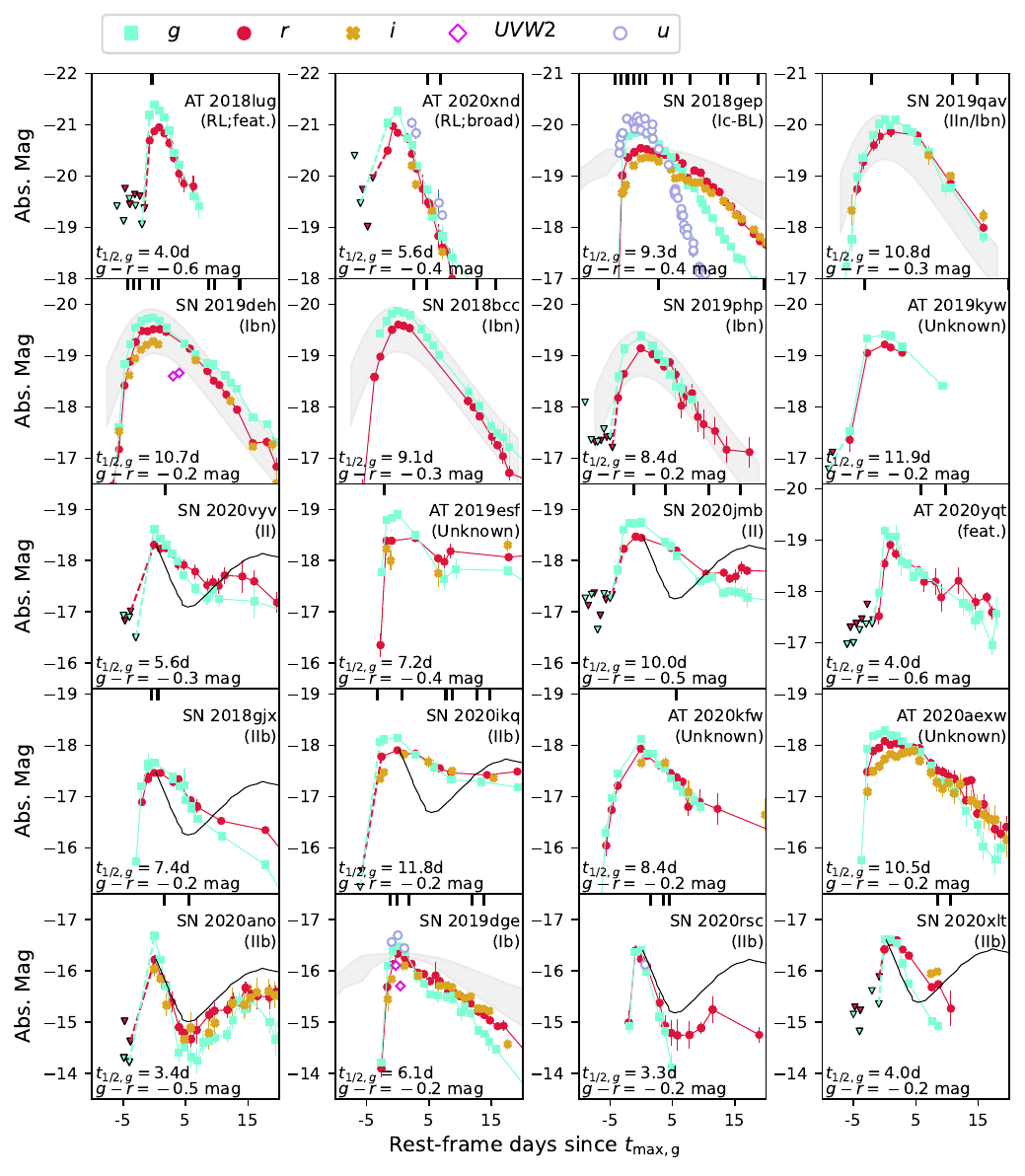}
    \caption{\edit2{Light curves of ZTF FBOTs with redshift measurements. FBOTs were selected on the basis of duration above half-maximum light and peak color: $1\,\days<t_{1/2}<12\,\days$ and $g-r\leq-0.2\,$mag.} Upper limits are indicated with triangles, and dashed lines connect non-detections to detections. Epochs of spectroscopy are indicated with vertical lines along the top of each panel.
    The contrast in each panel is 4\,mag along the $y$-axis and 30\,d along the $x$-axis.
    In panels with Type~Ibc and Type~Ibn SNe we show Type~Ibc and Type~Ibn light curve templates \citep{Drout2011,Hosseinzadeh2017} for reference, scaled to the peak of the light curve.
    The grey region of the Type~Ibc template is the standard. deviation of a set of interpolated light curves \citep{Drout2011}.
    The grey region of the Type~Ibn template contains 95\% of the photometric points in a light-curve sample \citep{Hosseinzadeh2017}.
    For the Type~II and Type~IIb SNe we show the \edit2{V-band} light curve of the Type~IIb SN\,1993J for reference \citep{Schmidt1993}.
    %\edit1{For the Type~Ibn (or IIn/Ibn) SNe we show a Ibn light-curve template for reference \citep{Hosseinzadeh2017}.}
    \edit1{In some cases light curves have been binned by day for clarity.}
    \edit2{Light curves of all remaining transients from Table~\ref{tab:sources_all} are shown in Figure~\ref{fig:lc-remaining-gold} in the Appendix.}
    }
    \label{fig:lc-representative}
\end{figure*}

\edit2{Previous searches (e.g., \citealt{Drout2014}) have found that a $1\,\days < t_{1/2} < 12\,$d cut primarily (but not exclusively) selects events with blue colors ($g-r\lesssim -0.2\,$mag) at peak light, leading to the term FBOT (e.g., \citealt{Margutti2019,Inserra2019}). 
To estimate the peak $g-r$ color of the ZTF transients in Table~\ref{tab:sources_all}, we used the $r$-band magnitude closest to the peak of the $g$-band light curve, which was within 1\,\days\ in all cases but one.
As shown in Table~\ref{tab:sources_all}, similarly to \citet{Drout2014} we find that most (\nfbot\ of the \ncand) objects have $g-r\lesssim -0.2\,$mag at peak light.
The remaining 10 objects have redder colors at peak and would not be referred to as FBOTs, such as the Type~Ic SN\,2020ano.
However, a strict peak-light color criterion will exclude FBOTs that are dust-extinguished.
Therefore, in this paper we present data for all \ncand\ events in Table~\ref{tab:sources_all},
but limit our use of the term FBOT to the \nfbot\ events with blue colors at peak light, which more closely resemble transients referred to as FBOTs in the literature.
\edit2{Light curves for the ZTF FBOTs with redshift measurements are shown in Figure~\ref{fig:lc-representative}. Light curves for all remaining objects in Table~\ref{tab:sources_all} are shown in Figure~\ref{fig:lc-remaining-gold} in Appendix~\ref{sec:appendix-lc}.}
}

\subsection{Literature Sample Selection}

%\startlongtable
\begin{deluxetable*}{lrrrrrrr}
\tablecaption{Literature FBOTs with redshift measurements and well-sampled light curves.\label{tab:literature}} 
\tablehead{
\colhead{Name} & 
\colhead{Redshift} & 
\colhead{Class} & 
\colhead{Filter} & 
\colhead{$M_{\mathrm{max}}^a$} & 
\colhead{$t_{1/2,\mathrm{rise}}^b$} & 
\colhead{$t_{1/2,\mathrm{fade}}^b$} & 
\colhead{Ref.}
\\[-0.3cm] & &  & & \colhead{(mag)} & \colhead{(days)} & \colhead{(days)} & } 
\startdata
SNLS04D4ec$^{c}$ & 0.593 & Unknown & $i$ & $-20.26\pm0.03$ & $<3.81$ & $8.60\pm0.43$ & [1] \\
PTF09uj$^{d}$ & 0.065 & IIn & $r$ & $-19.09\pm0.04$ & $2.04\pm0.76$ & $5.05\pm1.92$ & [2] \\
PS1-10ah$^{e}$ & 0.074 & Unknown & $g$ & $-17.5\pm0.11$ & $1.0\pm0.1$ & $6.3\pm0.6$ & [3] \\
PS1-10bjp & 0.113 & Unknown & $g$ & $-18.2\pm0.11$ & $1.0\pm0.1$ & $7.7\pm0.6$ & [3] \\
PS1-11qr & 0.324 & Unknown & $r$ & $-19.3\pm0.08$ & $2.9\pm0.1$ & $8.7\pm0.4$ & [3] \\
PS1-12bb & 0.101 & Unknown & $g$ & $-16.97\pm0.12$ &  $<1.8$ &  $6.3\pm0.3$ & [3] \\
PS1-12bv & 0.405 & Unknown & $r$ & $-19.1\pm0.11$ &  $<2.2$ &  3--9 & [3] \\
PS1-12brf & 0.275 & Unknown & $r$ & $-18.3\pm0.08$ &  $<1.0$ & $8.8\pm0.6$ & [3] \\
PS1-13dwm & 0.245 & Unknown & $r$ & $-17.5\pm0.13$ & $<3.0$ & 3--7 & [3] \\
PTF12ldy$^f$ & 0.106 & Ibn & $R$ & $-19.20\pm0.02$ & $3.34\pm0.17$ & $7.57\pm0.29$ & [4] \\
LSQ13ccw & 0.0603 & IIn/Ibn? & $g$ & $-18.4\pm0.2$ & $1.39\pm0.10$ & $3.86\pm0.31$ & [5] \\
iPTF14aki & 0.064 & Ibn & $R$ & $-19.30\pm0.03$ & $3.34\pm0.17$ & $7.58\pm0.30$ & [4] \\
iPTF15akq & 0.109 & Ibn & $R,r$ & $-18.62\pm0.31$ & $3.13\pm0.61$ & $8.86\pm0.80$ & [4] \\
iPTF15ul & 0.066 & Ibn? & $g$ & $-21.2 \pm 0.3$ & $1.53\pm0.05$ & $3.72\pm0.08$ & [4] \\
KSN2015K$^d$ & 0.090 & Unknown & \emph{Kepler} clear & $-18.78$ & $1.15$ & $5.54$ & [6] \\
DES16E2pv$^d$ & 0.73 & Unknown & $i$ & $-19.98 \pm 0.65$ & $0.71\pm0.42$ & $2.04\pm1.54$ & [7]\\
DES15S1fli & 0.45 & Unknown & $r$ & $-19.62 \pm 0.11$ & $<3.39$ & $8.60\pm1.42$ & [7]\\
DES17X3cds & 0.49 & Unknown & $i$ & $-19.09 \pm 0.06$ & 3.20--5.42 & $5.51\pm0.83$ & [7]\\
DES16C2ggt & 0.31 & Unknown & $r$ & $-18.12 \pm 0.08$ & $<2.93$ & $6.32\pm1.55$ & [7]\\
DES16C1cbd & 0.54 & Unknown & $i$ & $-19.38 \pm 0.10$ & $1.76\pm0.27$ & $5.84\pm0.77$ & [7]\\
DES13X3gmd & 0.78 & Unknown & $i$ & $-19.25 \pm 0.22$ & $<3.75$ & $7.37\pm4.14$ & [7]\\
DES14S2pli & 0.35 & Unknown & $r$ & $-18.64 \pm 0.06$ & $<3.83$ & $7.16\pm1.63$ & [7]\\
DES13X3gms & 0.65 & Unknown & $i$ & $-19.47 \pm 0.06$ & 1.85--6.22 & $9.94\pm1.76$ & [7]\\
DES15C3mgq & 0.23 & Unknown & $r$ & $-16.92 \pm 0.06$ & $<2.65$ & $8.37\pm0.35$ & [7]\\
DES17C3gen & 0.92 & Unknown & $z$ & $-19.55 \pm 0.22$ & 1.88--4.02 & $5.59\pm2.59$ & [7]\\
DES14C3tnz & 0.70 & Unknown & $i$ & $-19.16 \pm 0.16$ & 2.40--4.46 & $5.51\pm2.72$ & [7]\\
DES15E2nqh & 0.52 & Unknown & $i$ & $-19.22 \pm 0.24$ & 3.04--7.34 & $5.86\pm2.17$ & [7]\\
DES17S2fee & 0.24 & Unknown & $r$ & $-17.98 \pm 0.07$ & $<3.27$ & $5.87\pm1.72$ & [7]\\
DES16X3cxn & 0.58 & Unknown & $i$ & $-19.37 \pm 0.06$ & 2.80--5.93 & $6.62\pm0.42$ & [7]\\
DES16X1eho & 0.81 & Unknown & $z$ & $-21.02 \pm 0.14$ & 1.23--2.41 & $1.33\pm0.26$ & [7]\\
DES13X1hav & 0.58 & Unknown & $i$ & $-19.57 \pm 0.21$ & $<1.63$ & $5.90\pm3.11$ & [7]\\
iPTF16asu$^c$ & 0.187 & Ic-BL & $g$ & $-20.3\pm0.1$ & $1.14\pm0.13$ & $10.62\pm0.55$ & [8] \\
AT2018cow & 0.0141 & IIn/Ibn?RL$^{g}$ & $g$ & $-20.87\pm0.05$ & $1.10\pm0.04$ & $1.96\pm0.12$ & [9,10] \\
SN\,2018kzr$^h$ & 0.054 & Ic & $g$ & $-18.80\pm0.08$ & $<2.0$ & $1.6\pm0.2$ & [11] \\
SN\,2019bkc$^i$ & 0.0209 & Ic & $g$ & $-17.16\pm0.03$ & $5.28\pm0.38$ & $2.22\pm0.10$ & [12,13] \\
AT2018lqh & 0.05446 & Unknown & $g$ & $-16.96\pm0.18$ & $0.61\pm0.06$ & $1.54\pm0.23$ & [16] \\
SN\,2019ehk & 0.00524 & Ca-rich/IIb & $g$ & $-14.70\pm0.02$ & $1.31\pm0.04$ & $1.93\pm0.06$ & [15,24] \\
SN\,2019aajs & 0.0358 & Ibn & $g$ & $-18.86\pm0.03$ & $2.11\pm0.03$ & $6.05\pm0.24$ & [14] \\
SN\,2019rsq & 0.031 & IIb & $g$ & $-16.47\pm0.09$ & $4.7\pm0.4$ & $4.4\pm0.1$ & [20] \\
AT2020mrf & 0.1353 & UB;RL$^j$ & $g$ & $-20.0\pm0.1$ & $2.4\pm0.2$ & $4.8\pm0.2$ & [17] \\
SN\,1999cq & 0.0263 & Ic & $R$ & $-19.6$ & $<3.95$ & $9.6$ & [18] \\
HSC17bhyl & 0.750 & Unknown & $i$ & $-18.49\pm0.04$ & $2.38\pm0.31$ & $7.15\pm1.95$ & [19] \\
SN\,1885A & (In M31) & Unknown & $V$ & $-18.4\pm0.4$ & $\lesssim5$ & $\lesssim 5$ & [21,22] \\
SN\,2005ek & 0.01662 & Ic & $R$ & $-17.26\pm0.15$ & 2--4 & 5 & [23] \\
SN\,2021csp & 0.084 & Icn & $g$ & $-20.1$ & $2.5\pm0.5$ & $8.3\pm1.0$ & [25] \\
SN\,2019jc & 0.01948 & Icn & $g$ & $-17.2\pm0.1$ & $2.6\pm0.2$ & $3.1\pm0.1$ & [26] \\
SN\,2021ckj & 0.143 & Icn & $g$ & $-19.9\pm0.1$ & $3.0\pm0.5$ & $4.7\pm0.2$ & [26] 
\enddata
\tablenotetext{a}{Corrected for Galactic extinction, assuming zero host extinction in all cases except iPTF15ul.}
\tablenotetext{b}{Rest frame, measured using the light curve that most closely matches rest-frame $g$ band.}
\tablenotetext{c}{Measurements are from \citet{Ho2020b}.}
\tablenotetext{d}{Rise times, fade times, and peak luminosities (with approximate $K$-correction) calculated as part of this paper. DES light curves were provided by M. Pursiainen.}
\tablenotetext{e}{Luminosity and timescale measurements from Tab.~1 and Tab.~4 of \citet{Drout2014}, taking $K$-corrected values.}
\tablenotetext{f}{Peak magnitudes from Table~4 of \citet{Hosseinzadeh2017}. Rise and fade times calculated as part of this paper.}
\tablenotetext{g}{Strict spectroscopic definition based on presence of H and He emission features in optical spectrum. RL: radio loud.}
\tablenotetext{h}{Timescales calculated using $g$-band as well as one ATLAS $o$-band upper limit prior to peak.}
\tablenotetext{i}{Photometry from \citet{Chen2020}, with a ZTF datapoint added.}
\tablenotetext{j}{Radio loud, spectra mostly featureless, with an unidentified very broad feature.}
\tablereferences{
[1] \citet{Arcavi2016}, 
[2] \citet{Ofek2010}, 
[3] \citet{Drout2014}, 
[4] \citet{Hosseinzadeh2017}, 
[5] \citet{Pastorello2015}, 
[6] \citet{Rest2018}, 
[7] \citet{Pursiainen2018},
[8] \citet{Whitesides2017},
[9] \citet{Prentice2018},
[10] \citet{Perley2019cow}
[11] \citet{McBrien2019}, 
[12] \citet{Prentice2020}, 
[13] \citet{Chen2020},
[14] Kool et al. in prep,
[15] \citet{JacobsonGalan2020},
[16] \citet{Ofek2021},
[17] \citet{Yao2022},
[18] \citet{Matheson2000},
[19] \citet{Tampo2020},
[20] \citet{Perley2020_BTS},
[21] \citet{deVaucouleurs1985},
[22] \citet{Perets2011},
[23] \citet{Drout2013},
[24] \citet{De2021},
[25] \citet{Perley2022_SN2021csp},
[26] \citet{Pellegrino2022}
 %[16] \citet{Vinko2015}
}
\end{deluxetable*}

\begin{figure*}[htb!]
    \centering
    \includegraphics[width=\textwidth]{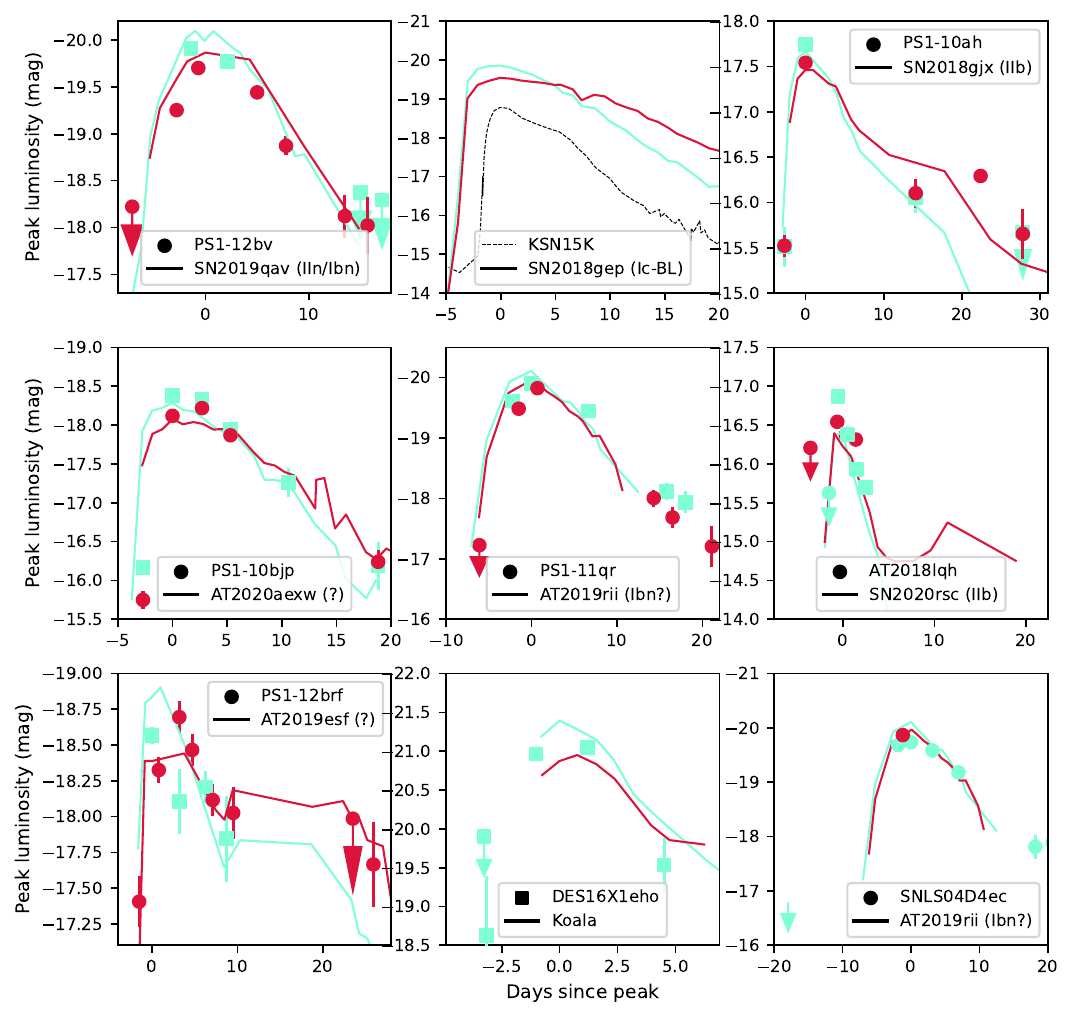}
    \caption{Light curves of FBOTs from the literature comparison sample, with the light curves of ZTF transients shown for comparison.
    The light curves of PS1 events are from \citet{Drout2014}. The light curve of DES16X1eho is from M. Pursiainen, private communication. The light curve of KSN2015K is from \citet{Rest2018}.
    The light curve of AT2018lqh is from \citet{Ofek2021}. Literature light curves were taken to be as close to $g$ (shown in cyan) and $r$ (shown in red) band in the rest-frame as possible, for a more direct comparison with the ZTF light curves.}
    \label{fig:lc-comparison}
\end{figure*}

We supplement the ZTF-selected events from Section~\ref{sec:ztf-sample-selection} with objects from the literature. The literature transients are listed in Table~\ref{tab:literature}, and a subset of their light curves are shown in Figure~\ref{fig:lc-comparison}.
To be consistent, we apply similar selection criteria: we require $1\,\days < t_{1/2} < 12\,$d in a filter as close to rest-frame $g$-band as possible, an observation in that filter 5.5\,d before the peak, and an observation in that filter within 5.5\,d after the peak.
We also require a redshift measurement.
We measured the duration in as close to rest-frame $g$-band as possible, since a number of the literature objects are at a significantly higher redshift than the ZTF objects. We estimate the peak absolute magnitude using

\begin{equation}
    M=m_\mathrm{obs}-5\,\log_{10}\left( \frac{D_L}{10\,\mathrm{pc}} \right) + 2.5\,\log_{10} (1+z).
    \label{eq:kcorrect}
\end{equation}

\edit2{
Table~\ref{tab:literature} includes the radio- and X-ray- loud transients AT2018cow \citep{Prentice2018} and AT2020mrf \citep{Yao2022}, as well as the PS1 \citep{Drout2014} and DES \citep{Pursiainen2018} samples, which are widely discussed as FBOTs in the literature \citep{Inserra2019,Margutti2019,Fox2019,Coppejans2020,Lyutikov2022}.
We include KSN2015K \citep{Rest2018}, SNLS04D4ec \citep{Arcavi2016}, and iPTF16asu \citep{Whitesides2017}, all of which have been referred to as FBOTs \citep{Inserra2019,Margutti2019,Coppejans2020}; as well as SN\,2019bkc \citep{Chen2020,Prentice2020}, described as an FBOT in \citet{Inserra2019} and \citet{Margutti2019}.
We include several interacting (Type~IIn/Ibn) SNe \citep{Ofek2010,Hosseinzadeh2017,Pastorello2015} whose similarity to FBOTs in terms of light curves and colors have been pointed out by others \citep{Fox2019,Margutti2019}.
We also include the Type~Ic SNe SN\,2018kzr \citep{McBrien2019}, SN\,1999cq \citep{Matheson2000}, and SN\,2005ek \citep{Drout2013}, which have been described as FBOTs \citep{Pursiainen2018,Coppejans2020,Wiseman2020,Chen2022}.
We include AT2018lqh \citep{Ofek2021} and HSC17bhyl \citep{Tampo2020}, described as FBOTs in \citet{Lyutikov2022} and \citet{Coppejans2020}.
We also include the fast transient SN\,1885A \citep{deVaucouleurs1985,Perets2011}.
In addition, the Calcium-rich transient SN\,2019ehk \citep{JacobsonGalan2020}, which has been referred to as Type~IIb; \citealt{De2021}, passes the selection criteria laid out in \S\ref{sec:ztf-sample-selection} because the shock-cooling peak is significantly brighter than the peak of the radioactively powered light curve.
}

Several objects referred to as FBOTs in the literature do not meet our $t_{1/2}<12\,$d criterion:
``Dougie'' \citep{Vinko2015,Inserra2019} and SN\,2015U \citep{Shivvers2016,Margutti2019} have too long of a duration.
In addition, we do not include
SN\,2002bj \citep{Poznanski2010,Margutti2019} because it did not have a sufficiently well-sampled peak; it was only detected on the decline.
The events from \citet{Tanaka2016} were fast-rising, luminous, and blue \citep{Fox2019,Inserra2019,Margutti2019} but were only detected during the rise phase---we therefore do not include them either.
The remaining four events from \citet{Tampo2020} either do not have a spectroscopic redshift measurement or do not have a sufficiently well sampled light curve for inclusion.

%. The exceptions are the fast transient ``Dougie'' \citep{Vinko2015} and the three fast transients linked to the Ca-rich family:  SN\,2002bj \citep{Poznanski2010}, and SN\,2005ek \citep{Drout2013}. These four objects either have durations slightly too long to meet our criteria (as in the case of Dougie and SN\,2010X) or have light curves that are not sufficiently well sampled.
%The majority of the objects in Table~\ref{tab:literature} have blue colors at peak light, with the exception of PS1-12bb ($g-r\approx0$; \citealt{Drout2014}). The Type~Ibn SNe iPTF14aki and iPTF15akq also had roughly flat colors near peak \citep{Hosseinzadeh2017}.

\subsection{Optical Photometry}
\label{sec:opt-phot}

We performed forced photometry on P48 images for all events using the pipeline developed by F. Masci and R. Laher\footnote{\url{http://web.ipac.caltech.edu/staff/fmasci/ztf/forcedphot.pdf}},
with the following additional steps:

\begin{enumerate}
    \item We removed data taken in bad observing conditions by discarding observations with $\texttt{scisigpix}$, $\texttt{zpmaginpsci}$, or $\texttt{zpmaginpscirms}$ exceeding five times the median of that value for the light curve.
    \item We removed observations with flux values or chisq from the difference image recorded as NaN.
    \item Following \citet{Yao2019}, we grouped observations by \texttt{fcqfID}, a combination of field ID, CCD ID, quadrant ID, and filter ID.
    \item For each group of \texttt{fcqfID},
    we checked whether the stack of images used to construct a reference image could overlap with images of the target, by seeing whether the final reference image was within 15\,days of the first ZTF alert issued. If so, we considered the data to be contaminated by the reference.
    \item If the data were contaminated by the reference\footnote{In principle a baseline should be subtracted for all events, but it has been found that this correction is very small, only $<0.1\%$ of transient flux values.} (as defined in the previous bullet), we checked whether there were sufficient (at least 30) images to subtract a baseline flux value.
    We obtained images prior to 15 days before the first detection, and after 100 days after the last detection.
    If there were at least 30 such images, we calculated the median \edit2{flux value} of all the baseline detections,
    %\edit2{We removed baseline data points with flux values exceeding 3 }
    %, rejected outliers greater than 3 times the median from the median, then re-calculated the median, 
    and subtracted that baseline value from the observations.
    If there were not sufficient baseline measurements, we excluded the observations.
    \item Following the Masci \& Laher documentation, we validated and rescaled the uncertainties on the flux values.
    \item \edit1{Following the Masci \& Laher documentation,} points with a S/N ratio greater than 3 were regarded as detections, and converted to magnitudes. Points with a lower S/N were regarded as upper limits, and reported as 5-$\sigma$.
    \item We corrected for Milky Way extinction \citep{Schlafly2011} using the \texttt{extinction} package\footnote{\url{https://github.com/kbarbary/extinction}} with $R_V=3.1$ and a \citet{Fitzpatrick1999} extinction law.
\end{enumerate}

When available, we added photometry obtained with other facilities:
the IO:O on the Liverpool Telescope (LT; \citealt{Steele2004}) and the Rainbow Camera on the automated 60-inch telescope at Palomar Observatory (P60; \citealt{Cenko2006}).
LT image reduction was provided by the basic IO:O pipeline. P60 and LT image subtraction
were performed following \citet{Fremling2016},
using PS1 images for $griz$ and SDSS for $u$ band.
The final combined light curves are provided in Table~\ref{tab:ztf-lc}. We discard points that are $>$20\,d after the last alert, or $<10\,$d before the first alert. Then, we discard points $>5\,$d after the last forced photometry detection or $<5$\,d before the first forced photometry detection.
\edit2{The ZTF FBOT light curves with redshift measurements are shown in Figure~\ref{fig:lc-representative};
the remainder are shown in the Appendix in Figure~\ref{fig:lc-remaining-gold}.}

\begin{deluxetable}{lrrrrrrr}[!h]
\tablecaption{Optical photometry of the ZTF transients in Table~\ref{tab:sources_all}.\label{tab:ztf-lc}} 
\tablewidth{0pt} 
\tablehead{ 
\colhead{Name} & \colhead{Filter} & \colhead{JD} & \colhead{Flux} & \colhead{eFlux} & \colhead{Mag} & \colhead{eMag} & \colhead{Tel} \\[-0.3cm]
 & & & \colhead{($\mu$Jy)} & \colhead{($\mu$Jy)} & &  & 
} 
\tabletypesize{\scriptsize} 
\startdata 
... & ... & ... & ... & ... & \\
SN2018bcc & $r$ & 2458224.93 & 12.36 & 4.55 & 20.51 & 99.00 & P48 \\ 
SN2018bcc & $r$ & 2458224.95 & 15.16 & 6.21 & 20.17 & 99.00 & P48 \\ 
SN2018bcc & $r$ & 2458226.84 & 103.84 & 3.71 & 18.86 & 0.04 & P48 \\ 
SN2018bcc & $r$ & 2458226.86 & 110.24 & 3.51 & 18.79 & 0.03 & P48 \\ 
... & ... & ... & ... & ... & \\
\enddata 
\tablenotetext{}{Non-detections are indicated with eMag$=$99. This table is published in its entirety in the machine-readable format. A portion is shown here for guidance regarding its form and content.}
\end{deluxetable} 

\subsection{Optical Spectroscopy}

% The duration-luminosity parameter space of the gold and silver samples is shown in the left panel of Figure~\ref{fig:lum-timescale}.
% \edit2{For context, in the right panel we show the FBOTs superimposed with a large sample of transients that were spectroscopically classified as part of the ZTF Bright Transient Survey \citep{Fremling2020_RCF,Perley2020_BTS}.
% It appears that most FBOTs are part of a continuum of established SN subtypes, particularly Type~IIb and Type~Ibn.
% As can be seen on the right panel, there are several BTS objects with $t_{1/2}<12\,$d that did not pass the criteria in Table~\ref{tab:search}; this is primarily due to the light curve not being sufficiently well-sampled in both $g$ and $r$. The BTS survey used the entire ZTF public 3-day survey, whereas the majority of our objects were discovered in the private high-cadence surveys.
% We emphasize that the objects in the left panel are not a subset of the objects in the right panel.
% }

\edit2{Of the \ncand\ ZTF transients in Table~\ref{tab:sources_all}, 20 were classified using optical spectroscopy, and an additional nine had redshift measurements but no classification. The redshift measurements and spectroscopic classifications are listed in Table~\ref{tab:sources_all}.}

\edit2{Spectra of the transients and their host galaxies were obtained using} a variety of telescopes: the Spectral Energy Distribution Machine (SEDM; \citealt{Blagorodnova2018,Rigault2019}),
the Alhambra Faint Object Spectrograph and Camera (ALFOSC\footnote{\url{http://www.not.iac.es/instruments/alfosc/}}) on the Nordic Optical Telescope (NOT; \citealt{Djupvik2010}),
the Double Beam Spectrograph (DBSP; \citealt{Oke1982}) on the 200-inch Hale telescope at Palomar Observatory,
the Spectrograph for the Rapid Acquisition of Transients (SPRAT; \citealt{Piascik2014}) on LT,
the Low Resolution Imaging Spectrometer (LRIS; \citealt{Oke1995}) on the Keck~I 10-m telescope, Binospec \citep{Fabricant2019} on the MMT,
the Optical System for Imaging and low-Intermediate-Resolution Integrated Spectroscopy (OSIRIS) on the Gran Telescopio Canarias (GTC; \citealt{Cepa2000}), \edit1{the Gemini Multi-object Spectrograph (GMOS; \citealt{Hook2004}) on Gemini North,}
and the Device Optimized for the LOw RESolution (DOLORES) on the Telescopio Nazionale Galileo (TNG).

The SEDM pipeline is described in \citet{Rigault2019}. NOT/ALFOSC spectra were reduced using \texttt{Foscgui}\footnote{Foscgui is a graphic user interface aimed at extracting SN spectroscopy and photometry obtained with FOSC-like instruments. It was developed by E. Cappellaro. A package description can be found at \url{http://sngroup.oapd.inaf.it/foscgui.html.}} The SPRAT pipeline is based on the FrodoSpec pipeline \citep{Barnsley2012}, the P200/DBSP pipeline is described in \citet{Bellm2016}, and the Keck/LRIS pipeline \texttt{Lpipe} is described in \citet{Perley2019lpipe}. Gemini, DOLORES, and GTC spectra were reduced using IRAF routines\footnote{IRAF is distributed by the National Optical Astronomy
Observatories, which is operated by the Association of Universities for
Research in Astronomy, Inc. (AURA) under cooperative agreement with the National
Science Foundation}. MMT/Binospec spectra were reduced using the standard pipeline \citep{Kansky2019}.

\edit1{The spectra of several transients in Table~\ref{tab:sources_all} have already been published: SN\,2018gep \citep{Ho2019gep,Pritchard2021}, SN\,2018gjx \citep{Prentice2020_SN2018gjx}, AT2018lug \citep{Ho2020b}, SN\,2019deh \citep{Pellegrino2022}, SN\,2019dge \citep{Yao2019}, SN\,2020oi \citep{Horesh2020,Rho2021_ZTF20aaelulu,Gagliano2022}, and AT2020xnd \citep{Perley2021}.
We provide a log of non-published}
spectroscopic observations (Appendix~\ref{sec:appendix-spec}, Table~\ref{tab:spec-log}),
and plot the spectroscopic evolution of each object (Appendix~\ref{sec:appendix-spec}, figures~\ref{fig:spec-events-1}, \ref{fig:spec-events-2}, \ref{fig:spec-events-3},
and \ref{fig:spec-events-4}).

\subsection{UV and X-ray Observations}

\edit1{Six} of the transients in Table~\ref{tab:sources_all}
were observed with the UV/optical (UVOT; \citealt{Roming2005}) and X-Ray Telescope (XRT; \citealt{Burrows2005}) on board the Neil Gehrels \emph{Swift} observatory \citep{Gehrels2004}.
The UVOT and XRT observations of three of these transients have not yet been published, and are provided in
Table~\ref{tab:uvot-lc} and Table~\ref{tab:xrt-lc}.\footnote{The published events are
SN\,2019dge \citep{Yao2020},
SN\,2018gep \citep{Ho2019gep}, and SN\,2020oi \citep{Horesh2020}.} \edit1{We also provide UVOT/XRT data for a rapidly evolving Type~Ibn SN\,2019aajs that will be published as part of a ZTF Ibn sample paper (Kool et al. in prep); SN\,2019aajs did not pass our cuts because its rapidly evolving nature was only revealed by non-ZTF photometry}.
None of \edit1{these objects} were detected by XRT.

\startlongtable 
\begin{deluxetable*}{lrrrr} 
\tablecaption{UVOT photometry. Epochs are given with respect to the first ZTF             detection.\label{tab:uvot-lc}} 
\tablewidth{0pt} 
\tablehead{ \colhead{Name (SN)} & \colhead{Date (JD)} & \colhead{$\Delta t$ (d)} & \colhead{Filter} & \colhead{AB Mag} } 
\tabletypesize{\scriptsize} 
\startdata 
2019aajs & 2458547.90 & 8.04 & UVW1 & $ 18.57 \pm 0.08 $ \\ 
2019aajs & 2458547.90 & 8.04 & U & $ 18.04 \pm 0.09 $ \\ 
2019aajs & 2458547.90 & 8.04 & B & $ 17.93 \pm 0.1 $ \\ 
2019aajs & 2458547.90 & 8.04 & UVW2 & $ 19.05 \pm 0.09 $ \\ 
2019aajs & 2458547.90 & 8.04 & V & $ 18.45 \pm 0.23 $ \\ 
2019aajs & 2458547.90 & 8.04 & UVM2 & $ 18.71 \pm 0.07 $ \\ 
2019aajs & 2458551.15 & 11.29 & UVW1 & $ 19.33 \pm 0.11 $ \\ 
2019aajs & 2458551.15 & 11.29 & U & $ 18.60 \pm 0.11 $ \\ 
2019aajs & 2458551.15 & 11.29 & B & $ 18.23 \pm 0.12 $ \\ 
2019aajs & 2458551.16 & 11.29 & UVW2 & $ 19.81 \pm 0.11 $ \\ 
2019aajs & 2458551.16 & 11.30 & V & $ 18.37 \pm 0.22 $ \\ 
2019aajs & 2458551.16 & 11.30 & UVM2 & $ 19.54 \pm 0.09 $ \\ 
2019aajs & 2458553.60 & 13.74 & UVW1 & $ 19.38 \pm 0.11 $ \\ 
2019aajs & 2458553.61 & 13.74 & U & $ 18.81 \pm 0.12 $ \\ 
2019aajs & 2458553.61 & 13.75 & B & $ 18.81 \pm 0.17 $ \\ 
2019aajs & 2458553.61 & 13.75 & UVW2 & $ 20.24 \pm 0.13 $ \\ 
2019aajs & 2458553.61 & 13.75 & V & $ 18.62 \pm 0.25 $ \\ 
2019aajs & 2458553.61 & 13.75 & UVM2 & $ 20.08 \pm 0.12 $ \\ 
2019aajs & 2458569.30 & 29.44 & UVW1 & $ 20.42 \pm 0.19 $ \\ 
2019aajs & 2458569.30 & 29.44 & U & $ 19.81 \pm 0.25 $ \\ 
2019aajs & 2458569.30 & 29.44 & B & $ >20.0$  \\ 
2019aajs & 2458569.30 & 29.44 & UVW2 & $ 21.29 \pm 0.24 $ \\ 
2019aajs & 2458569.30 & 29.44 & V & $ >19.0$  \\ 
2019aajs & 2458569.30 & 29.44 & UVM2 & $ 21.06 \pm 0.19 $ \\ 
2019aajs & 2458575.20 & 35.34 & UVW1 & $ 20.80 \pm 0.23 $ \\ 
2019aajs & 2458575.20 & 35.34 & U & $ 19.92 \pm 0.22 $ \\ 
2019aajs & 2458575.20 & 35.34 & B & $ >20.0$  \\ 
2019aajs & 2458575.21 & 35.34 & UVW2 & $ 21.30 \pm 0.23 $ \\ 
2019aajs & 2458575.21 & 35.35 & V & $ >19.0$  \\ 
2019aajs & 2458575.21 & 35.35 & UVM2 & $ 21.15 \pm 0.2 $ \\ 
2019deh & 2458584.98 & 3.16 & UVW1 & $ 17.55 \pm 0.07 $ \\ 
2019deh & 2458584.98 & 3.16 & U & $ 17.28 \pm 0.07 $ \\ 
2019deh & 2458584.98 & 3.16 & B & $ 17.60 \pm 0.09 $ \\ 
2019deh & 2458584.98 & 3.17 & UVW2 & $ 18.41 \pm 0.08 $ \\ 
2019deh & 2458584.98 & 3.17 & V & $ 17.63 \pm 0.15 $ \\ 
2019deh & 2458584.99 & 3.18 & UVM2 & $ 18.01 \pm 0.06 $ \\ 
2019deh & 2458585.68 & 3.87 & V & $ 17.55 \pm 0.15 $ \\ 
2019deh & 2458585.69 & 3.87 & UVM2 & $ 18.06 \pm 0.07 $ \\ 
2019deh & 2458586.01 & 4.20 & UVW1 & $ 17.62 \pm 0.06 $ \\ 
2019deh & 2458586.01 & 4.20 & U & $ 17.27 \pm 0.07 $ \\ 
2019deh & 2458586.01 & 4.20 & B & $ 17.34 \pm 0.07 $ \\ 
2019deh & 2458586.02 & 4.20 & UVW2 & $ 18.34 \pm 0.08 $ \\ 
2019qav & 2458755.88 & 18.24 & UVW1 & $ 21.59 \pm 0.33 $ \\ 
2019qav & 2458755.88 & 18.24 & U & $ 21.06 \pm 0.49 $ \\ 
2019qav & 2458755.88 & 18.24 & B & $ 21.64 \pm 1.55 $ \\ 
2019qav & 2458755.88 & 18.24 & UVW2 & $ 22.41 \pm 0.33 $ \\ 
2019qav & 2458755.88 & 18.25 & V & $ 20.20 \pm 0.78 $ \\ 
2019qav & 2458755.89 & 18.25 & UVM2 & $ 22.22 \pm 0.27 $ \\ 
2020rsc & 2459087.78 & 6.87 & UVW1 & $ >21.0$  \\ 
2020rsc & 2459092.83 & 11.92 & U & $ >20.0$  \\ 
\enddata 
\end{deluxetable*}

\startlongtable 
\begin{deluxetable*}{lrrrrr} 
\tablecaption{\emph{Swift} XRT observations. Flux is given as unabsorbed flux. Conversions from count rate to flux assume a photon index $\Gamma=2$ and values of $n_H$ are taken from \citet{Willingale2013}. Upper limits are 3-$\sigma$. Epochs are given with respect to the first ZTF detection.\label{tab:xrt-lc}} 
\tablewidth{0pt} 
\tablehead{ \colhead{Name (SN)} & \colhead{Date (JD)} & \colhead{$\Delta t$ (d)} & \colhead{Count Rate ($10^{-3}$\,\psec)} & \colhead{Flux ($10^{-13}$\,\erg\,\pcmsq\,\psec)} & \colhead{Luminosity ($10^{41}$\,\erg\,\psec)} } 
\tabletypesize{\scriptsize} 
\startdata 
2019aajs & 2458547.90 & 8.04 & $<5.98$ & $<2.24$ & $<7.08$ \\ 
2019aajs & 2458551.16 & 11.30 & $<4.52$ & $<1.69$ & $<5.36$ \\ 
2019aajs & 2458553.61 & 13.75 & $<5.78$ & $<2.16$ & $<6.84$ \\ 
2019aajs & 2458569.30 & 29.44 & $<5.39$ & $<2.02$ & $<6.39$ \\ 
2019aajs & 2458575.21 & 35.34 & $<4.02$ & $<1.50$ & $<4.76$ \\ 
2019deh & 2458584.99 & 3.17 & $<5.52$ & $<1.92$ & $<14.56$ \\ 
2019deh & 2458586.01 & 4.20 & $<8.33$ & $<2.89$ & $<21.97$ \\ 
2019qav & 2458755.88 & 18.25 & $<4.05$ & $<1.37$ & $<71.12$ \\ 
2020rsc & 2459087.78 & 6.87 & $<7.08$ & $<2.91$ & $<7.00$ \\ 
2020rsc & 2459092.83 & 11.92 & $<22.96$ & $<9.43$ & $<22.68$ \\ 
\enddata 
\end{deluxetable*}

The brightness in the UVOT filters was measured with UVOT-specific tools in the HEAsoft version 6.26.1. Source counts were extracted from the images using a circular $3^{\prime\prime}$-radius aperture. The background was estimated over a significantly larger area close to the SN position. The count rates were obtained from the images using the \swift\ tool \texttt{uvotsource}. They were converted to magnitudes using the UVOT photometric zeropoints \citep{Breeveld2011} and the UVOT calibration files from September 2020.  All magnitudes were transformed into the AB system using \citet{Breeveld2011}. If the transient was affected by the host, we made use of archival UVOT observations or obtained templates after the SN faded.
XRT data were reduced using the online tool\footnote{\url{https://www.swift.ac.uk/user_objects/}} from the \emph{Swift} team \citep{Evans2007,Evans2009},
using hydrogen column density values from \citet{Willingale2013}.

\subsection{Millimeter and Radio Observations}
\label{sec:radio-obs}

Four transients in Table~\ref{tab:sources_all} have published millimeter and/or radio observations \citep{Yao2020,Ho2019gep,Horesh2020,Ho2020b,Perley2021}.
We observed an additional four objects
with the IRAM Northern Extended Millimeter Array (NOEMA), the Submillimeter Array (SMA),
and the Karl G. Jansky Very Large Array (VLA; \citealt{Perley2011}): \edit1{three from Table~\ref{tab:sources_all} and one (SN\,2019aajs) from Table~\ref{tab:literature}.}
Observations are listed in Table~\ref{tab:radio};
all resulted in non-detections.

\begin{deluxetable*}{lcccccccc}[htb!]
\tablecaption{Millimeter and radio observations. Upper limit given as 3$\times$ the image RMS. Time given since first ZTF detection.
\label{tab:radio}} 
\tablewidth{0pt} 
\tablehead{ 
\colhead{Object Name} & Instrument & Program ID (PI) & \colhead{Start Date} & \colhead{$\Delta t$} & \colhead{$\nu$} &  \colhead{$f_\nu$} &  \colhead{$L_\nu$} \\ 
& & & \colhead{(JD)} & \colhead{(d)} & \colhead{(GHz)} & \colhead{($\mu$Jy)} &  \colhead{(\erg\,\psec\,\phz)}
} 
%\rotate 
\tabletypesize{\scriptsize} 
\startdata 
SN\,2019aajs & SMA & 2018B-S047 (Ho) & 2458564.19 & 24.33 & 230 & $<840$ & $<2.7 \times 10^{28}$ \\
SN\,2019aajs & VLA & 18B-242 (Perley) & 2458563.99 & 24.13 & 10 & $<15$ & $< 4.8\times10^{26}$ \\
SN\,2019myn & VLA & 18B-242 (Perley) & 2458712.78 & 10.83 & 10 & $<16$ & $<4.3\times10^{27}$ \\
SN\,2019qav & NOEMA & S19BC (Ho) & 2458753.20 & 15.56 & 90 & $<90$ & $<4.8\times10^{28}$ \\
SN\,2019qav & VLA & 18B-242 (Perley) & 2458765.42 & 27.78 & 10 & $<18$ & $<9.6\times10^{27}$ \\
SN\,2019qav & VLA & 20A-374 (Ho) & 2458922.12 & 184.48 & 10 & $<27$ & $<1.4\times10^{28}$  \\
SN\,2020rsc & VLA & 20A-374 (Ho) & 2459100.73 & 19.82 & 10 & $<15$ & $<3.6\times10^{26}$ \\
\enddata 
\end{deluxetable*}

We observed SN\,2019aajs, SN\,2019myn, SN\,2019qav, and SN\,2020rsc with the VLA.
Data were calibrated using the automated pipeline available in the Common Astronomy Software Applications (CASA; \citealt{McMullin2007}) with additional flagging applied manually,
then imaged using the CLEAN algorithm \citep{Hogbom1974}.

SN\,2019qav was observed with NOEMA under conditions of excellent atmospheric stability and transparency.
Data calibration and analysis was done within the GILDAS\footnote{\url{http://www.iram.fr/IRAMFR/GILDAS}} software package using CLIC for calibration and MAPPING for $uv$-plane analysis and imaging of the data.
The absolute flux calibration accuracy is estimated to be better than 10\%.
The upper limit reported in Table~\ref{tab:radio} is from combining the two sidebands.

SN\,2019aajs was observed with the SMA in the Extended configuration, using all eight antennas, under excellent conditions.
Both receivers were tuned to local oscillator (LO) frequency of 225.5 GHz.
Data were calibrated in IDL using the Millimeter Interferometer Reduction (MIR) package then exported for additional analysis and imaging using the \emph{Miriad} package \citep{Sault1995}.
No obvious detection was seen in the dirty image, so no CLEANing was attempted.

In addition, we also queried ongoing radio surveys to determine whether any objects had been serendipitously observed. To query the VLA Sky Survey \citep[VLASS;][]{Lacy2020}, which observes at 3\,GHz, we used the same approach as \citet{Ho2020b}. Twenty-seven of the sources in our sample were observed by VLASS, but none are detected. Table~\ref{tab:radio_vlass} lists the sources, the date they were observed and the associated RMS values.

\begin{deluxetable}{lrrrr}[htb!]
\tablecaption{Serendipitous observations of the transients in Table~\ref{tab:sources_all} by the VLA Sky Survey (3\,GHz), the Rapid ASKAP Continuum Survey (888\,MHz), and phase one of the VAST Pilot survey (888\,MHz).\label{tab:radio_vlass}} 
\tablewidth{0pt} 
\tablehead{ \colhead{Name} & \colhead{Survey} & \colhead{MJD} & \colhead{$\Delta t$ [d]} & \colhead{RMS [$\mu$Jy]} } 
\tabletypesize{\scriptsize} 
\startdata 
SN\,2018bcc & RACS & 58595 & 364 &   248 \\
SN\,2019dge & VLASS & 59072 & 488 & 156 \\ 
AT\,2018lwd & VLASS & 59074 & 755 & 141 \\ 
SN\,2018gep & VLASS & 58607 & 232 & 134 \\ 
-- & RACS &  58595 & 221 &   535 \\
SN\,2018ghd & VLASS & 59070 & 692 & 157 \\ 
-- & RACS &  58598 & 221 &   353 \\
AT\,2018lug & VLASS & 58551 & 176 & 133 \\ 
-- & RACS &  58602 & 228 &   332 \\
SN\,2018gjx & VLASS & 58568 & 188 & 171 \\ 
-- & RACS &  58595 & 216 &   257 \\
AT\,2019dcm & VLASS & 58609 & 36 & 110 \\ 
SN\,2019deh & VLASS & 58611 & 23 & 138 \\ 
-- & RACS &  58598 & 11 &    248 \\
AT\,2019aajt & VLASS & 59111 & 525 & 140 \\ 
-- & RACS & 58595 & 10 &    270 \\
AT\,2019lbr & VLASS & 59090 & 414 & 217 \\ 
AT\,2019kyw & VLASS & 59084 & 407 & 124 \\ 
AT\,2019aajv & VLASS & 59074 & 353 & 143 \\ 
SN\,2019php & VLASS & 59084 & 353 & 123 \\ 
SN\,2019qav & VLASS & 59072 & 332 & 166 \\ 
AT\,2019qwx & VLASS & 59091 & 336 & 161 \\ 
SN\,2019rta & VLASS & 59135 & 375 & 175 \\ 
AT\,2019scr & VLASS & 59067 & 303 & 162 \\ 
SN\,2020ano & VLASS & 59063 & 191 & 139 \\ 
AT\,2020bdh & VAST & 59090 & 215 &    380 \\
SN\,2020ikq & VLASS & 59109 & 137 & 130 \\ 
AT\,2020kfw & VLASS & 59093 & 101 & 130 \\ 
AT\,2020aexw & VLASS & 59072 & 20 & 141 \\ 
AT\,2020yqt & VLASS & 59103 & 22 & 136 \\
%SN\,2019aajs & VLASS & 59094 & 551 & 109 \\ 
%SN\,2020ntt & VLASS & 59090 & 56 & 151 \\ 
\enddata 
\end{deluxetable} 

We also searched for radio counterparts in two surveys that are being undertaken with the Australian Square Kilometre Array Pathfinder \citep[ASKAP;][]{Hotan2021}: the Rapid ASKAP Continuum Survey \citep[RACS:][]{McConnell2020} and phase one of the Variables And Slow Transients Pilot survey \citep[VAST-P1;][]{VAST2013}. RACS covers $\sim 35000\,\deg^2$ at 888\,MHz to a typical RMS noise of $\sim 250\,\mu$Jy, while VAST-P1 targets 113 RACS fields with identical observing parameters, covering $\sim 5000\,\deg^2$. There are 12 VAST epochs in total with each field covered at least 5 times, and 7 on average. Nine of the sources in our sample were observed by RACS and none had any associated radio emission. 
%However, after manual inspection of the image and comparison to archival radio imaging from the NRAO VLA Sky Survey \citep{NVSSpaper} and the FIRST survey \citep{FIRSTpaper} we conclude that the emission originates from the transient host galaxy. 
One source (AT\,2020bdh) has additional coverage in VAST, and no emission is detected. Table \ref{tab:radio_vlass} lists the observation details for sources in RACS and VAST.

\subsection{Host Galaxy Photometry}
\label{sec:host-photometry}

\edit2{We obtained host-galaxy photometry for all transients with redshift measurements, provided} in Table~\ref{tab:host_phot}.
We retrieved science-ready coadded images from the \textit{Galaxy Evolution Explorer} (\galex) general release 6/7 \citep{Martin2005}, the Sloan Digital Sky Survey data release 9 (SDSS DR9; \citealt{Ahn2012}), PS-1 Data Release 1 \citep{Chambers2016}, the Two Micron All Sky Survey \citep[2MASS;][]{Skrutskie2006}, and preprocessed \wise\ images \citep{Wright2010} from the unWISE archive \citep{Lang2014a}\footnote{\href{http://unwise.me}{http://unwise.me}}. The unWISE images are based on the public \wise\ data and include images from the ongoing NEOWISE-Reactivation mission R3 \citep{Mainzer2014a, Meisner2017a}. The hosts of two objects (SN\,2019php, AT\,2020xnd) were too faint, so we retrieved deeper optical images from the DESI Legacy Imaging Surveys \citep[LS;][]{Dey2019} DR8. We measured the brightness of the host using LAMBDAR\footnote{\href{https://github.com/AngusWright/LAMBDAR}{https://github.com/AngusWright/LAMBDAR}} \citep[Lambda Adaptive Multi-Band Deblending Algorithm in R;][]{Wright2016} and the methods described in \citet{Schulze2021}. The 2MASS and unWISE photometry were converted from the Vega system to the AB system using the offsets reported by \citet{Blanton2007} and \citet[][their Table 3 in Section 4.4h]{Cutri2013}.

In addition to this, we use the UVOT observations of SN\,2018gep that were obtained after the transient faded. The brightness in the UVOT filters was measured with UVOT-specific tools in the HEAsoft\footnote{\href{https://heasarc.gsfc.nasa.gov/docs/software/heasoft}{https://heasarc.gsfc.nasa.gov/docs/software/heasoft}} version 6.26.1. Source counts were extracted from the images using large apertures, to measure the total flux of the hosts. The background was estimated from regions close to the SN position. The count rates were obtained from the images using the \swift\ tool uvotsource. They were converted to magnitudes using the UVOT calibration file from September 2020. All magnitudes were transformed into the AB system using \citet{Breeveld2011}.

\begin{table}
\scriptsize
\caption{Photometry of the host galaxies.}\label{tab:host_phot}
\begin{tabular}{cccc}
\hline
Object & Survey/Telescopes/ & Filter & Brightness \\
        & Instrument        &        & (mag) \\
\hline
SN\,2018bcc & GALEX & $FUV$ & $20.00\pm0.17$ \\
SN\,2018bcc & GALEX & $NUV$ & $19.90\pm0.06$ \\
SN\,2018bcc & SDSS & $g$ & $18.53\pm0.03$ \\
SN\,2018bcc & SDSS & $i$ & $17.96\pm0.09$ \\
SN\,2018bcc & SDSS & $r$ & $18.24\pm0.06$ \\
SN\,2018bcc & SDSS & $u$ & $19.55\pm0.11$ \\
SN\,2018bcc & SDSS & $z$ & $17.93\pm0.09$ \\
SN\,2018bcc & WISE & $W1$ & $18.75\pm0.14$ \\
SN\,2018bcc & WISE & $W2$ & $19.52\pm0.12$ \\
\hline
\end{tabular}
\tablecomments{All measurements are reported in the AB system and not corrected for reddening. This table is published in its entirety in the machine-readable format. A portion is shown here for guidance regarding its form and content.}
\end{table}

\section{Analysis of ZTF and Literature Transients}
\label{sec:analysis}

In \S\ref{sec:sample-overview}, we selected transients with well-sampled light curves and durations of $1\,\days<t_{1/2}<12\,\days$ from both ZTF and the literature.
In this section, we analyze the photometric and spectroscopic evolution of the transients from \S\ref{sec:sample-overview}.
We show that based on timescales, colors, and luminosities, many of the ZTF transients can be securely classified as FBOTs.

\subsection{Photometric Evolution}
\label{sec:photometric-evolution}

\startlongtable
\begin{deluxetable*}{lrrrrrrrr}
\tablecaption{Light curve properties of the ZTF transients in Table~\ref{tab:sources_all}. \label{tab:lc-properties}} 
\tablewidth{0pt} 
\tablehead{ \colhead{IAU Name} & \colhead{$m_{\mathrm{g,max}}$} & \colhead{$M_{\mathrm{g,max}}^d$} & \colhead{$t_{\mathrm{g,1/2,rise}}^c$} & \colhead{$t_{\mathrm{g,1/2,fade}}$} & \colhead{$m_{\mathrm{r,max}}$} & \colhead{$M_{\mathrm{r,max}}$} & \colhead{$t_{\mathrm{r,1/2,rise}}$} & \colhead{$t_{\mathrm{r,1/2,fade}}$} \\
 & \colhead{(mag)} & \colhead{(mag)} & \colhead{(d)} & \colhead{(d)} & \colhead{(mag)} & \colhead{(mag)} & \colhead{(d)} & \colhead{(d)}} 
\tabletypesize{\scriptsize} 
\startdata 
SN\,2018bcc & $17.46\pm0.04$ & $-19.82\pm0.04$ & $3.20\pm0.08^a$ & $5.87\pm0.37$ & $17.74\pm0.03$ & $-19.55\pm0.03$ & $3.19\pm0.09$ & $7.84\pm0.17$ \\ 
%AT2018lwc & $20.02\pm0.06$ & -- & 1.92--4.0$^{b}$ & $9.83\pm1.64$ & $20.26\pm0.09$ & -- & $3.47\pm0.36$ & $10.01\pm2.75$ \\ 
%AT2018cow & $13.11\pm0.05$ & $-20.87\pm0.05$ & $1.10\pm0.04$ & $1.96\pm0.12$ & $13.60\pm0.10$ & $-20.38\pm0.10$ & $<5.75$ & $2.56\pm0.30$ \\ 
SN\,2019dge & $18.40\pm0.02$ & $-16.49\pm0.02$ & $1.98\pm0.04$ & $4.14\pm0.17$ & $18.57\pm0.01$ & $-16.31\pm0.01$ & $1.80\pm0.07$ & $7.68\pm0.78$ \\ 
AT2018lwd & $19.55\pm0.05$ & -- & 2.02--3.0 & $4.27\pm0.85$ & $19.75\pm0.07$ & -- & 2.06--3.01 & $6.29\pm1.02$ \\ 
SN\,2018gep & $15.91\pm0.00$ & $-19.84\pm0.00$ & $3.27\pm0.02$ & $6.00\pm0.17$ & $16.23\pm0.00$ & $-19.51\pm0.00$ & $3.21\pm0.05$ & $10.90\pm0.55$ \\ 
SN\,2018ghd & $18.49\pm0.03$ & $-17.73\pm0.03$ & $2.49\pm0.10$ & $7.03\pm0.92$ & $18.58\pm0.04$ & $-17.64\pm0.04$ & $2.26\pm0.16$ & $10.69\pm4.12$ \\ 
AT2018lug & $19.34\pm0.05$ & $-21.17\pm0.05$ & $1.12\pm0.03$ & $2.92\pm0.14$ & $19.82\pm0.06$ & $-20.69\pm0.06$ & 1.51--2.34 & $2.75\pm0.34$ \\ 
SN\,2018gjx & $15.58\pm0.01$ & $-17.65\pm0.01$ & $2.32\pm0.01$ & $5.05\pm0.08$ & $15.78\pm0.01$ & $-17.45\pm0.01$ & 1.97--4.0 & $8.22\pm0.19$ \\ 
%ZTF19aakssbm & $17.16\pm0.03$ & $-18.86\pm0.03$ & $2.11\pm0.03$ & $6.05\pm0.24$ & $17.43\pm0.04$ & $-18.59\pm0.04$ & $2.01\pm0.06$ & $5.88\pm0.22$ \\ 
AT2019dcm & $19.09\pm0.04$ & -- & 4.02--5.0 & $5.24\pm0.31$ & $19.17\pm0.04$ & -- & 4.0--5.04 & $7.30\pm0.40$ \\ 
SN\,2019deh & $17.22\pm0.02$ & $-19.73\pm0.02$ & $4.35\pm0.07$ & $6.33\pm0.66$ & $17.43\pm0.05$ & $-19.52\pm0.05$ & $5.00\pm0.15$ & $6.57\pm0.42$ \\ 
AT2019aajt & $19.49\pm0.05$ & -- & $1.45\pm0.04$ & $4.22\pm0.68$ & $19.75\pm0.05$ & -- & $1.36\pm0.07$ & $5.40\pm0.47$ \\ 
AT2019aaju & $19.41\pm0.02$ & -- & $<3.1$ & $8.80\pm1.98$ & $19.69\pm0.06$ & -- & $<4.02$ & $15.10\pm5.15$ \\ 
AT2019esf & $18.84\pm0.03$ & $-18.82\pm0.03$ & $2.31\pm0.24$ & $4.92\pm0.59$ & $19.20\pm0.05$ & $-18.47\pm0.05$ & $1.38\pm0.09$ & $21.97\pm7.30$ \\ 
AT2019lbr & $19.09\pm0.04$ & -- & $2.25\pm0.15$ & $6.93\pm0.79$ & $19.35\pm0.07$ & -- & $4.46\pm0.20$ & $6.53\pm1.85$ \\ 
AT2019kyw & $18.28\pm0.04$ & $-19.34\pm0.04$ & $4.35\pm0.09$ & $7.54\pm0.47$ & $18.48\pm0.05$ & $-19.14\pm0.05$ & $4.26\pm0.14$ & $>2.6$ \\ 
SN\,2019myn & $18.84\pm0.02$ & $-19.44\pm0.02$ & $3.48\pm0.06$ & $6.04\pm0.84$ & $18.91\pm0.03$ & $-19.37\pm0.03$ & $3.45\pm0.15$ & $5.45\pm0.68$ \\ 
AT2019aajv & $19.48\pm0.03$ & -- & $0.80\pm0.07$ & $2.58\pm0.50$ & $19.86\pm0.08$ & -- & $0.78\pm0.12$ & $1.77\pm0.62$ \\ 
SN\,2019php & $18.68\pm0.06$ & $-19.29\pm0.06$ & $3.64\pm0.11$ & $4.71\pm0.42$ & $18.92\pm0.04$ & $-19.05\pm0.04$ & $3.21\pm0.17$ & $5.98\pm0.28$ \\ 
SN\,2019qav & $18.99\pm0.06$ & $-19.96\pm0.06$ & $3.39\pm0.28$ & $7.38\pm0.38$ & $19.22\pm0.10$ & $-19.73\pm0.10$ & $4.51\pm0.39$ & $8.21\pm0.42$ \\ 
AT2019qwx & $19.03\pm0.04$ & -- & 3.0--4.0 & $6.41\pm0.51$ & $19.29\pm0.08$ & -- & 2.91--4.0 & $8.87\pm2.83$ \\ 
SN\,2019rii & $18.75\pm0.02$ & $-20.00\pm0.02$ & $4.36\pm0.16$ & $5.61\pm0.38$ & $18.89\pm0.03$ & $-19.86\pm0.03$ & $3.96\pm0.96$ & $5.37\pm1.61$ \\ 
SN\,2019rta & $17.88\pm0.02$ & $-17.53\pm0.02$ & $1.09\pm0.03$ & $5.74\pm0.43$ & $17.98\pm0.03$ & $-17.42\pm0.03$ & $1.20\pm0.04$ & $8.99\pm1.09$ \\ 
AT2019scr & $18.91\pm0.05$ & -- & $<3.0$ & $2.03\pm0.24$ & $19.48\pm0.13$ & -- & 13.94--5.92 & $1.20\pm1.51$ \\ 
AT2019van & $18.54\pm0.11$ & -- & $<3.31$ & $4.75\pm3.85$ & $18.75\pm0.18$ & -- & $1.36\pm0.23$ & $8.01\pm1.93$ \\ 
SN\,2020oi & $14.06\pm0.12$ & $-17.75\pm0.12$ & $2.92\pm0.33$ & $8.05\pm0.46$ & $13.74\pm0.12$ & $-18.07\pm0.12$ & $6.07\pm0.39$ & $7.41\pm0.76$ \\ 
SN\,2020ano & $19.06\pm0.03$ & $-16.65\pm0.03$ & $<3.9$ & $1.46\pm0.07$ & $19.52\pm0.06$ & $-16.19\pm0.06$ & $<3.91$ & $2.02\pm0.79$ \\ 
AT2020bdh & $18.60\pm0.03$ & $-17.72\pm0.03$ & $<2.87$ & $7.50\pm0.81$ & $18.74\pm0.06$ & $-17.58\pm0.06$ & $<2.85$ & $9.03\pm2.21$ \\ 
AT2020bot & $19.46\pm0.04$ & $-20.33\pm0.04$ & $1.19\pm0.33$ & $2.53\pm0.14$ & $19.54\pm0.17$ & $-20.26\pm0.17$ & $3.38\pm0.30$ & $0.55\pm0.24$ \\ 
SN\,2020ikq & $18.27\pm0.03$ & $-18.10\pm0.03$ & 2.88--5.8 & $7.47\pm1.11$ & $18.51\pm0.05$ & $-17.86\pm0.05$ & 2.73--5.67 & $27.88\pm2.11$ \\ 
SN\,2020jmb & $18.51\pm0.03$ & $-18.68\pm0.03$ & $3.61\pm0.07$ & $6.44\pm0.37$ & $18.78\pm0.03$ & $-18.41\pm0.03$ & 1.92--3.78 & $11.98\pm1.87$ \\ 
SN\,2020jji & $19.50\pm0.09$ & $-16.64\pm0.09$ & $4.02\pm0.56$ & $6.76\pm0.88$ & $18.89\pm0.17$ & $-17.25\pm0.17$ & $3.13\pm0.53$ & $10.55\pm1.00$ \\ 
AT2020mlq & $19.71\pm0.06$ & -- & $4.59\pm0.42$ & $6.44\pm0.74$ & $19.75\pm0.06$ & -- & $3.71\pm0.28$ & $14.61\pm2.53$ \\ 
AT2020kfw & $19.05\pm0.03$ & $-18.07\pm0.03$ & $3.83\pm0.12$ & $4.54\pm0.22$ & $19.25\pm0.04$ & $-17.87\pm0.04$ & $3.67\pm0.20$ & $6.75\pm0.33$ \\ 
%SN\,2020ntt & $18.61\pm0.09$ & $-19.01\pm0.09$ & $4.54\pm0.67$ & $8.68\pm1.17$ & $18.29\pm0.09$ & $-19.33\pm0.09$ & $2.87\pm0.57$ & $6.93\pm1.69$ \\ 
AT2020aexw & $19.39\pm0.03$ & $-18.21\pm0.03$ & $3.07\pm0.04$ & $7.44\pm0.35$ & $19.59\pm0.05$ & $-18.01\pm0.05$ & $2.87\pm0.10$ & $9.30\pm0.95$ \\ 
AT2020yqt & $19.17\pm0.11$ & $-19.08\pm0.11$ & $0.66\pm0.07$ & $3.38\pm0.96$ & $19.44\pm0.09$ & $-18.81\pm0.09$ & $1.35\pm0.12$ & $5.58\pm0.83$ \\ 
SN\,2020rsc & $19.36\pm0.07$ & $-16.37\pm0.07$ & $1.62\pm0.04$ & $1.72\pm0.28$ & $19.36\pm0.11$ & $-16.36\pm0.11$ & $0.50\pm0.27$ & $2.75\pm1.53$ \\ 
AT2020xnd & $19.24\pm0.04$ & $-21.03\pm0.04$ & 1.6--4.81 & $2.39\pm0.30$ & $19.54\pm0.06$ & $-20.73\pm0.06$ & 0.88--3.26 & $3.60\pm0.18$ \\ 
SN\,2020vyv & $18.68\pm0.03$ & $-18.55\pm0.03$ & $1.49\pm0.07$ & $4.13\pm0.31$ & $18.98\pm0.05$ & $-18.25\pm0.05$ & $2.08\pm0.13$ & $8.21\pm1.24$ \\ 
SN\,2020xlt & $19.59\pm0.04$ & $-16.58\pm0.04$ & $<0.91$ & $3.52\pm0.31$ & $19.61\pm0.07$ & $-16.56\pm0.07$ & 2.06--2.87 & $4.58\pm0.39$ \\ 
\enddata 
\tablenotetext{a}{Based on $r$-band points}
\tablenotetext{b}{A range indicates a rise or fade that was not resolved by detections.}
\tablenotetext{c}{Times are given in the rest-frame.}
\tablenotetext{d}{Peak absolute magnitudes have an approximate $K$-correction applied.}
\end{deluxetable*} 

\begin{figure*}[!htb]
    \centering
    \includegraphics[width=0.8\textwidth]{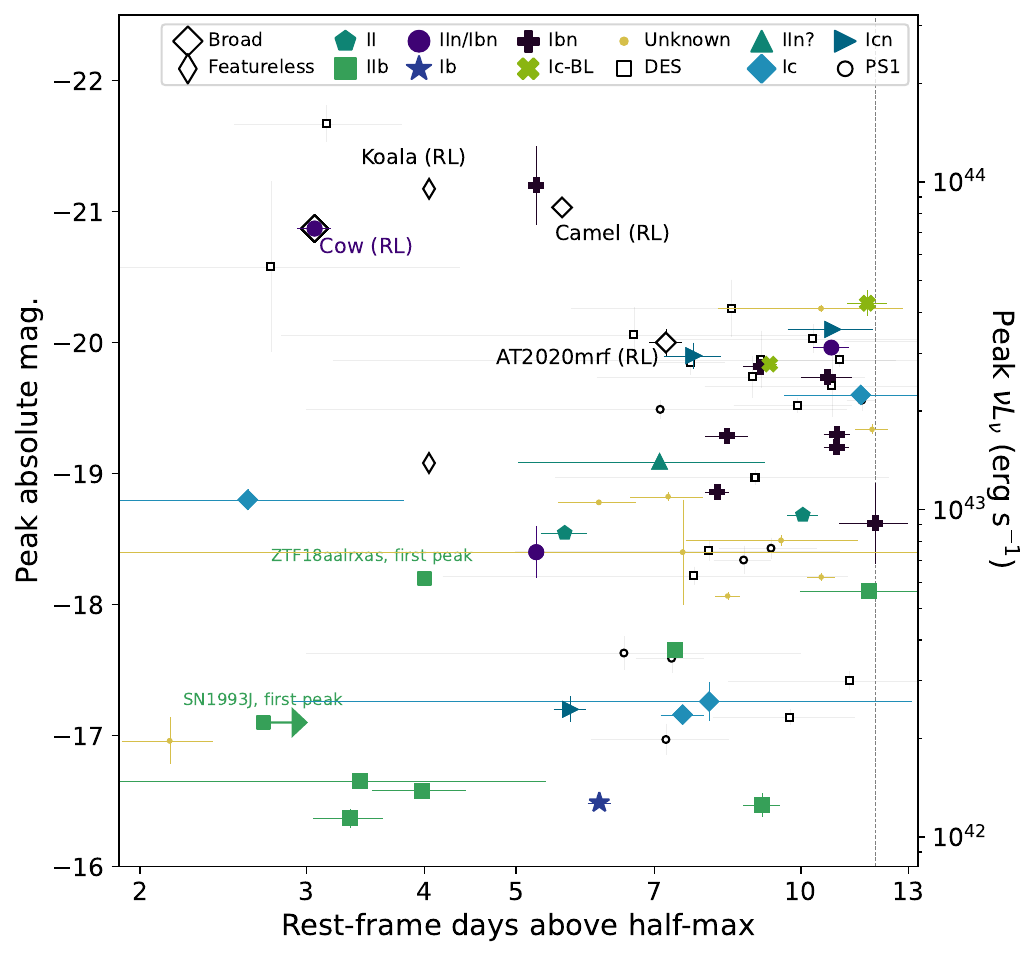}
    \caption{\edit2{The luminosity-duration parameter space of ZTF and literature FBOTs. 
    Measurements are in $g$-band and rest-frame when possible.
    The vertical line indicates the duration cut of $t_{1/2}=12\,$d that is commonly used to define FBOTs in the literature \citep{Drout2014,Inserra2019}.
    Events with luminous radio emission (RL)---Cow/AT2018cow, Koala/AT2018lug, Camel/AT2020xnd, AT2020mrf---have particularly fast and luminous optical light curves, and predominantly featureless spectra at peak light (with a very broad $v>0.1c$ unidentified feature in some cases).
    Most spectroscopically classified FBOTs are members of established core-collapse supernova types. For clarity we do not plot SN\,2019ehk, because the peak luminosity is significantly fainter ($-14.7\,$mag.)}}
    \label{fig:lum-timescale}
\end{figure*}

\edit2{The photometric properties of FBOTs are summarized in \citet{Inserra2019}: peak magnitudes ranging from faint core-collapse SNe to superluminous SNe, faster rise times than decline times, a wide range of decline timescales implying a variety of powering sources, and blue colors at peak light with reddening over time. The combination of a fast rise time and high peak luminosity rules out a radioactive decay power source in some cases; as summarized in \citet{Margutti2019}, two popular alternatives are shock-interaction with circumstellar matter and long-lived energy injection by a central engine.}

\edit2{The redshifts and peak magnitudes of the ZTF transients are reported in Table~\ref{tab:sources_all} and Table~\ref{tab:lc-properties}, respectively.
Eleven objects had spectroscopic host-galaxy redshifts measured prior to the discovery of the transient; for the remaining eighteen, the redshifts were measured via observations of the transient or the host galaxy post-discovery.
}

We linearly interpolated the $g$- and $r$-band light curves in flux space to estimate a rise time $t_{1/2,\mathrm{rise}}$ and fade time $t_{1/2,\mathrm{fade}}$ from the half-maximum of the observed peak in each filter.
We estimated error bars by performing a Monte Carlo with 600 realizations of the light curve.
We estimated the absolute magnitude using Equation~\ref{eq:kcorrect}.
The measured timescales are provided in Table~\ref{tab:lc-properties}.

Figure~\ref{fig:lum-timescale} shows the parameter space of duration and \edit2{peak luminosity for the ZTF and literature FBOTs}, color coded by spectroscopic type (discussed in \S\ref{sec:spectroscopic-evolution}).
%For reference, we also show the parameter space of SNe from a magnitude-limited survey \citep{Perley2020_BTS,Fremling2020_RCF} on the right-hand side of Figure~\ref{fig:lum-timescale},
%and the parameter space of literature FBOTs in Figure~\ref{fig:phase-space-lit}.
The peak absolute magnitudes of the ZTF transients span $M=-16\,$mag to $M=-22\,$mag,
similar to literature FBOT samples \citep{Drout2014,Pursiainen2018},
and similar to core-collapse SNe.
\edit2{For reference, we also show the first peak of two double peaked Type~IIb SNe, SN\,1993J and ZTF18aalrxas.}

% There are two predominant classes of events at the shortest durations:
% Type~IIb events at low luminosities,
% and events similar to AT\,2018cow (in terms of their radio emission) at high luminosities.
% There is also one unusual, very luminous and fast-evolving unclassified event (AT\,2020bot), although unlike AT\,2018cow it rose to a plateau or second peak.
% In addition, SN\,2018kzr \citep{McBrien2019} had an overall very short duration and an intermediate luminosity.
% Only two objects in the comparison sample have such short durations: both were from DES, and very luminous.
% So, it appears that the fast-subluminous Type~IIb SNe are a previously unrecognized group of objects,
% and that---without knowing the redshift of the host galaxy a priori---they represent the primary extragalactic contaminant in the search for events similar to AT\,2018cow.

\begin{figure*}[!htb]
    \centering
    \includegraphics[width=1.5\columnwidth]{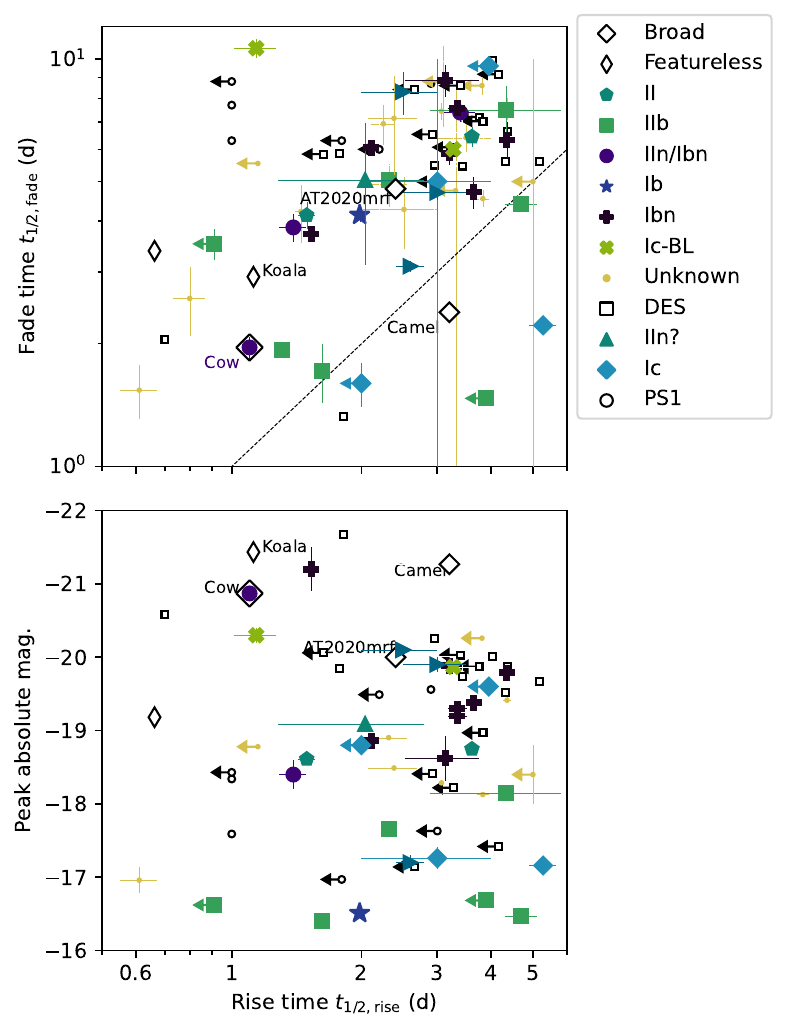}
    \caption{Top panel: The rise and fade times of ZTF FBOTs
    and well-observed FBOTs in the literature.
    The dashed line in the top panel indicates equal rise and fade times.
    Both ZTF FBOTs and unclassified FBOTs in the literature typically have longer fade times than rise times.
    Bottom panel: Rise time vs. peak luminosity of ZTF and literature FBOTs.
    Timescales and luminosities are measured in $g$ band, in the rest-frame, from half-peak to peak.
    Radio-loud FBOTs (Cow/AT2018cow, Koala/AT2018lug, Camel/AT2020xnd, and AT2020mrf) have particularly high peak luminosities.
    }
    \label{fig:rise-fade}
\end{figure*}

\edit2{The top panel of} Figure~\ref{fig:rise-fade} compares the rise time to the fade time of the ZTF and literature FBOTs.
Similar to literature FBOTs, 
the ZTF objects have a slower fade time than rise time,
and a wide range of fade times.
\edit2{The bottom panel of Figure~\ref{fig:rise-fade} shows the rise time and peak luminosity of the ZTF and literature FBOTs.}

%The exceptions appear to be the two Type~Ic events, SN\,2018kzr \citep{McBrien2019} and SN\,2019bkc \citep{Chen2020,Prentice2020}.

%We include several Type~IIP SNe from the ZTF BTS \citep{Fremling2020_RCF,Perley2020_BTS} to show that they can also contaminate the search for fast-evolving transients during the rise phase.

\begin{figure*}[!htb]
    \centering
    \includegraphics[width=0.914\textwidth]{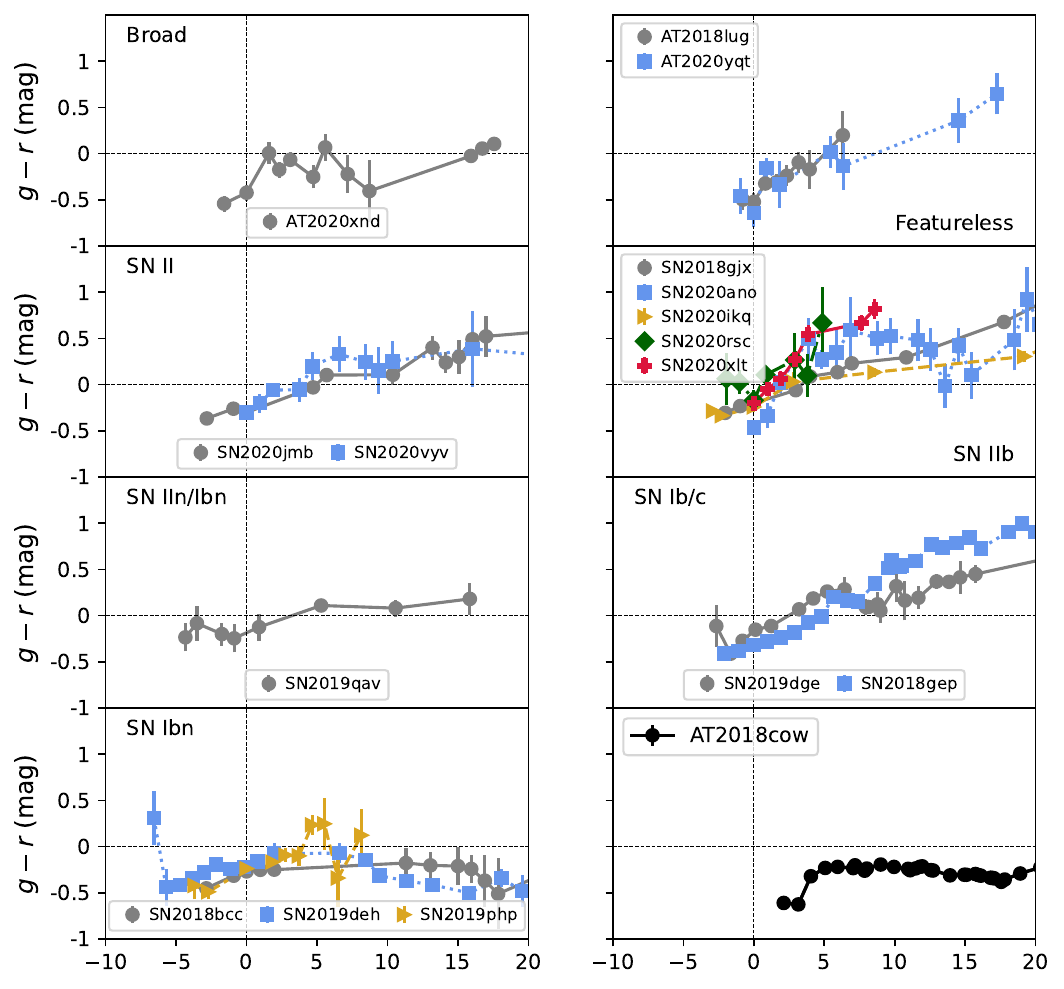}
    \caption{Color evolution of ZTF FBOTs, with AT2018cow shown on the lower right for comparison. Unlike AT2018cow, most events redden after peak,
    with the exception of the Type~Ibn SNe and potentially the radio-loud event AT\,2020xnd}.
    \label{fig:gr}
\end{figure*}

We calculate the $g-r$ color on nights where observations were acquired in both filters (not correcting for host reddening).
\edit2{The peak-light colors are reported in Table~\ref{tab:sources_all}. Similar to literature FBOTs,}
most transients are blue at maximum light and redden with time.
There are exceptions, however, most notably the Type~Ibn SNe and events with persistent interaction-dominated spectra (AT\,2018cow and SN\,2019qav).
\edit2{The evolution of $g-r$ color over time for each ZTF FBOT is shown in Figure~\ref{fig:gr}.}

\subsection{Spectroscopic Classification of FBOTs}
\label{sec:spectroscopic-evolution}

One of the challenges in spectroscopically classifying \edit2{FBOTs} is that the peak-light spectra often appear relatively featureless \citep{Drout2014,Inserra2019,Karamehmetoglu2021,Perley2019cow,Ho2019gep,Ho2020b,Perley2021}.
Some have weak features from interaction with circumstellar material (CSM), such as PTF09uj \citep{Ofek2010}, but such features have been difficult to discern in the low-S/N spectra often obtained for events at high redshift \citep{Drout2014}.
Furthermore, by the phase at which SN features tend to become most distinguishable (two weeks after peak light; \citealt{Williamson2019}) a rapidly fading event is difficult to observe.
The advantage of a high-cadence and shallow survey like ZTF is that objects are discovered young and relatively nearby, respectively:
\edit2{for several of the ZTF FBOTs}
we were able to obtain spectra within 2--3 days of peak light with sufficiently high S/N to discern even weak CSM interaction features, as well as late-time spectra that enabled spectroscopic classifications.
For our analysis here, we only consider spectra obtained with instruments other than the SEDM, due to its low resolution ($R\sim100$).

\begin{figure*}[!ht]
    \centering
    \includegraphics[width=0.85\textwidth]{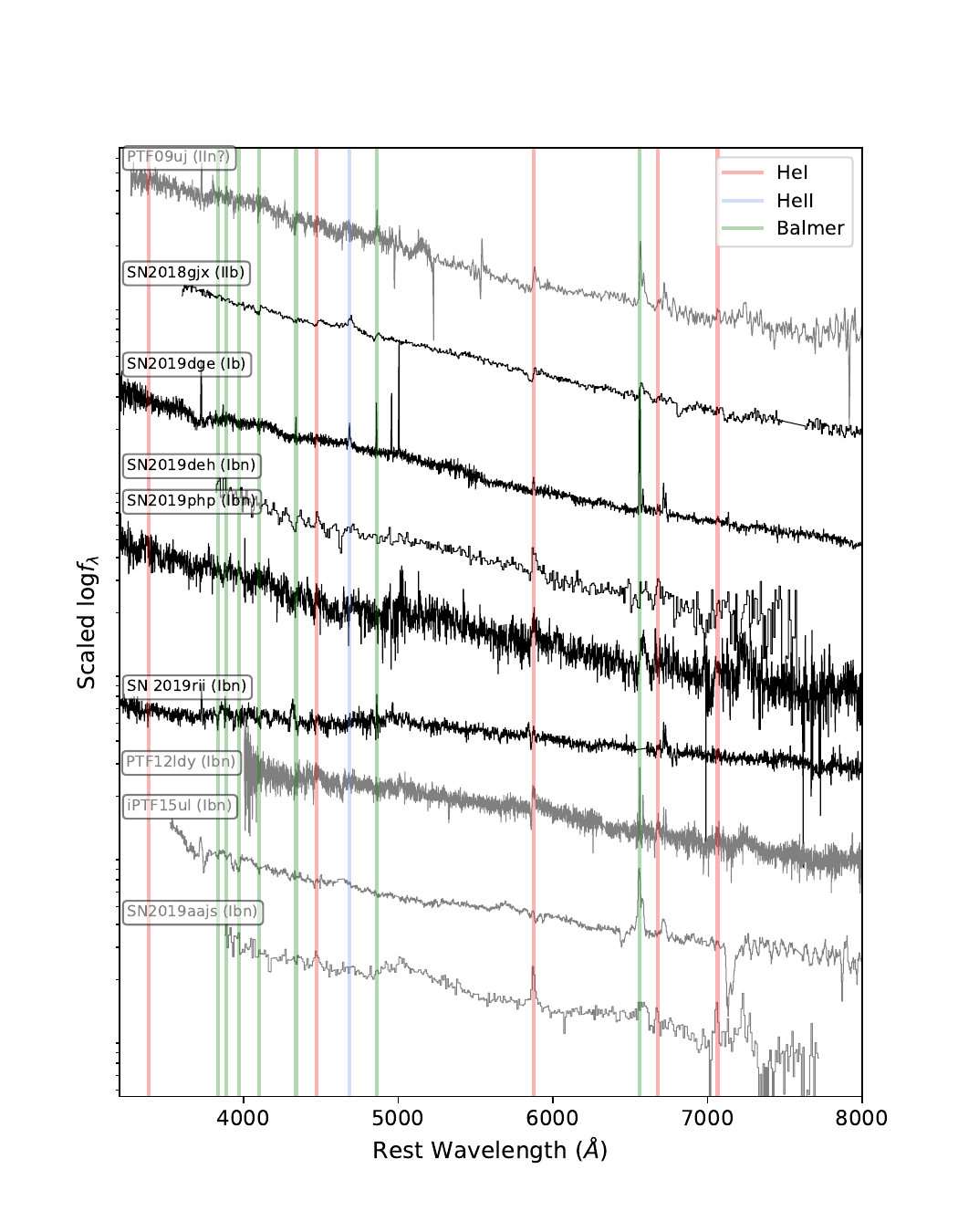}
    \caption{Peak-light spectra of well-observed FBOTs that show narrow emission features.
    \edit2{ZTF FBOTs} are shown in black.
    The spectrum of SN\,2018gjx was downloaded from the TNS \citep{Gromadzki2018_ZTF18abwkrbl}.
    Comparison-sample FBOTs are shown in grey.
    The comparison-sample spectra were initially presented in \citet{Ofek2010} and \citet{Hosseinzadeh2017},
    and the spectrum of SN2019dge was initially presented in \citet{Yao2020}.
    }.
    \label{fig:spec-peak-narrow}
\end{figure*}

\begin{figure*}[!ht]
    \centering
    \includegraphics[width=0.85\textwidth]{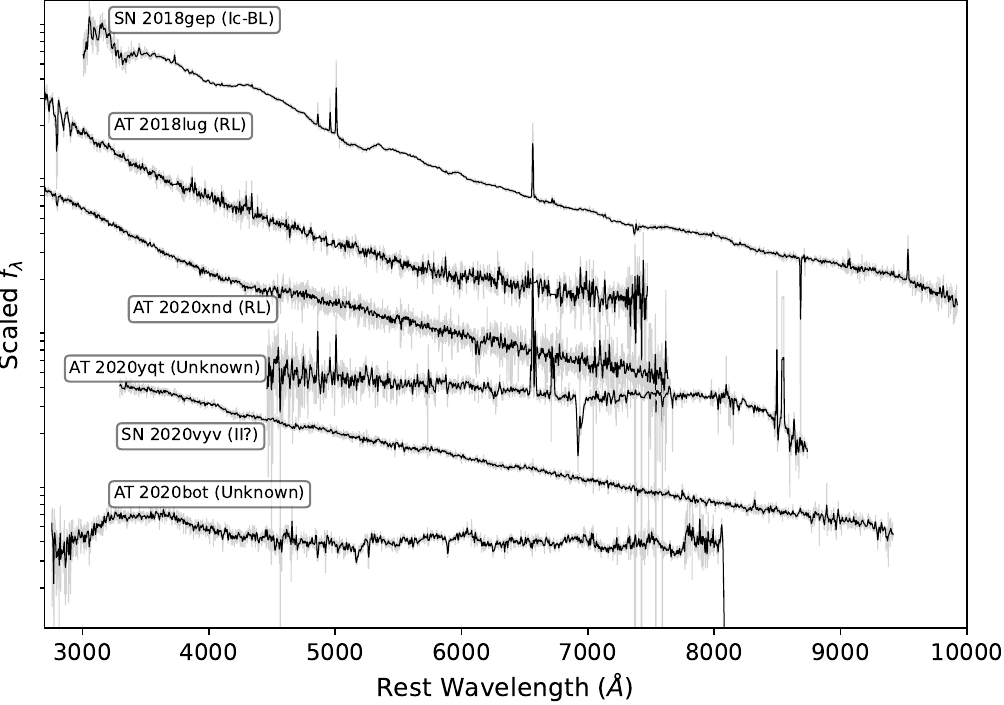}
    \caption{Peak-light spectra of well-observed ZTF FBOTs that are entirely featureless or show broad absorption features. The spectra of SN\,2018gep, AT\,2018lug, and AT\,2020xnd were originally published in \citet{Ho2019gep}, \citet{Ho2020b}, and \citet{Perley2021}, and the spectrum of SN\,2020vyv was downloaded from the TNS \citep{Siebert2020}. AT\,2020bot is not strictly an ``FBOT'' because its peak-light color is $g-r=-0.1$\,mag.}
    \label{fig:spec-peak-broad}
\end{figure*}

The most common behavior at peak light is a spectrum dominated by a blue continuum, as has been found for previous FBOT samples \citep{Drout2014,Inserra2019}.
\edit1{Some events show
narrow (width hundreds of km\,s$^{-1}$) emission features of helium and hydrogen (Figure~\ref{fig:spec-peak-narrow}), while others show entirely featureless spectra or spectra with very broad absorption features ($v>0.1c$; Figure~\ref{fig:spec-peak-broad}).
The events with very broad features or entirely featureless spectra include the most luminous objects in the ZTF sample (AT\,2020xnd, SN\,2018gep, AT\,2018lug), suggesting that high velocities may link the high luminosity of the light curve and the broad features of the spectra.}

\edit1{SN\,2019rii warrants particular note}: it has \ion{He}{1} $\lambda\lambda$3389, $\lambda\lambda$4471 and $\lambda\lambda$5876 (though neither $\lambda\lambda$6678 nor $\lambda\lambda$7065), with weak narrow emission at $v=0$, narrow absorption at $v=900\,\km\,\psec$ for $\lambda\lambda$5876 and $600\,\km\,\psec$ for $\lambda\lambda$4471 and $\lambda\lambda$3889.
We tentatively classify it as a Type~Ibn on the basis of this spectrum (\edit1{and the late-time spectrum of the transient is dominated by the host galaxy}) but note that this classification is not fully secure.
\edit1{We include this object in Figure~\ref{fig:spec-peak-narrow} but note that it is not strictly an FBOT, because its peak-light color is $g-r=-0.1$\,mag.}

SN\,2020vyv has a TNS classification of Type~II SN based on a tentative broad H$\alpha$ feature in the peak-light spectrum. As there is no definitive late-time spectrum we list the classification as II? in Table~\ref{tab:sources_all}. SN\,2020rsc had a peak-light spectrum that already showed prominent P-Cygni features, enabling the Type~IIb classification.

Due to a lack of spectra obtained after peak,
previous \edit2{FBOT samples} have not been able to conclude whether the objects were hydrogen-rich or hydrogen-poor \citep{Drout2014,Pursiainen2018}.
By 1--3 weeks after peak light
the spectra of most \edit2{of the ZTF FBOTs} began to exhibit P-Cygni features from optically thin ejecta, enabling their spectroscopic classification as SNe.
The compositions range from H-rich (Type~II/IIb), H-poor (Type~Ib), to fully stripped (Type~Ic-BL).

\begin{figure*}[htb!]
    \centering
    \includegraphics[width=0.7\textwidth]{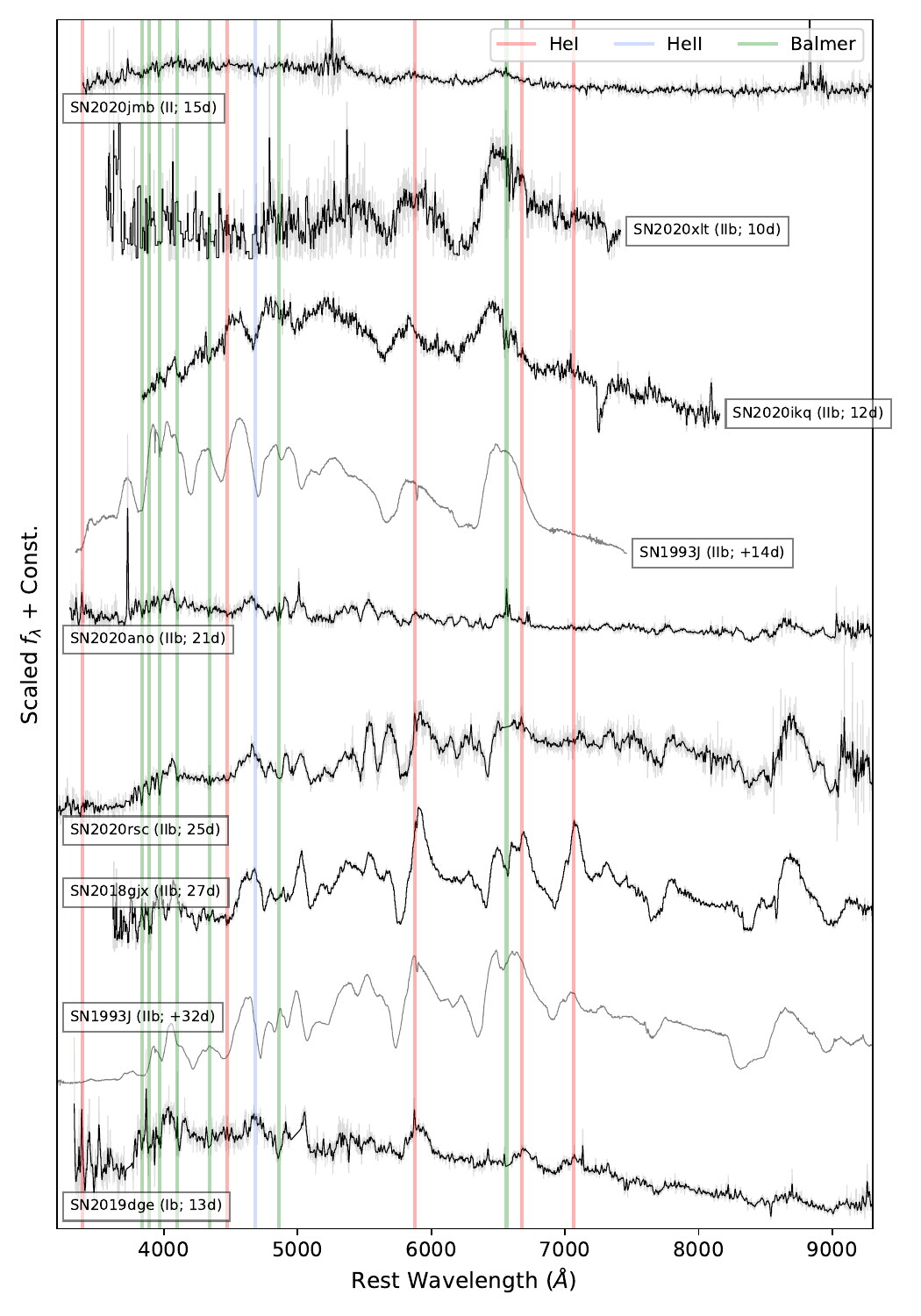}
    \caption{Post-peak spectra of the \edit2{ZTF FBOTs classified} as Type~IIb or Type~Ib SNe based on their H and He P-Cygni features at late times.
    For comparison we show spectra of the Type~IIb SN\,1993J, obtained from WISeREP and originally from the UCB SN database \citep{Silverman2012}.
    The phase of the SN\,1993J spectra is given with respect to the peak of the first (shock-cooling) peak, 30 March 1993. 
    For the ZTF objects,
    epochs are given with respect to the maximum of the $g$-band light curve;
    raw spectra are shown in light grey, with smoothed spectra overlaid in black; and in some cases we have clipped host emission lines for clarity.
    }.
    \label{fig:iib}
\end{figure*}

The subluminous \edit2{FBOTs} ($M>-18\,$mag) most commonly evolve into Type~II, Type~IIb, and Type~Ib SNe,
as shown in Figure~\ref{fig:iib}.
We note that the distinction between these classes can be subtle when spectroscopic coverage is limited.
For example, SN\,2020jji and SN\,2020jmb have spectra at two weeks after peak light that resemble both Type~IIP and Type~IIb objects, and we use a Type~II classification to be more generic.
The full sample of short-duration Type~II and Type~IIb events will be presented and modeled in a separate paper by Fremling et al.

\begin{figure*}[htb!]
    \centering
    \includegraphics[width=0.8\textwidth]{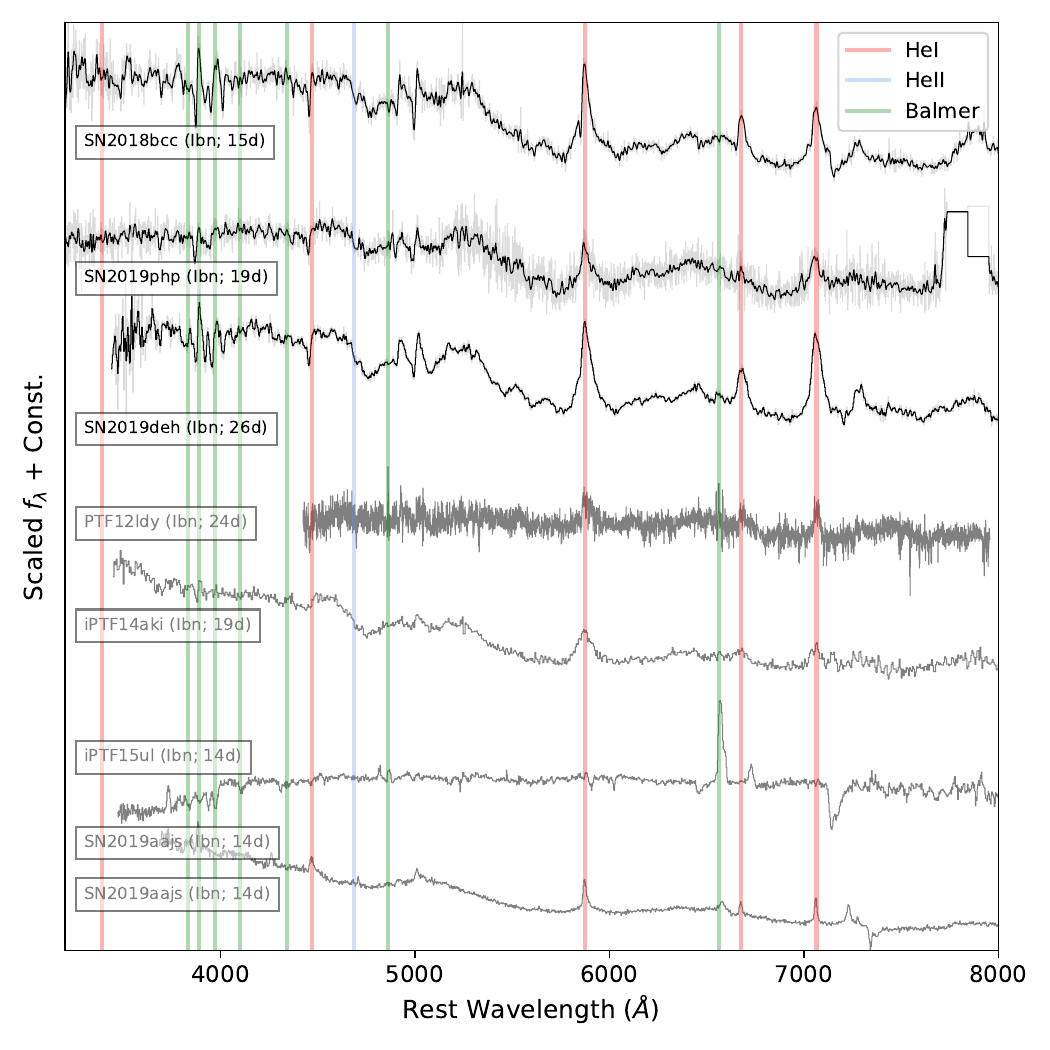}
    \caption{Post-peak spectra of well-observed ZTF FBOTs classified as Type~Ibn based on their He P-Cygni features at late times, together with the post-peak spectra of FBOTs classified as Type~Ibn in the literature.  Spectra of PTF\,12ldy and iPTF\,15ul were obtained from WISeREP and are originally from \citet{Hosseinzadeh2017}.
    }.
    \label{fig:ibn}
\end{figure*}

The luminous ($M<-18\,$mag) and somewhat longer duration ($>6\,$d) \edit2{FBOTs} most commonly evolve into Type~Ibn SNe.
We show the Type~Ibn post-peak spectra in Figure~\ref{fig:ibn},
together with spectra of the literature comparison sample objects that were also classified as Type~Ibn \citep{Pastorello2015,Hosseinzadeh2017}.
Type~Ibn SNe are named for the strong and relatively narrow ($\sim2000\,\km\,\psec$) \ion{He}{1} emission lines in their early spectra \citep{Pastorello2008,Smith2017,GalYam2017}.
The detailed properties of the Type~Ibn SNe observed in ZTF will be presented in a separate paper by Kool et al.

\begin{figure*}[htb!]
    \centering
    \includegraphics[width=0.6\textwidth]{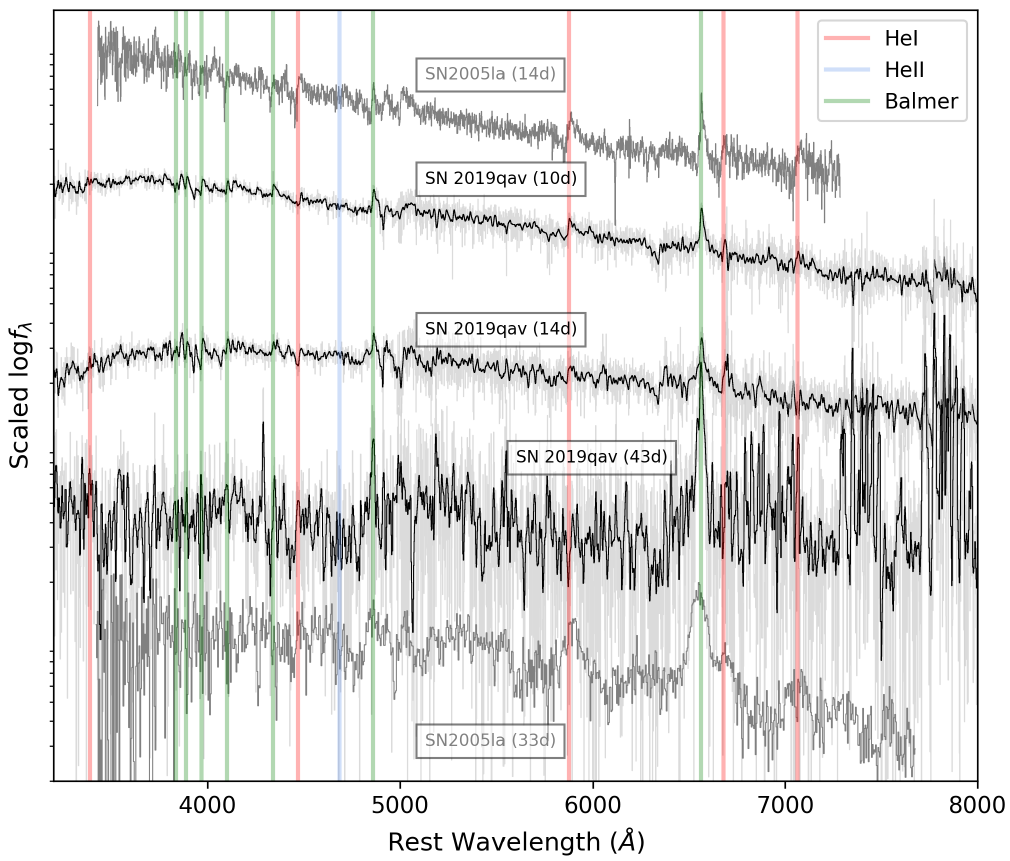}
    \caption{Post-peak evolution of SN\,2019qav, which we classify as a Type~IIn/Ibn transitional object due to its similarity with SN\,2005la. Spectra of SN\,2005la were obtained from WiseREP and are originally from \citet{Modjaz2014} and \citet{Pastorello2008}.
    }.
    \label{fig:ztf19abyjzvd}
\end{figure*}

Finally, some events have post-peak spectra that remained dominated by a blue continuum 
with narrow emission lines,
with no nebular emission from optically thin inner ejecta.
In particular, SN\,2019qav evolved in a similar fashion to the Type~IIn/Ibn transition object SN\,2005la \citep{Pastorello2008,Smith2012}, as we show in Figure~\ref{fig:ztf19abyjzvd}.
Similarly, AT\,2018cow had \ion{He}{2} emission lines that emerged after one week, and Balmer emission lines that emerged one week after that,
but never developed P-Cygni features.
The only transient in the PS1 sample with a post-peak spectrum, PS1-12bb (+33\,d), also had a persistently continuum-dominated spectrum,
although weak features would not have been detectable at this low S/N.
\citet{Drout2014} noted that a persistent continuum was unusual for rapidly declining SNe.
\edit2{Both SN\,2019qav and PS1-12bb were slightly redder than the $g-r\leq-0.2\,$mag of FBOTs.}

In conclusion, a picture is emerging in which the \edit2{FBOT spectroscopic classification is strongly correlated with} its luminosity, with interacting SNe dominating the most luminous ($M<-18\,$mag) events,
and Type~IIb and Type~Ib dominating the subluminous ($M>-18\,$mag) events.
Some of the most luminous events have broad absorption features from high velocities, suggesting that the high velocities are related to the high luminosity, \edit2{and perhaps also to the presence of luminous radio emission.}

\subsection{X-ray and Radio Emission}
\label{sec:radio}

There is considerable interest in understanding to what extent AT\,2018cow is part of a continuum that extends into other parts of the fast-transient parameter space, and to what extent it is a distinct class.
In the literature AT\,2018cow is often described as \edit2{an FBOT, but it has not been clear to what extent its properties are representative of the parameter space in Figure~\ref{fig:lum-timescale}}.
In this section we discuss to what extent the luminous X-ray, millimeter, and radio emission of AT2018cow, can be ruled out in other parts of the parameter space of Figure~\ref{fig:lum-timescale}.

\begin{figure*}[htb!]
    \centering
    \includegraphics[width=0.8\textwidth]{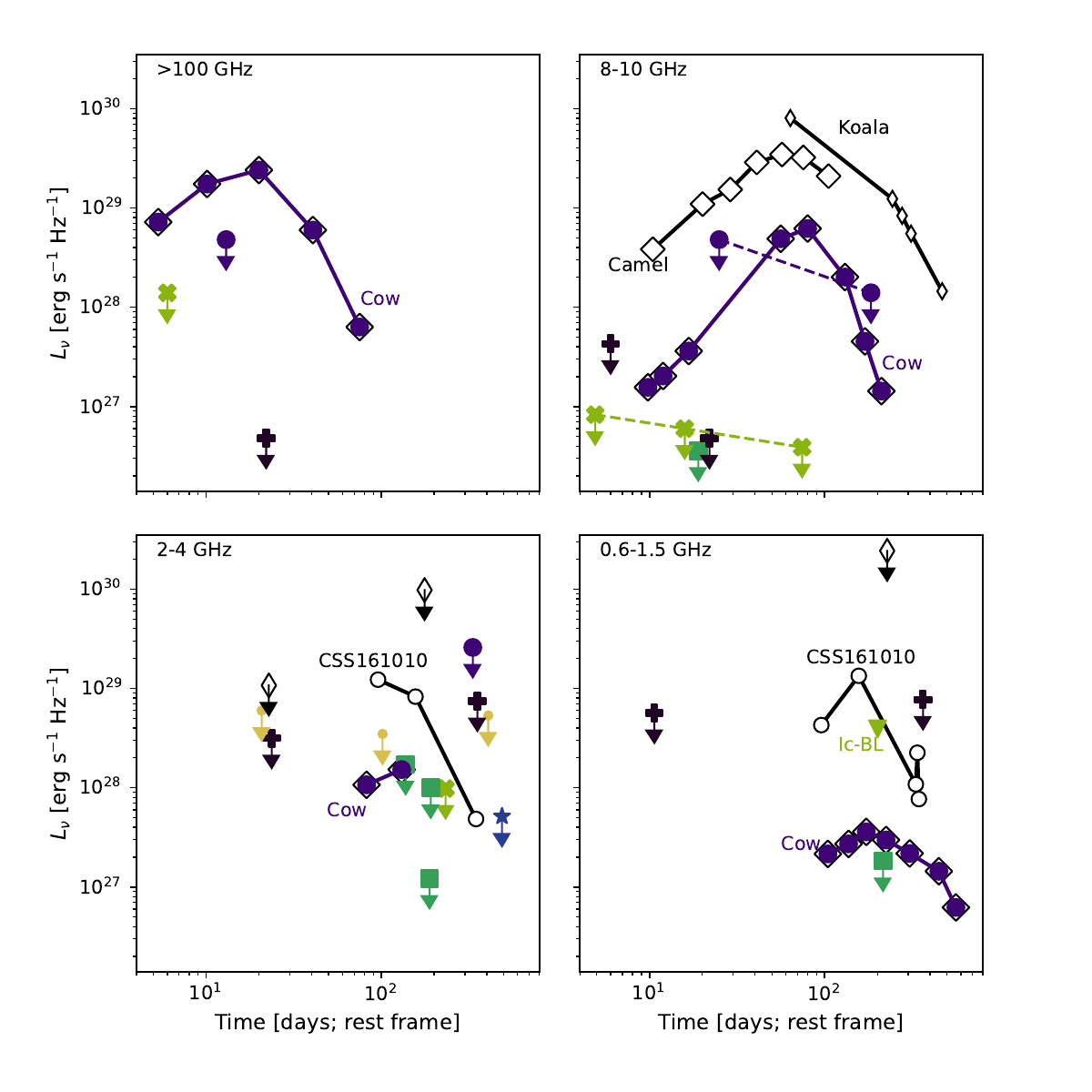}
    \caption{Millimeter and radio observations of \edit2{ZTF FBOTs, with light curves of radio-loud FBOTs such as the Cow/AT2018cow and CSS161010 shown for comparison.}
    The CSS161010 light curves are from \citet{Coppejans2020}.
    The light curves of the Cow, the Koala/AT\,2018lug, and SN\,2018gep (Ic-BL) at 10\,GHz were taken from the literature \citep{Ho2019gep,Margutti2019,Bietenholz2020,Ho2020b,Coppejans2020}.
    The 10\,GHz light curve of the Camel/AT\,2020xnd is from \citet{Ho2021_20xnd}.
    The 0.75\,GHz light curve of the Cow is from \citet{Nayana2021}.
    Additional observations at $>100\,$GHz and 8--10\,GHz are from this work.
    Limits at 2--4\,GHz are from VLASS.
    Limits at 888\,MHz are from the RACS and VAST (see text).
    In the bottom-right panel, the positions of the Ic-BL and Ib markers have been shifted slightly for clarity.
    The only objects with robust detections of luminous ($>10^{28}\,\erg\,\psec\,\phz$) millimeter and radio emission appear to be the shortest-duration, highest-luminosity optical transients: the Cow, the Koala, and the Camel.
    }
    \label{fig:radio}
\end{figure*}

In Figure~\ref{fig:radio} we show the millimeter and radio upper limits presented in Section~\ref{sec:radio-obs} compared to the light curve of AT\,2018cow. The only events with similar millimeter and radio behavior---AT\,2020xnd and AT\,2018lug---also have very similar optical light curves to AT\,2018cow.
SN\,2019qav (Type~IIn/Ibn) also had a high luminosity and spectra persistently dominated by interaction;
yet X-ray, millimeter, and radio observations rule out emission similar to that of AT\,2018cow.
SN\,2020rsc (Type~IIb) had a light curve similar to AT\,2018cow in its duration (albeit significantly less luminous), yet we can also rule out \edit2{X-ray/mm/radio} emission similar to AT\,2018cow by orders of magnitude.
Finally, SN\,2019deh was a rapidly evolving and luminous Type~Ibn SN that remained persistently blue, with a relatively constant effective temperature---millimeter and radio observations also resulted in non-detections, ruling out emission similar to AT\,2018cow by orders of magnitude.

To our knowledge, only two Type~Ibn SNe have X-ray detections, and both were nearby: SN\,2006jc \citep{Immler2008} and SN\,2010al \citep{Ofek2013}.
Although these two events had a similar late-time luminosity to that of AT\,2018cow ($\sim10^{40}\,\erg\,\psec$), the early-time luminosity was orders of magnitude smaller.
SN\,2006jc took 100\,d to rise to peak luminosity in X-rays, whereas AT\,2018cow rose to peak light in X-rays within three days.

So, although we cannot rule out AT2018cow-like X-ray, millimeter, and radio emission for all of the events in our sample, it appears that neither a high luminosity, nor persistent interaction, nor a constant blue color, is predictive of this behavior \edit2{on its own}.
Such emission is only seen in events that
also have a rapidly fading light curve.
This supports the idea that AT2018cow-like FBOTs are a distinct class, and that a single term is too vague for a part of parameter space that includes events as diverse as AT\,2018cow, subluminous Type~IIb SNe with shock-cooling peaks, and the well-established class of Type~Ibn SNe.

\subsection{Host Galaxies}
\label{sec:host-analysis}

In this section we present the host-galaxy properties of the objects in our sample.
In the Appendix we describe the modeling procedure, and provide a table of the
fit parameters (Table~\ref{t:sed_input})
as well as the host properties (Table~\ref{tab:host_sed}).

\begin{figure}[htb!]
\centering
\includegraphics[width=1\columnwidth]{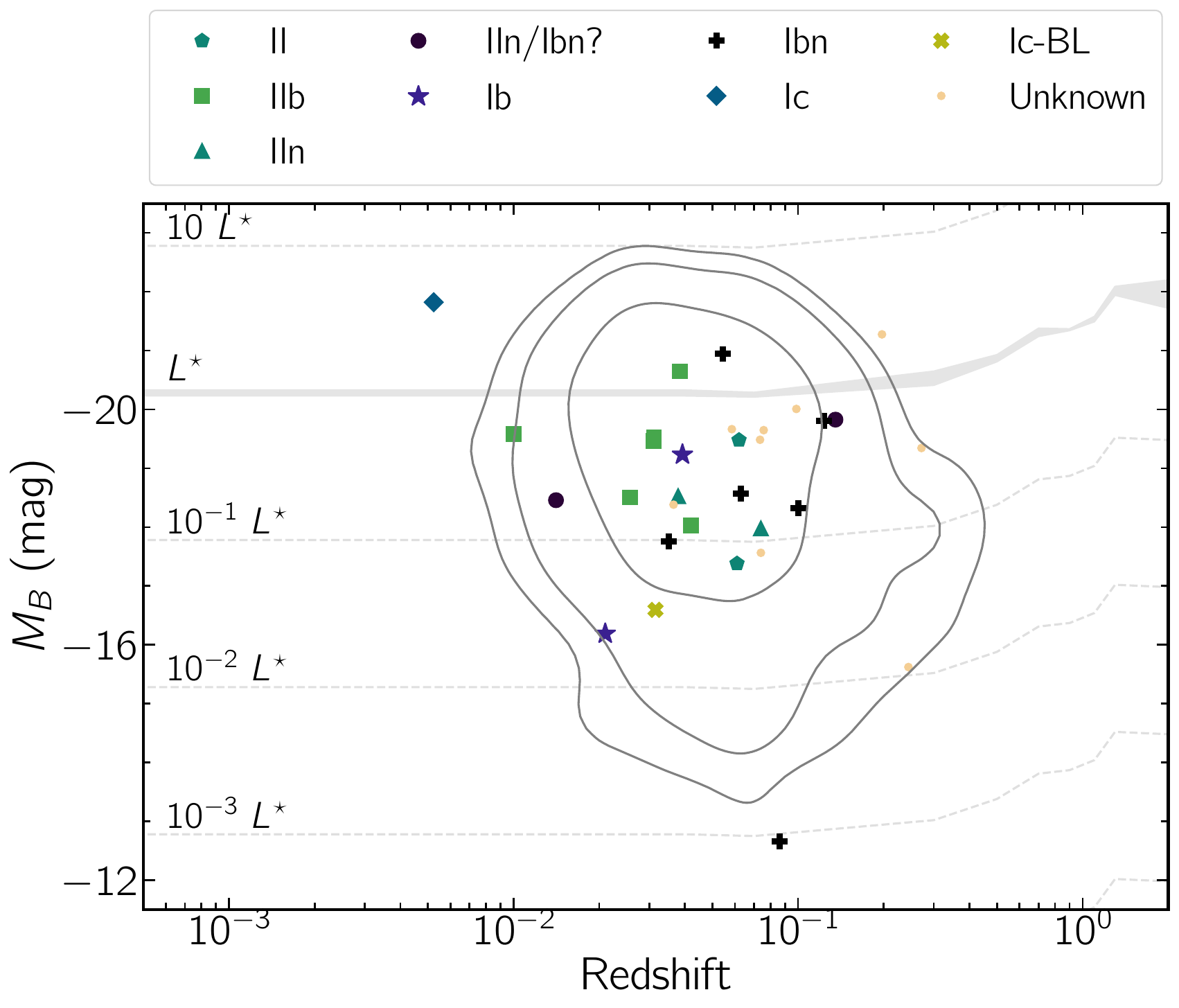}
\caption{The absolute $B$ magnitude of the host galaxies for \edit2{the ZTF FBOTs} as a function of redshift. Our events are found in the least-luminous ($10^{-3}~L^\star$) to the most luminous star-forming galaxies ($\lesssim10^{-3}~L^\star$) ($L^\star$ is the characteristic luminosity of the $B$-band luminosity function of star-forming galaxies). Most hosts have luminosities of $10^{-2}$ to a few $L^\star$, similar to regular CC SNe (indicated by the contours encircling 68, 90 and 95\% of the PTF+iPTF CC SN sample). We indicate the $L^\star$ presented in \citet{Faber2007} and multiples of it in gray.
}
\label{fig:host_MB}
\end{figure}

Figure~\ref{fig:host_MB} shows the $B$-band luminosities of the hosts of the ZTF FBOTs, which span $M_B\approx-12.7$~mag to $M_B\approx-21.8$~mag. The distribution is similar to that of regular CC SNe, which we illustrate with contours encircling 68, 90 and 95\% of the PTF+iPTF CC SN sample \citep{Schulze2021}, which includes 888 objects spanning all major CC SN classes.

One noteworthy object is the Type~Ibn SN\,2019php. We detect a $g\sim25.5\pm0.3$~mag object approximately $1''$ South-East of the transient position in Legacy Survey images. If this is indeed the host, its luminosity is $M_B\sim-12.7$~mag. Such faint galaxies are very rare but not unheard of for CC SN host galaxies \citep[e.g.,][]{Gutierrez2018a, Schulze2021}. If the marginally detected object is an image artefact, the SN\,2019php host galaxy would be even fainter and pushing into the regime of the faintest and least-massive star-forming galaxies \citep{McConnachie2012a}. It could also point to an extremely low-surface brightness galaxy \citep[e.g.,][]{vanDokkum2015a}.

\begin{figure}[htb!]
\centering
\includegraphics[width=1\columnwidth]{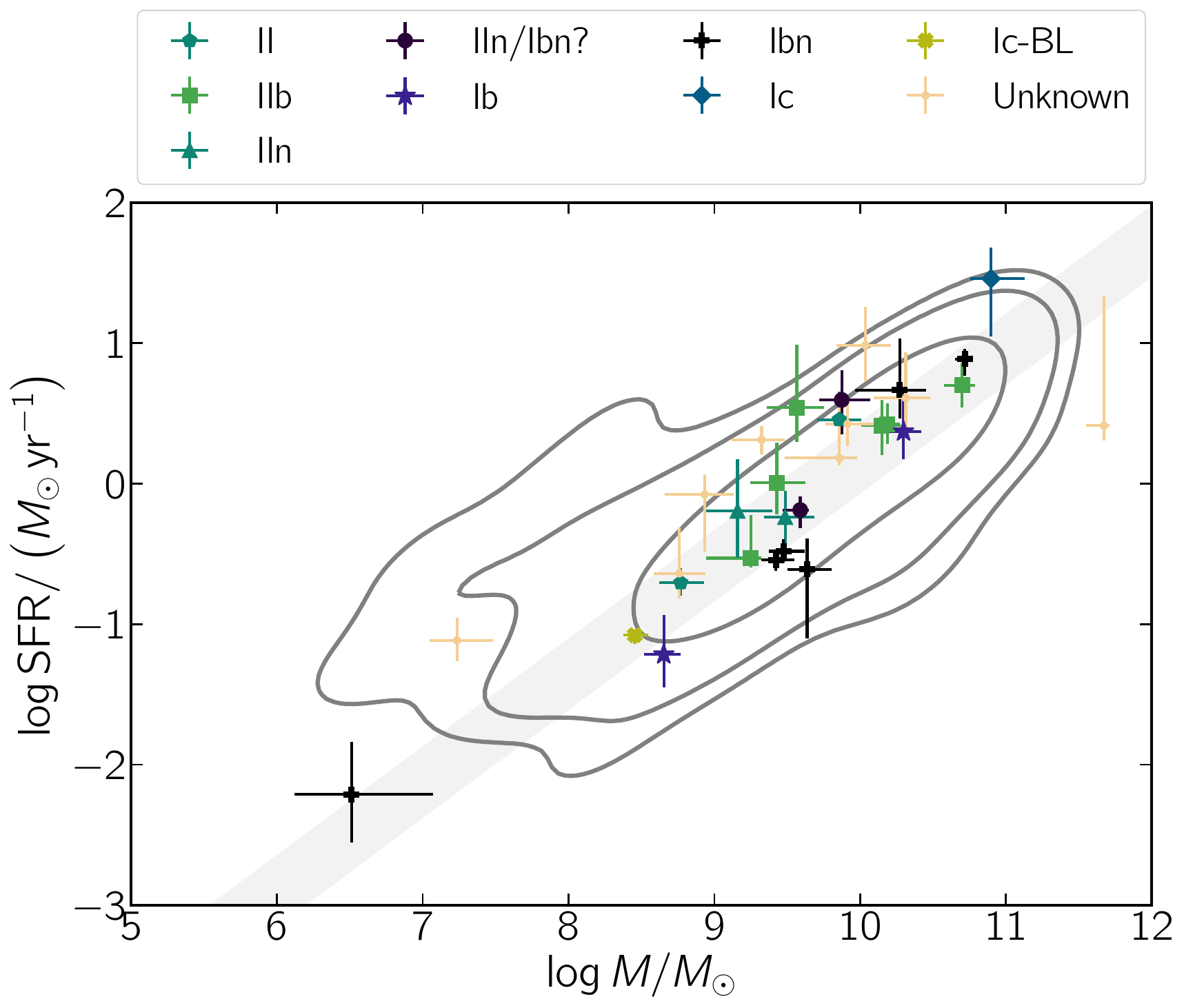}
\caption{Host galaxies for \edit2{ZTF FBOTs} in the mass-SFR plane. Almost all exploded in star-forming galaxies. This is illustrated by their location with respect to the main sequence of star-forming galaxies (grey shaded region). The only exception is AT\,2020bot, which exploded $\sim10$~kpc from the center of an elliptical galaxy. Moreover, the overwhelming majority of hosts have also properties consistent with those of CC SNe from the PTF+iPTF surveys (grey contours indicate the region encircling 68, 90 and 95\% of the sample).}
\label{fig:sfr_mass}
\end{figure}

Figure~\ref{fig:sfr_mass} shows the host properties in the mass-SFR plane.
The hosts are located along the so-called main sequence of star-forming galaxies (indicated by the grey shaded region; based on Eq. 5 in \citealt{Elbaz2007a}). A small minority of objects occurred in galaxies that lie above the galaxy main sequence and are experiencing a starburst. This phenomenon is not exclusive to a particular spectroscopic subtype. Our results are similar to \citet{Wiseman2020} who studied the hosts of rapidly evolving transients between $z=0.2$ and $z=0.85$. As in Figure~\ref{fig:host_MB}, we overlay the 68, 90 and 95\% contours of the PTF CC SN host sample. The hosts of regular CC SNe occupy the same parameter space, including the starburst regime \citep[e.g,][]{Taggart2021}.

An outstanding object is AT\,2020bot. It exploded $\approx10$~kpc from the center of an early-type galaxy. The GalaxyZoo Project classified the host morphology as elliptical \citep{Lintott2008a, Lintott2011a}. The SDSS spectrum shows no emission lines. Such an environment is extreme for any type of transient originating from the explosion of a massive star, but it is not unheard of for CC SNe \citep{Sanders2013,Irani2019,Hosseinzadeh2019,Irani2022}.
We discuss the implications in Section~\ref{sec:discussion}.

\section{The Rate of AT2018cow-like Transients}
\label{sec:rates}

\edit2{The transient AT2018cow is widely referred to in the literature as an FBOT (e.g., \citealt{Margutti2019}), and FBOTs have been reported to have a volumetric rate of 1\% of the core-collapse SN rate \citep{Drout2014,Pursiainen2018}.
However, our work shows that transients with similar properties to AT2018cow (X-ray, radio, unusual optical spectra, and rapidly fading optical light curves) are only a small subset of FBOTs\footnote{The term ``luminous FBOT'' (LFBOT) has recently been adopted for such events (e.g., \citealt{Metzger2022}). In this paper we use ``AT2018cow-like.''},
motivating a revised estimate of their rate,
which is in turn an important clue to their progenitor system.
}

We estimate the rate \edit2{of AT2018cow-like transients} using two systematic ZTF classification efforts: the volume-limited survey \edit2{(the Census of the Local Universe, or CLU; \citealt{De2020})} and the magnitude-limited survey \edit2{(the Bright Transient Survey, or BTS; \citealt{Fremling2020_RCF,Perley2020_BTS}).}
\edit2{CLU and BTS used different selection criteria from the search we performed in this paper, and both samples include AT2018cow.}

CLU aims to classify all transients down to $r=20.0\,$mag within 200\,Mpc, using data from all survey streams.
Over the timescale of our search,
CLU classified 429 CC SNe brighter than $M=-16$\,mag within 150\,Mpc.
At this distance AT\,2018cow would peak at $16\,$mag and remain over the $r=20\,$mag threshold for over two weeks, so CLU can be expected to be reasonably complete.
The primary limitation is the use of a galaxy redshift catalog \citep{Cook2019}, so we caution that our rate is only valid for the types of galaxies well represented in this catalog.
Given the detection of a single AT\,2018cow-like object (AT\,2018cow itself), and accounting for the fact that half of CC SNe are fainter than $M=-16\,$mag \citep{Li2011,Perley2020_BTS},
we find a rate of 0.1\% the CC SN rate, with a 95\% confidence interval from binomial counting statistics of [0.003\%, 0.6\%].
In absolute terms, this corresponds to a volumetric rate of 70\,\pyr\,\pgpccub.

We can also estimate the rate using the Bright Transient Survey (BTS; \citealt{Fremling2020_RCF,Perley2020_BTS}),
which aims to classify all transients down to $r=18.5\,$mag in the public survey (15,000\,\degsq).
We consider a volume of 250\,Mpc, out to which BTS should be quite complete for events like AT\,2018cow.
Using the BTS Survey Explorer\footnote{\url{https://sites.astro.caltech.edu/ztf/bts/explorer.php}}, and applying a quality cut, we find that there were 68 CC SNe classified in this volume brighter than $M=-18.5\,$mag ,
and AT\,2018cow itself.
Correcting for the SN luminosity function (1--3\% are \edit1{brighter} than this; \citealt{Perley2020_BTS})
we find a rate of 0.01\% with a 95\% confidence interval of [0.0004\%, 0.08\%].

To be conservative, we take the lower limit from the BTS and the upper limit from CLU, and estimate that the rate is \edit2{0.0004\% to 0.6\% of the local CC SN rate,
or 0.3--420\,\pyr\,\pgpccub.}
\edit2{Our rate estimate is consistent with the finding of \citet{Coppejans2020}, using data from the Palomar Transient Factory, that the rate of events with light curves identical to AT\,2018cow is $<0.4\%$ of the local CC~SN rate.}

Finally, although a measurement of the overall ``FBOT'' volumetric rate does not have a straightforward interpretation due to their heterogeneity (Figure~\ref{fig:lum-timescale}), we estimate the rate of $1<t_{1/2}<12\,$d transients in ZTF as a comparison to the high rates quoted in the literature (1\% in \citealt{Pursiainen2018}, 4--7\% in \citealt{Drout2014}).
For the faintest object in our sample ($M=-16.3\,$mag),
we conservatively estimate that the BTS would be complete out to 70\,Mpc; at this distance, the source would be brighter than the BTS threshold for one week.
There have been 55 CC SNe classified in this volume brighter than $-16.3$\,mag, of which only SN\,2018gjx would be called an FBOT (Table~\ref{tab:sources_all}), and additionally AT\,2018cow itself.
Correcting for the SN luminosity function (50\% are more luminous than $-16.3\,$mag), we find a rate of 7\% with a 95\% confidence interval of [0.9\%, 30\%]. Our lower limit of $\approx1\%$ is consistent with previous results in the literature \citep{Pursiainen2018,Drout2014}, and is dominated by events at lower luminosities (predominantly Type~IIb SNe).

% \edit1{Next, we address the rate of events similar to AT\,2018cow.
% Motivated by the high luminosity of AT\,2018cow-like objects, \citet{Coppejans2020} estimated a rate for only the brightest ($M<-19\,$mag) events and found 1--2\%.}
% For events with light curves identical to AT\,2018cow, they found a rate of $<0.4\%$ of the (local) CC~SN rate.
% Our classifications and additional data at radio, X-ray, and mm bands (\S\ref{sec:radio}) provide physical motivation for only considering the shortest-duration ($3\,$d) and most luminous ($M=-21\,$mag) events.

\section{Discussion}
\label{sec:discussion}

\begin{figure}[htb!]
    \centering
    \includegraphics[width=\columnwidth]{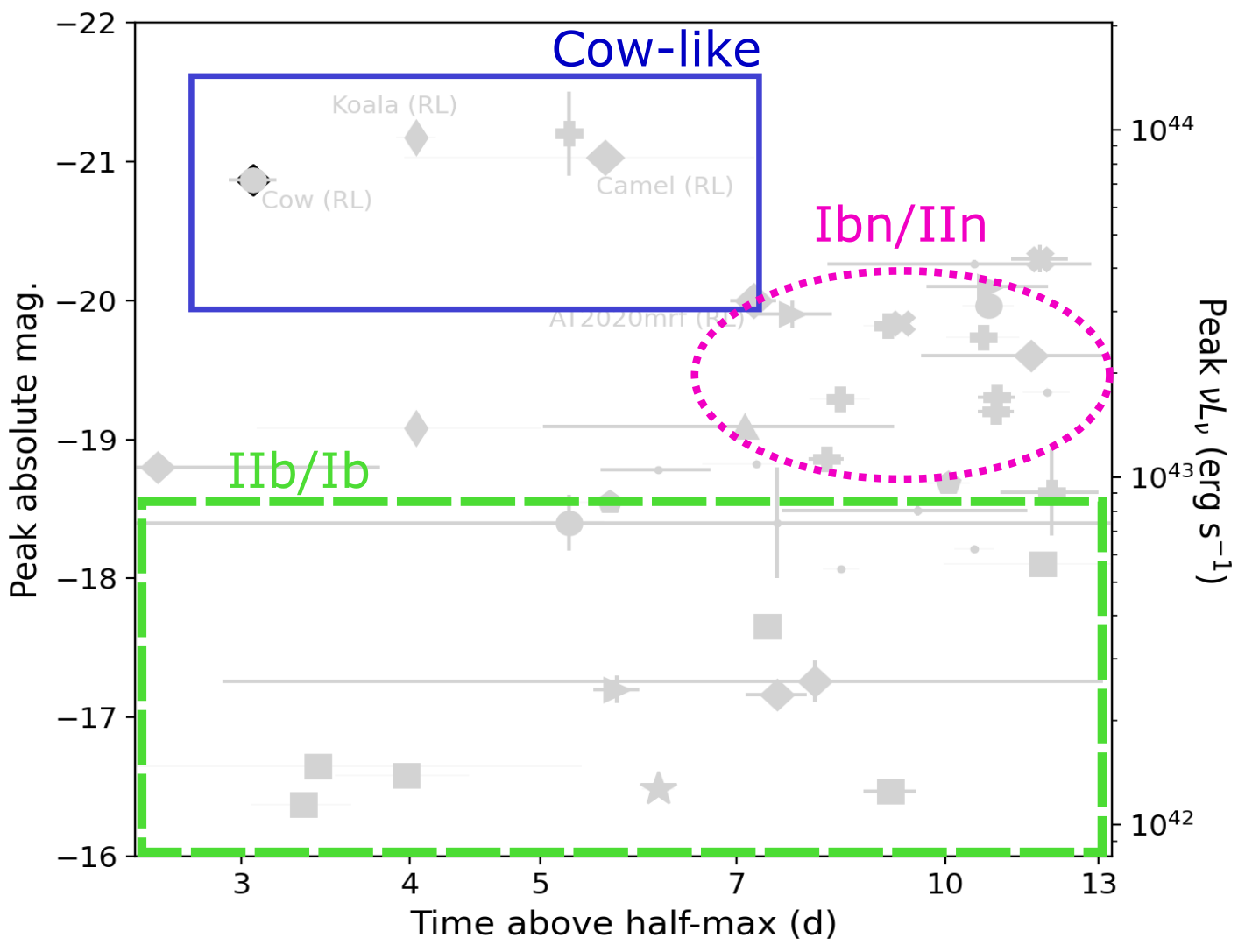}
    \caption{Same as Figure~\ref{fig:lum-timescale}, this time delineating predominant spectroscopic subgroups of FBOTs.}
    \label{fig:real-cartoon}
\end{figure}

\edit2{We have shown that ``FBOTs,'' a class previously defined primarily by photometric properties, have multiple spectroscopic subtypes, summarized in Figure~\ref{fig:real-cartoon}.}
In this section we discuss the implications of our findings for the progenitors and the powering mechanism for the optical light curves,
\edit2{and make suggestions for how to more effectively select rare exotic objects such as AT2018cow.}
%referring to our summary figure (Figure~\ref{fig:finalfig}) as a guide.

% \begin{figure*}[!p]
%     \centering
%     \includegraphics[width=0.9\textwidth]{f19.png}
%     \caption{The rise time vs. peak luminosity of the ZTF objects in our sample, the literature comparison events that meet our criteria, and several additional events from the literature that do not strictly meet our search criteria due to a second peak of comparable luminosity but which are clearly related phenomena.
%     We include two Type~Ic-BL SNe with shock-cooling peaks, SN\,2006aj \citep{Campana2006} and SN\,2020bvc \citep{Ho2020c}.
%     We include three Type~IIb SNe with shock-cooling peaks: SN\,2016gkg \citep{Bersten2018}, ZTF18aalrxas \citep{Fremling2019_rxas}, and SN\,1993J \citep{Schmidt1993}.
%     We include the first peak of the double-peaked Type~Ib SN\,2008D \citep{Modjaz2006} and the first peak of the
%     Type~Ic SN iPTF14gqr, argued to be an ultra-stripped SN \citep{De2018}.
%     Finally, we include the Type~Icn SN\,2021csp \citep{Perley2021_Icn_disc}.
%     Measurements are in the rest-frame and as close to $g$-band as possible.
%     }
%     \label{fig:finalfig}
% \end{figure*}

First, \edit2{FBOT} progenitors appear to be predominantly massive-star explosions, and most events fall into established SN spectroscopic classes.
At the subluminous ($M>-18.5\,$mag) end, the most common subtype is Type~IIb SNe.
The light curve durations, luminosities, and colors are reminiscent of the shock-cooling peaks seen in double-peaked Type~IIb SNe such as SN\,1993J \citep{Schmidt1993} and ZTF18aalrxas \citep{Fremling2019_rxas}, which are
included in Figure~\ref{fig:lum-timescale}.
In fact, we see a distinct second peak in SN\,2020ano, which is significantly less luminous than the first peak.
By analogy, it seems reasonable to conclude that shock-cooling emission plays a key role in powering our events.
We suggest that Type~IIb SNe simply have a range of relative brightness of the shock-cooling peak and nickel-powered peak, and \edit2{some FBOTs} reflect cases where the former is significantly brighter than the latter.
This could arise from material at particularly large radii (CSM), events with very low nickel masses, or both;
we defer modeling \edit2{of the Type~IIb FBOTs} to a forthcoming paper by Fremling et al.

% Early shock-cooling peaks have also been seen in Type~Ic-BL SNe, with \citep{Campana2006} and without \citep{Ho2020c} gamma-ray bursts;
% in Type~Ic SNe, argued to be an ultra-stripped SN \citep{De2018} or simply arising from enhanced pre-SN mass-loss \citep{Taddia2016};
% in a Type~Ib SNe with an X-ray flash \citep{Modjaz2006};
% and in several Type~IIb SNe,
% the subtype in which this phenomenon is most well-established \citep{Schmidt1993,Arcavi2011,Bersten2018,Fremling2019_rxas}.
% We show the first peak of several classes of stripped-envelope SNe in Figure~\ref{fig:finalfig}.

%There have been a variety of efforts to estimate the rates of rapidly evolving transients, which have typically had one of two goals:
%measure the overall rate as a clue to the underlying physical mechanism \citep{Drout2014,Pursiainen2018},
%or measure the rate of events similar to AT\,2018cow \citep{Ho2020b,Coppejans2020}.
% \citet{Drout2014} estimated a remarkably high rate of 4--7\% of the CC~SN rate for events spanning the full range of $-16>M>-20\,$mag, implying that the mechanism producing fast-luminous light curves, perhaps CSM interaction (and therefore end-of-life mass-loss in massive stars) was very common.

\edit2{FBOTs} with peak luminosities between $M_g=-20\,$mag and $M_g=-18.5$\,mag are dominated by interacting SNe, particularly those of Type~Ibn.
% In Figure~\ref{fig:finalfig} we include the rise time and peak luminosity of the Ibn light curve template from \citet{Hosseinzadeh2017}.
% We have no reason to believe that our events are significantly different from the general Type~Ibn population:
The connection of Type~Ibn SNe to fast-evolving transients has already been pointed out \citep{Karamehmetoglu2021,Fox2019,Xiang2021}, and
as discussed in \citet{Karamehmetoglu2021}
the rise time of most Type~Ibn SNe has not been well sampled, so their true duration is relatively uncertain.
Type~Ibn light curves are generally thought to be powered by CSM interaction,
with material much more extended than that involved in the shock-cooling peaks we have discussed previously.

At the highest luminosities ($M\lesssim-20\,$mag) lie the radio-loud events AT\,2018cow (the Cow), AT\,2020xnd (the Camel), and AT\,2018lug (the Koala),
\edit2{as well as the unusual Type~Ic-BL SNe SN\,2018gep and iPTF16asu.}
In the shock-interaction picture, a fast rise time and high peak luminosity arise from a fast shock speed---i.e., a significant amount of energy is coupled to ejecta traveling at high velocities.
This is likely a distinguishing characteristic of \edit2{these events.}
Indeed, these events are the only objects that show very broad absorption features in their optical spectra.

However, despite sharing several characteristics with other events---a high luminosity, a fast rise, persistent interaction and blue colors---it appears that only the AT2018cow-like events, the fastest-fading most luminous transients, are accompanied by luminous millimeter, X-ray, and radio emission.
One possible explanation is that these events are engine-powered, \edit1{while established classes of interacting SNe (like Type Ibn SNe) are not.}
\edit1{A central engine has been suggested to power AT2018cow-like events on the basis of the luminous long-lived X-ray emission, which is in excess of what would be expected from circumstellar interaction \citep{Ho2019cow,Margutti2019,Coppejans2020,Ho2021_20xnd,Bright2022}}.
\edit1{The fact that they also stand out in duration-luminosity parameter space supports a distinct origin for the AT2018cow-like events.}
\edit1{It is not yet clear whether they represent a distinct progenitor entirely.}%, or the extreme of a continuum of phenomena extending to ``ordinary'' interacting SNe.}

\edit2{It has been suggested that FBOTs as a whole have a high rate, 1\% of the CC SN rate \citep{Drout2014,Pursiainen2018}. Clearly, this rate refers to multiple classes of objects, many of which are not distinct classes but rather a subset of broader classes (Type~IIb SNe, Type~Ibn SNe).
Our work suggests that events similar to AT2018cow---the most luminous, fastest transients---are very rare, less than 0.1\% of the CC SN rate.}

% One event in our sample is particularly puzzling.
% As discussed in Section~\ref{sec:host-analysis},
% AT\,2020bot is luminous and rapidly evolving but located near an early-type galaxy.
% The host-galaxy spectrum shows no sign of star formation, and there is no visible dwarf galaxy in deep imaging.
% It is possible that some of the white-dwarf progenitor models invoked for AT\,2018cow (e.g., accretion-induced collapse; \citealt{Metzger2009,Lyutikov2019}) may be applicable here.

\edit2{By laying out the spectroscopic diversity of FBOTs classified as part of ZTF,
our work illustrates the limitations of the simple cut on duration that has has been used in the literature.
Figure~\ref{fig:lum-timescale} shows that an essential metric for selecting exotic events such as AT2018cow and SN\,2018gep is the transient's peak luminosity.
Currently, the completeness of galaxy redshift catalogs is only $\approx50\%$ even at $z=0.05$ \citep{Fremling2020_RCF}.
This fraction will be substantially improved by massively multiplexed spectroscopic surveys such as the Dark Energy Spectroscopic Instrument \citep{DESICollab}.
Transient luminosity estimates will also benefit from improved photometric redshifts from the Vera Rubin Observatory \citep{Graham2018}.
}

\section{Summary}
\label{sec:summary}

We present a systematically selected sample of FBOTs with spectroscopic classifications.
The objects in our sample are similar to unclassified events in the literature in terms of their photometric evolution, host-galaxy properties,
and continuum-dominated spectra at peak light.
By several weeks after peak light,
the objects typically redden in color,
and develop spectra classifiable as traditional classes of CC SNe.

Our work supports suggestions in the literature that the dominant physical mechanisms at work in FBOTs are shock-interaction with extended material, such as in the first peak of Type~IIb SNe, and interaction, such as in Type~Ibn SNe.
\edit2{Furthermore, we find that AT2018cow-like transients are a rare ($<0.1\%$ of CC SNe) subset of objects referred to broadly as FBOTs in the literature.}

%Subluminous ($M>-18.5\,$mag) transients dominate the high rates previously quoted in the literature, likely powered by shock-cooling emission, and the most common type is Type~IIb SNe.

%Our Figure~\ref{fig:finalfig} suggests how to move forward.
%A major limitation of our work is the arbitrary selection criterion of $t_{1/2}<12\,$d.

\edit1{Identifying exotic FBOTs still requires substantial human intervention (Table~\ref{tab:search}). Events in a similar region of luminosity-timescale space to established SN classes (e.g., the ultra-stripped SN\,2019dge) must currently be identified through brute-force spectroscopic classification. Some exotic classes (e.g., AT2018cow-like objects or the Ic-BL SN\,2018gep) stand out due to their high luminosity. However, current galaxy redshifts catalogs are highly incomplete, meaning that brute-force spectroscopic observations are still required to measure the redshift and distinguish exotic luminous objects at higher redshifts from more ordinary objects (e.g., IIb SNe) at lower redshifts. The completeness of galaxy redshifts in the local universe will increase in the next few years due to massively multiplexed surveys such as DESI, but more targeted approaches will likely be required for the intermediate-redshift ($z~0.1$--0.3) galaxies in which most AT2018cow-like transients are currently being found.}

\edit1{In addition to the need for more complete galaxy redshift catalogs, we note that} AT2018cow-like objects are primarily ultraviolet, not optical, transients. The Koala and the Camel have some of the bluest peak-light colors ($g-r=-0.6\,$mag and $g-r=-0.4$, respectively) of the transients in Table~\ref{tab:sources_all}.
Such objects might therefore be more effectively discovered using wide-field ultraviolet time-domain surveys, such as ULTRASAT \citep{Sagiv2014} and the Ultraviolet Explorer \citep{UVEX}.

\edit1{The code used to produce the plots in this paper is available in a public Github repository\footnote{\url{https://github.com/annayqho/fbot\_survey}}.}

\vspace{5mm}
\facilities{Hale, Swift, EVLA, VLA, Liverpool:2m, PO:1.2m, PO:1.5m, NOT, GTC, Sloan, AAVSO, ASKAP, Keck:1, IRAM:NOEMA, SMA, MMT, TNG, ASKAP, GALEX, PS1, CTIO:2MASS, FLWO:2MASS, WISE, NEOWISE, Blanco, Gemini:North}

\software{{\tt CASA} \citep{McMullin2007},
          {\tt astropy} \citep{Astropy2013,Astropy2018},
          {\tt matplotlib} \citep{Hunter2007},
          {\tt scipy} \citep{Virtanen2020},
          {\tt ztfquery} (Rigault 2018),
          {\tt extinction}, {\tt penquins}
}

\acknowledgements

The authors would like to thank the anonymous referees for detailed comments that greatly improved the clarity of the paper.
A.Y.Q.H. would like to thank Schuyler van Dyk for a thorough reading of the manuscript; Eliot Quataert, Dan Kasen, and Peter Nugent for useful discussions;
and Miika Pursiainen for generously sharing the data for the DES objects.
A.G.Y.'s research is supported by the EU via ERC grant No. 725161, the ISF GW excellence center, an IMOS space infrastructure grant and BSF/Transformative and GIF grants, as well as The Benoziyo Endowment Fund for the Advancement of Science, the Deloro Institute for Advanced Research in Space and Optics, The Veronika A. Rabl Physics Discretionary Fund, Minerva, Yeda-Sela and the Schwartz/Reisman Collaborative Science Program;  A.G.Y. is the recipient of the Helen and Martin Kimmel Award for Innovative Investigation.
R.L. acknowledges support from a Marie Sk\l{}odowska-Curie Individual Fellowship within the Horizon 2020 European Union (EU) Framework Programme for Research and Innovation (H2020-MSCA-IF-2017-794467). 

D.K. is supported by NSF grant AST-1816492.
E.C.K. acknowledges support from the G.R.E.A.T research environment funded by {\em Vetenskapsr\aa det}, the Swedish Research Council, under project number 2016-06012, and support from The Wenner-Gren Foundations.
A.A.M.~is funded by the Large Synoptic Survey Telescope Corporation (LSSTC), the Brinson Foundation, and the Moore Foundation in support of the LSSTC Data Science Fellowship Program; he also receives support as a CIERA Fellow by the CIERA Postdoctoral Fellowship Program (Center for Interdisciplinary Exploration and Research in Astrophysics, Northwestern University).
E.O.O. acknowledges support from the Israeli Science Foundation, The Israeli Ministry of Science, The Bi-National Science foundation, and Minerva.
A.J.C.T. acknowledges Y.-D. Hu and A. F. Azamat for their assistance regarding the GTC observation.
L.T. acknowledges support from MIUR (PRIN 2017 grant 20179ZF5KS).

Based on observations obtained with the Samuel Oschin Telescope 48-inch and the 60-inch Telescope at the Palomar Observatory as part of the Zwicky Transient Facility project. ZTF is supported by the National Science Foundation under Grant No. AST-1440341 and a collaboration including Caltech, IPAC, the Weizmann Institute for Science, the Oskar Klein Center at Stockholm University, the University of Maryland, the University of Washington, Deutsches Elektronen-Synchrotron and Humboldt University, Los Alamos National Laboratories, the TANGO Consortium of Taiwan, the University of Wisconsin at Milwaukee, and Lawrence Berkeley National Laboratories. Operations are conducted by COO, IPAC, and UW.
The ZTF forced-photometry service was funded under the Heising-Simons Foundation grant \#12540303 (PI: Graham).
The GROWTH Marshal was supported by the GROWTH project funded by the National Science Foundation under Grant No 1545949.

SED Machine is based upon work supported by the National Science Foundation under Grant No. 1106171.
The data presented here were obtained in part with ALFOSC, which is provided by the Instituto de Astrofisica de Andalucia (IAA) under a joint agreement with the University of Copenhagen and NOT.
Based on observations made with the Gran Telescopio Canarias (GTC), installed at the Spanish Observatorio del Roque de los Muchachos of the Instituto de Astrofísica de Canarias, in the island of La Palma.
Observations reported here were obtained at the MMT Observatory, a joint facility of the University of Arizona and the Smithsonian Institution.
The Liverpool Telescope is operated on the island of La Palma by Liverpool John Moores University in the Spanish Observatorio del Roque de los Muchachos of the Instituto de Astrofisica de Canarias with financial support from the UK Science and Technology Facilities Council.
Based on observations made with the Italian Telescopio Nazionale Galileo (TNG) operated on the island of La Palma by the Fundación Galileo Galilei of the INAF (Istituto Nazionale di Astrofisica) at the Spanish Observatorio del Roque de los Muchachos of the Instituto de Astrofisica de Canarias.

Some of the data presented herein were obtained at the W. M. Keck Observatory, which is operated as a scientific partnership among the California Institute of Technology, the University of California and the National Aeronautics and Space Administration. The Observatory was made possible by the generous financial support of the W. M. Keck Foundation.
The authors wish to recognize and acknowledge the very significant cultural role and reverence that the summit of Maunakea has always had within the indigenous Hawaiian community.  We are most fortunate to have the opportunity to conduct observations from this mountain.

This work made use of data supplied by the UK Swift Science Data Centre at the University of Leicester.
The National Radio Astronomy Observatory is a facility of the National Science Foundation operated under cooperative agreement by Associated Universities, Inc.
The Submillimeter Array is a joint project between the Smithsonian Astrophysical Observatory and the Academia Sinica Institute of Astronomy and Astrophysics and is funded by the Smithsonian Institution and the Academia Sinica.
This work is based on observations carried out under project number
S19BC with the IRAM NOEMA Interferometer. IRAM is supported by
INSU/CNRS (France), MPG (Germany) and IGN (Spain).
The Australian SKA Pathfinder is part of the Australia Telescope National Facility which is managed by CSIRO. Operation of ASKAP is funded by the Australian Government with support from the National Collaborative Research Infrastructure Strategy. ASKAP uses the resources of the Pawsey Supercomputing Centre. Establishment of ASKAP, the Murchison Radio-astronomy Observatory and the Pawsey Supercomputing Centre are initiatives of the Australian Government, with support from the Government of Western Australia and the Science and Industry Endowment Fund. We acknowledge the Wajarri Yamatji people as the traditional owners of the Observatory site.
Parts of this research were conducted by the Australian Research Council Centre of Excellence for Gravitational Wave Discovery (OzGrav), through project number CE170100004.

This research made use of Astropy, a community-developed core Python package for Astronomy \citep{Astropy2013,Astropy2018}.
The ztfquery code was funded by the European Research Council (ERC) under the European Union's Horizon 2020 research and innovation programme (grant agreement n°759194 - USNAC, PI: Rigault).

The Legacy Surveys consist of three individual and complementary projects: the Dark Energy Camera Legacy Survey (DECaLS; Proposal ID \#2014B-0404; PIs: David Schlegel and Arjun Dey), the Beijing-Arizona Sky Survey (BASS; NOAO Prop. ID \#2015A-0801; PIs: Zhou Xu and Xiaohui Fan), and the Mayall z-band Legacy Survey (MzLS; Prop. ID \#2016A-0453; PI: Arjun Dey). DECaLS, BASS and MzLS together include data obtained, respectively, at the Blanco telescope, Cerro Tololo Inter-American Observatory, NSF’s NOIRLab; the Bok telescope, Steward Observatory, University of Arizona; and the Mayall telescope, Kitt Peak National Observatory, NOIRLab. The Legacy Surveys project is honored to be permitted to conduct astronomical research on Iolkam Du’ag (Kitt Peak), a mountain with particular significance to the Tohono O’odham Nation.

This project used data obtained with the Dark Energy Camera (DECam), which was constructed by the Dark Energy Survey (DES) collaboration. 
Funding for the DES Projects has been provided by the U.S. Department of Energy, the U.S. National Science Foundation, the Ministry of Science and Education of Spain, the Science and Technology Facilities Council of the United Kingdom, the Higher Education Funding Council for England, the National Center for Supercomputing Applications at the University of Illinois at Urbana-Champaign, the Kavli Institute of Cosmological Physics at the University of Chicago, Center for Cosmology and Astro-Particle Physics at the Ohio State University, the Mitchell Institute for Fundamental Physics and Astronomy at Texas A\&M University, Financiadora de Estudos e Projetos, Fundacao Carlos Chagas Filho de Amparo, Financiadora de Estudos e Projetos, Fundacao Carlos Chagas Filho de Amparo a Pesquisa do Estado do Rio de Janeiro, Conselho Nacional de Desenvolvimento Cientifico e Tecnologico and the Ministerio da Ciencia, Tecnologia e Inovacao, the Deutsche Forschungsgemeinschaft and the Collaborating Institutions in the Dark Energy Survey. The Collaborating Institutions are Argonne National Laboratory, the University of California at Santa Cruz, the University of Cambridge, Centro de Investigaciones Energeticas, Medioambientales y Tecnologicas-Madrid, the University of Chicago, University College London, the DES-Brazil Consortium, the University of Edinburgh, the Eidgenossische Technische Hochschule (ETH) Zurich, Fermi National Accelerator Laboratory, the University of Illinois at Urbana-Champaign, the Institut de Ciencies de l'Espai (IEEC/CSIC), the Institut de Fisica d'Altes Energies, Lawrence Berkeley National Laboratory, the Ludwig Maximilians Universitat Munchen and the associated Excellence Cluster Universe, the University of Michigan, NSF's NOIRLab, the University of Nottingham, the Ohio State University, the University of Pennsylvania, the University of Portsmouth, SLAC National Accelerator Laboratory, Stanford University, the University of Sussex, and Texas A\&M University.
The Legacy Survey team makes use of data products from the Near-Earth Object Wide-field Infrared Survey Explorer (NEOWISE), which is a project of the Jet Propulsion Laboratory/California Institute of Technology. NEOWISE is funded by the National Aeronautics and Space Administration.
The Legacy Surveys imaging of the DESI footprint is supported by the Director, Office of Science, Office of High Energy Physics of the U.S. Department of Energy under Contract No. DE-AC02-05CH1123, by the National Energy Research Scientific Computing Center, a DOE Office of Science User Facility under the same contract; and by the U.S. National Science Foundation, Division of Astronomical Sciences under Contract No. AST-0950945 to NOAO.

\appendix

\section{Details of individual events}
\label{sec:discovery-details}

\edit1{Here we provide details on the discovery and follow-up observations of events in our sample that have not yet been published elsewhere.}

\subsection{SN2018ghd / ZTF18abvkmgw / ATLAS18vew }

SN\,2018ghd was detected by ATLAS \citep{Tonry2018,Smith2020} on 2018 September 14 and reported to TNS the same day \citep{Tonry_ZTF18abvkmgw}.
It was first detected in ZTF data as ZTF18abvkmgw on 2018 September 12 as part of the Caltech 1DC survey at $g=20.48\pm0.22\,$mag, and saved by an alert-stream scanner on September 13 as part of a filter for rapidly evolving transients.
It was in a galaxy with an SDSS spectrum and known redshift of $z=0.0385$.
On September 15 it was saved by the CLU filter, and by the public BTS survey on September 16.
As part of CLU and BTS, it received a series of SEDM spectra, with the first obtained on September 14.
These spectra were not definitive for classification.
It was classified as a Type~II SN based on an SEDM spectrum on September 21 \citep{Fremling_ZTF18abvkmgw}, then reclassified as a Type~Ib SN with an LRIS spectrum on November 10.

\subsection{SN2018gjx / ZTF18abwkrbl / ATLAS18vis / Gaia18csc / kait-18ao / PS19do / PSP18C}

SN\,2018gjx was discovered by the Xingming Observatory Sky Survey (XOSS) as PSP18C on 2018 September 15, and reported to TNS on September 16 \citep{Zhang_ZTF18abwkrbl}.
The source was coincident with NGC 865 ($z=0.00999$).
The first ZTF detection was also on 2018 September 15, at $g=17.91\pm0.06\,$mag as part of the Caltech 1DC survey.
The source was saved by alert-stream scanners on 2018 September 17 as part of the infant SN and CLU programs, and an SEDM spectrum was triggered which showed flash features; an SEDM spectrum obtained the next day showed that the features had disappeared. It was also saved as part of BTS on September 17, as it exceeded the 19th magnitude threshold in an image obtained as part of the public survey ($r=16.16\pm0.04\,$mag).

ePESSTO \citep{Smartt2015} classified the source as SN~II based on a September 18 spectrum obtained with the ESO Faint Object Spectrograph and Camera (EFOSC2) on the 3.6m New Technology Telescope (NTT) at La Silla \citep{Gromadzki2018_ZTF18abwkrbl}.
Based on an October 12 SEDM spectrum the classification was revised to SN~IIb \citep{Dahiwale2020}.

\subsection{SN2019aajs / ZTF19aakssbm}

SN\,2019aajs was discovered by ZTF on 2019 Feb. 25 at $r=19.10\pm0.17\,$mag in an image obtained as part of the high-cadence partnership survey.
It was saved on Feb. 26 as part of a search for rapidly evolving transients,
because it rose 1.5 mag in 1 day.
This led to an extensive sequence of follow-up observations, including imaging, spectroscopy, millimeter, and radio.
The object was classified as a Type~Ibn SN using an LT spectrum taken on 2019 Mar. 02.

\subsection{SN2019deh / ZTF19aapfmki / ATLAS19gez / PS19aaq}

SN\,2019deh was first detected in a ZTF public-survey image on 2019 Apr. 7 at $r=20.75\pm0.28\,$mag,
and again the same night as part of the 1DC survey.
It was reported to the TNS on Apr. 10 \citep{Nordin2019_ZTF19aapfmki} by
the alert management, photometry, and evaluation of light curves (AMPEL) system \citep{Nordin2019_AMPEL,Soumagnac2018}.
It was classified as a Type~Ibn SN by SPRAT using a spectrum obtained on 2019 Apr. 12 \citep{Prentice2019_ZTF19aapfmki}.

\subsection{AT2019esf / ZTF19aatoboa / PS19afa}

AT\,2019esf was discovered by ZTF in an image obtained 
on 2019 May 3 at $g=19.97\pm0.16\,$mag as part of the public survey, and detected the next night as part of both the high-cadence and 1DC surveys.
It was saved on May 4 by filters for infant supernovae and fast transients.
As part of the rapidly evolving transients program it received a P60 spectrum at peak light that did not show distinct features.
It was uploaded to TNS by AMPEL on May 6 \citep{Nordin2019_ZTF19aatoboa}.
The host galaxy redshift was measured with a Keck spectrum on 2020 Feb 17.

\subsection{AT2019kyw / ZTF19abfarpa}

AT\,2019kyw was discovered by ZTF \citep{Fremling2019_ZTF19abfarpa}, first detected at 
$g=20.18\pm0.31\,$mag in an image obtained on 2019 Jul 6 as part of the public survey.
It was saved as part of BTS, infant supernovae, and CLU, and received an inconclusive SEDM spectrum on July 9 as part of routine classification efforts.
It received additional DBSP spectra on Aug 1 and Aug 9 that led to a redshift measurement but no conclusion about the transient itself.

\subsection{SN2019myn / ZTF19abobxik / PS19eop}

SN\,2019myn was discovered by ZTF \citep{Nordin2019}, first detected at $g=21.24\pm0.31\,$mag in an image obtained on 2019 Aug 05 as part of the high-cadence partnership survey. It was saved by a filter for rapidly evolving transients on Aug 11,
and SEDM and LT were triggered for spectroscopy and imaging.
Given the rapid evolution, the VLA was triggered and the observation took place on Aug 17. % Dan's program
An LRIS spectrum on Aug 31 led to the Type~Ibn classification.
The source will be included in a Type~Ibn sample paper by Kool et al.

\subsection{SN2019php / ZTF19abuvqgw / ATLAS19ufu}

SN\,2019php was discovered by ATLAS on September 2 and reported to TNS that day \citep{Tonry2019_ZTF19abuvqgw}.
The first ZTF detection was on 2019 Aug 31 as part of the public survey,
and passed the AMPEL filter \citep{Nordin2019_AMPEL}.
It was detected the next night (September 1) as part of the Caltech 1DC Survey and passed a filter for fast transients.
As part of the fast-transient program, it received a spectrum with DBSP on September 9 that was relatively featureless.
It received an additional spectrum on September 23 with LRIS that led to the Type~Ibn classification.

\subsection{SN2019qav / ZTF19abyjzvd / PS19fbn}

SN\,2019qav was discovered by Pan-STARRS1 \citep{Chambers2016} on September 11 and reported to TNS on September 12 \citep{Chambers2019_ZTF19abyjzvd}.
The first ZTF detection was at $r=20.31\pm0.24\,$mag on 2019 September 8 as part of the partnership high-cadence survey.
It was saved on September 12 by the infant SN filter,
and SEDM was triggered for a spectrum.
On September 14 it was recognized that the rise was unusually fast.
An LRIS spectrum on September 24 showed H and He features and led to the measurement of $z=0.137$.
Given the unusual spectrum, it was thought that this might be an analog to AT\,2018cow, and as a result a variety of facilities were triggered: Swift, NOEMA, and the VLA.
We obtained a spectrum of the host galaxy with LRIS on 2021 Apr 14, leading to a more precise redshift ($z=0.1353$) from strong starforming emission lines.

\subsection{SN2019rii / ZTF19acayojs / ATLAS19wqu}

SN\,2019rii was first identified in the ZTF public stream by ALeRCE broker \citep{Forster2021},
and reported to TNS on September 28.
The first ZTF detection was on 2019 September 25 at $g=20.30\pm0.21\,$mag as part of the high-cadence partnership survey.
It was saved on October 2 as part of a filter for rapidly evolving transients.
It was observed the same night with DBSP, leading to the redshift measurement of $z=0.1234$ from narrow emission lines from the host galaxy.
An additional spectrum was obtained on October 26 with LRIS, which showed distinct \ion{He}{1} lines.
Given the He features and rapid evolution it was tentatively classified as a Type~Ibn.

\subsection{SN2019rta / ZTF19accjfgv} 

SN\,2019rta was first detected on 2019 October 3 at $g=17.96\pm0.07\,$mag in the ZTF public survey, and reported to TNS by AMPEL \citep{Nordin2019_AMPEL} the same day \citep{Nordin2019_ZTF19accjfgv}. The source was saved on October 3 as part of the BTS,
and SEDM was triggered for a spectrum.
It was saved again on October 5 as part of CLU,
and a DBSP spectrum obtained that night showed a blue continuum and strong emission lines from the host galaxy.
A final spectrum was obtained with LRIS on October 27 that led to the Type~IIb classification \citep{Dahiwale2019_ZTF19accjfgv}.

\subsection{SN2020ano / ZTF20aahfqpm}

SN\,2020ano was first detected on 2020 January 23 at $i=19.93\pm0.21\,$mag in an image obtained as part of the ZTF Uniform Depth Survey (ZUDS).
It was also detected in a public image, and reported to TNS the same day by ALeRCE \citep{Forster2020_ZTF20aahfqpm}.
It was saved by the AmpelRapid filter, and SEDM was triggered.
The spectrum showed primarily a blue continuum.
It was also saved that day by a filter for fast transients.

The next day (January 24) it was saved by scanners as part of the CLU experiment due to its proximity to a galaxy at $z=0.0311$.
By this day, it was clear that it was fading quickly.
From the blue colors and rapid behavior, it was thought to perhaps be a foreground CV.
A GMOS-N spectrum was obtained on Jan 29, and by Jan 31 it was clear from ZUDS photometry that it was rising again in all three filters---this became clear in the regular alerts by Feb 6.
An LRIS spectrum obtained on Feb 18 showed a good match to SN1993J, leading to the Type~IIb classification.

\subsection{AT2020bdh / ZTF20aaivtof / ATLAS20elz}

AT\,2020bdh was discovered by ALeRCE on 2020 Jan 27 using the ZTF public stream and reported to TNS the same day \citep{Forster2020_ZTF20aaivtof}.
The magnitude at discovery was $g=18.69\pm0.07\,$mag.
It was also detected the next night as part of the 1DC survey.
It was saved on Jan 29 as part of CLU, on Feb 2 as part of a search for rapidly evolving transients, and on Feb 3 as part of the BTS.
SEDM was triggered but observations were unsuccessful.
A broad H$\alpha$ feature was noted in a report to TNS \citep{Smith2020_ZTF20aaivtof} from an ePESSTO spectrum obtained on Feb 13.
A DBSP spectrum obtained as part of routine ZTF classification led to a redshift measurement but not a definitive classification.
The spectrum also tentatively showed a broad emission feature around H$\alpha$.

\subsection{AT2020bot / ZTF20aakypiu / PS20va}

AT\,2020bot was discovered by ZTF at $r=20.41\pm0.24\,$mag in an image obtained as part of ZUDS on 2020 Jan 30.
On Feb 1 it passed the infant SN filter.
Its proximity to an SDSS galaxy of known redshift ($z=0.197$) implied a high luminosity;
together with the fast rise, this led us to initiate follow-up observations, including LT and P60 imaging.
DBSP and GMOS-N spectra were inconclusive, although the GMOS-N spectrum was noted to have broad features somewhat similar to young Ic-BL SNe.
Attempts at follow-up spectroscopy with DBSP and Keck were unsuccessful.

\subsection{SN2020ikq / ATLAS20lfu / ZTF20aaxhzhc / PS20ctw}

SN\,2020ikq was discovered by ATLAS and reported to the TNS on 2020 April 28 \citep{Tonry2020_ZTF20aaxhzhc}.
It was first detected by ZTF and saved by a scanner on April 29 at $g=18.42\pm0.08\,$mag in public-survey data, as part of BTS and CLU.
SN\,2020ikq was classified as a Type~IIb SN by the NOT on 2020 May 15 \citep{Angus2020}.

\subsection{SN2020jmb / ZTF20aayrobw / ATLAS20lwn}

SN\,2020jmb was first detected on 
2020 May 08 at $g=19.52\pm0.21\,$mag in the Caltech one-day cadence survey.
The source was saved by alert-stream scanners on 2020 May 10 as part of a search for rapidly evolving transients, and separately as part of the CLU experiment due to its proximity to a galaxy at $z=0.032$ (the transient later proved unassociated).
It was reported to TNS as part of CLU \citep{De2020_ZTF20aayrobw_ZTF20aazchcq}.
The source was saved as part of BTS on 2020 May 11, when it exceeded the 19th magnitude threshold in an image obtained as part of the public survey ($g=18.56\pm0.07\,$mag).
A spectrum was obtained with the SEDM on 2020 May 11 under the CLU program, which showed no distinct features.
Additional spectra were obtained with the LT on May 16 and the SEDM on May 23, neither of which showed distinct features.
Finally, a spectrum was obtained on May 27 with DBSP on the P200 for the rapidly evolving transients program, which showed a prominent H$\alpha$ feature and narrow emission lines from the host galaxy consistent with $z=0.061$, leading to the classification as a Type~II SN \citep{Dahiwale2020_ZTF20aayrobw}.

\subsection{SN2020jji / ZTF20aazchcq / ATLAS20mfw / PS20czx}

SN\,2020jji was first detected in the ZTF public survey on 2020 May 1 at $r=20.53\pm0.30\,$mag. 
It was saved on May 10 as part of the AMPEL and CLU filters, and reported to TNS that day \citep{De2020_ZTF20aayrobw_ZTF20aazchcq}.
It received a follow-up spectrum by the SEDM that night, which was inconclusive.
On May 16 it was saved by a scanner as part of the fast transients program, and received a DBSP spectrum as part of that effort. The DBSP spectrum showed a Type~IIn classification.

\subsection{AT2020kfw / ZTF20ababxjv / ATLAS20nfg}

AT\,2020kfw was first detected by ZTF on 2020 May 17 at $r=20.53\pm0.20\,$mag in an image obtained as part of the public survey,
and reported to TNS by ALeRCE \citep{Forster2020_ZTF20ababxjv}.
It was detected later that night in the 1DC survey.
On May 23 it had peaked and started fading, and as a result passed a filter for fast transients.
The resulting DBSP spectrum was of low quality and not conclusive.

\subsection{AT2020aexw / ZTF20abmocba}

AT\,2020aexw was discovered by ZTF, first detected on 2020 July 18 at $r=20.52\pm0.21\,$mag as part of ZUDS.
It was saved on July 19 by AMPEL and on July 26 by the rapidly evolving transients program.
LT follow-up imaging was acquired.
A NOT spectrum was attempted but not successful.
A DBSP spectrum was obtained on Aug 12 resulting in a redshift measurement from very strong emission lines.

\subsection{AT2020yqt / ZTF20abummyz / PS20ksm}

AT\,2020yqt was discovered by Pan-STARRS1 on 2020 Aug 22 and reported to TNS on 2020 November 1 \citep{Chambers2020_ZTF20abummyz}.
It was first detected as part of the ZTF 1DC survey on 2020 Aug 19 at $r=19.95\pm0.17\,$mag, and saved by a scanner on Aug 20 as part of a filter for rapidly evolving transients.
The host galaxy had an SDSS spectrum which classified it as a starburst, with a redshift $z=0.09855\pm0.00001$.
Due to the fast evolution, we triggered a Gemini ToO program (PI: A. Miller) and obtained a GMOS-N spectrum on Aug 25 that was primarily featureless.
Subsequent spectra with P200 and LRIS did not show obvious supernova features and were dominated by host-galaxy light.

\subsection{SN2020rsc / ZTF20aburywx / ATLAS20xxj}

SN\,2020rsc was first detected on 2020 Aug 19 as part of the public survey at $g=19.58\pm0.17\,$mag,
and reported to TNS by ALeRCE \citep{Forster2020_ZTF20aburywx}.
It was also detected that night as part of the Caltech 1DC survey, and
saved on 2020 Aug 20 by a filter for rapidly evolving transients.
The next day, it was noted that it was already fading, and a Gemini ToO was triggered.
The GMOS-N spectrum on Aug 22 showed \ion{He}{1} at 8000\,\km\,\psec.
On 2020 Aug 22 the source passed the CLU filter.
An additional spectrum with MMT+Binospec was obtained on Aug 24, and a GTC spectrum was obtained on Aug 25.
Swift was triggered and observed on Aug 26, and the VLA was triggered on Sept 2 with the observation taking place on Sept 9.
A final Keck spectrum on Sept 15 led to the Type~IIb classification.

\subsection{SN2020vyv / ZTF20acigusw / ATLAS20bdhi / PS20kra}

SN\,2020vyv was first detected in the ZTF public survey on 2020 October 12 at $r=19.17\pm0.10\,$mag.
It was saved the same day by AMPEL and the BTS survey, and reported to TNS \citep{Fremling2020_ZTF20acigusw}.
It was classified as a Type~II SN with a Keck/LRIS spectrum on 2020-10-14 \citep{Siebert2020}.

\subsection{SN2020xlt / ZTF20aclfmwn}

SN\,2020xlt was first detected on 2020 October 19 in public-survey data at $r=19.93\pm0.22$,
and was reported to TNS that day by ALeRCE \citep{Forster2020_ZTF20aclfmwn}.
It was also detected that night as part of the Caltech 1DC survey, and the next night as part of the high-cadence partnership survey.
It was saved on October 19 as part of AMPEL \citep{Nordin2019_AMPEL} and the infant supernova program, and SEDM was triggered: the spectrum was featureless.
Despite the proximity to a bright extended galaxy, the redshift was unknown.
The transient was noted to be fading on October 22, and on October 26 it passed a filter for rapidly evolving transients, leading to a GTC+OSIRIS spectrum on 2020 October 30 that enabled the redshift measurement of $z=0.0389$ and the classification as a Type~IIb.

\section{Photometric evolution of individual events}
\label{sec:appendix-lc}

\begin{figure}[!h]
    \centering
    \includegraphics[width=0.9\textwidth]{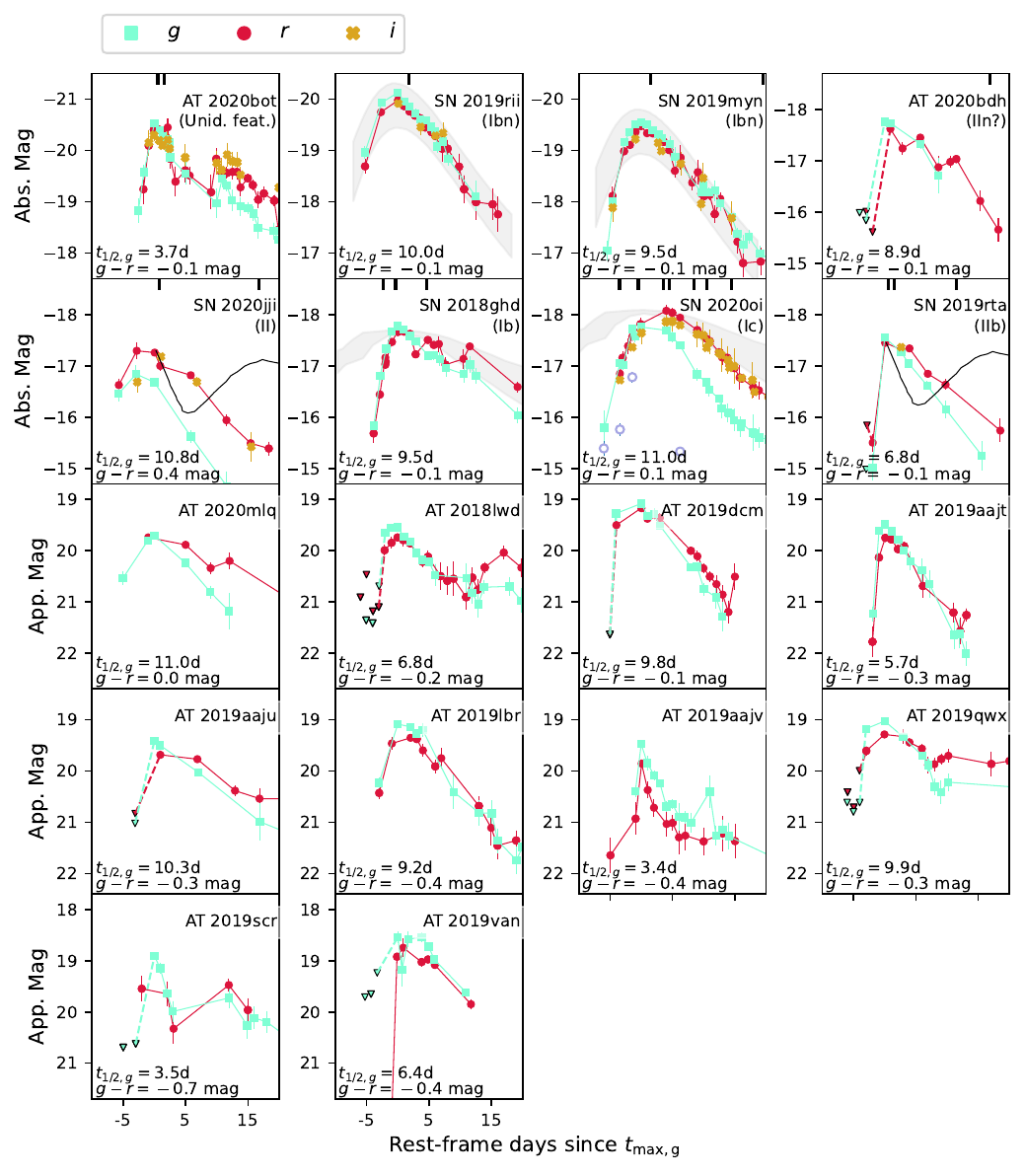}
    \caption{Light curves for \edit2{ZTF transients in Table~\ref{tab:sources_all} that are not shown in Figure~\ref{fig:lc-representative}.} In panels with H-poor SNe we show the Type~Ibc template from \citet{Drout2011} for reference. In the Type~II and Type~IIb panels we show the $V$-band light curve of SN\,1993J \citep{Schmidt1993} for reference. \edit1{In some cases light curves have been binned by day for clarity.}
}
    \label{fig:lc-remaining-gold}
\end{figure}

In Table~\ref{fig:lc-remaining-gold} we provide light curves for the Table~\ref{tab:sources_all} transients that are not included in Figure~\ref{fig:lc-representative}.

\newpage

\section{Log of Optical Spectra}
\label{sec:appendix-spec}

In Table~\ref{tab:spec-log} we provide the full log of optical spectra, some of which were obtained from TNS.
We do not report observations for objects with extensive spectroscopic observations previously presented in the literature: AT\,2018cow \citep{Perley2019cow}, SN\,2018gep \citep{Ho2019gep}, SN\,2019dge \citep{Yao2020}, and SN\,2020oi.

\startlongtable 
\begin{deluxetable*}{lrrrr}
\tablecaption{Log of spectroscopic observations of objects presented in this paper. Phase given with respect to the observed maximum of the $g$-band light curve.\label{tab:spec-log}} 
\tablewidth{0pt} 
\tablehead{ \colhead{Name} & \colhead{UT Date} & \colhead{Target} & \colhead{Phase} & \colhead{Telescope + Instrument} } 
\tabletypesize{\scriptsize} 
\startdata 
SN\,2018ghd & 20180914 & Transient & $-2$ & P60+SEDM \\ 
SN\,2018ghd & 20180916 & Transient & $0$ & P60+SEDM \\ 
SN\,2018ghd & 20180921 & Transient & $4$ & P60+SEDM \\ 
SN\,2018ghd & 20181110 & Transient & $54$ & Keck1+LRIS \\ 
SN\,2018ghd & 20190105 & Transient & $110$ & Keck1+LRIS \\ 
SN\,2018gjx & 20180918 & Transient & $0$ & P60+SEDM \\ 
SN\,2018gjx & 20180918 & Transient & $0$ & EFOSC2+NTT [1] \\ 
SN\,2018gjx & 20180919 & Transient & $0$ & P60+SEDM \\ 
SN\,2018gjx & 20181012 & Transient & $23$ & P60+SEDM \\ 
SN\,2018gjx & 20181016 & Transient & $27$ & NOT+ALFOSC \\ 
SN\,2018gjx & 20181110 & Transient & $52$ & Keck1+LRIS \\ 
SN\,2018gjx & 20181130 & Transient & $72$ & TNG+DOLORES \\ 
SN\,2018gjx & 20190105 & Transient & $108$ & Keck1+LRIS \\ 
SN\,2018gjx & 20190706 & Transient & $290$ & Keck1+LRIS \\ 
SN\,2019aajs & 20190227 & Transient & $-1$ & LT+SPRAT \\ 
SN\,2019aajs & 20190302 & Transient & $1$ & LT+SPRAT \\ 
SN\,2019aajs & 20190304 & Transient & $3$ & NOT+ALFOSC \\ 
SN\,2019aajs & 20190315 & Transient & $14$ & NOT+ALFOSC \\ 
SN\,2019aajs & 20190406 & Transient & $36$ & Keck1+LRIS \\ 
SN\,2019deh & 20190410 & Transient & $-4$ & P60+SEDM \\ 
SN\,2019deh & 20190410 & Transient & $-4$ & LT+SPRAT \\ 
SN\,2019deh & 20190411 & Transient & $-3$ & LT+SPRAT \\ 
SN\,2019deh & 20190412 & Transient & $-2$ & LT+SPRAT \\ 
SN\,2019deh & 20190414 & Transient & $0$ & LT+SPRAT \\ 
SN\,2019deh & 20190415 & Transient & $0$ & P60+SEDM \\ 
SN\,2019deh & 20190423 & Transient & $8$ & NOT+ALFOSC \\ 
SN\,2019deh & 20190423 & Transient & $8$ & P60+SEDM \\ 
SN\,2019deh & 20190424 & Transient & $9$ & P200+DBSP \\ 
SN\,2019deh & 20190428 & Transient & $13$ & P60+SEDM \\ 
SN\,2019deh & 20190511 & Transient & $26$ & NOT+ALFOSC \\ 
AT\,2019esf & 20190504 & Transient & $-2$ & P60+SEDM \\ 
AT\,2019esf & 20200218 & Host & $287$ & Keck1+LRIS \\ 
AT\,2019kyw & 20190709 & Transient & $-3$ & P60+SEDM \\ 
AT\,2019kyw & 20190801 & Transient & $19$ & P200+DBSP \\ 
AT\,2019kyw & 20190809 & Transient & $27$ & P200+DBSP \\ 
SN\,2019myn & 20190813 & Transient & $1$ & P60+SEDM \\ 
SN\,2019myn & 20190831 & Transient & $19$ & Keck1+LRIS \\ 
SN\,2019php & 20190907 & Transient & $2$ & P200+DBSP \\ 
SN\,2019php & 20190924 & Transient & $19$ & Keck1+LRIS \\ 
SN\,2019qav & 20190911 & Transient & $-2$ & P60+SEDM \\ 
SN\,2019qav & 20190924 & Transient & $10$ & Keck1+LRIS \\ 
SN\,2019qav & 20190928 & Transient & $14$ & Keck1+LRIS \\ 
SN\,2019qav & 20191027 & Transient & $43$ & Keck1+LRIS \\ 
SN\,2019rii & 20191003 & Transient & $1$ & P200+DBSP \\ 
SN\,2019rii & 20191027 & Transient & $25$ & Keck1+LRIS \\ 
SN\,2019rta & 20191004 & Transient & $0$ & P60+SEDM \\ 
SN\,2019rta & 20191005 & Transient & $1$ & P200+DBSP \\ 
SN\,2019rta & 20191015 & Transient & $11$ & P60+SEDM \\ 
SN\,2019rta & 20191027 & Transient & $23$ & Keck1+LRIS \\ 
SN\,2020ano & 20200125 & Transient & $1$ & P60+SEDM \\ 
SN\,2020ano & 20200129 & Transient & $5$ & Gemini+GMOS \\ 
SN\,2020ano & 20200214 & Transient & $21$ & P200+DBSP \\ 
SN\,2020ano & 20200218 & Transient & $25$ & Keck1+LRIS \\ 
AT\,2020bdh & 20200213 & Transient & $16$ & EFOSC2+NTT [2] \\ 
AT\,2020bdh & 20200226 & Transient & $29$ & P200+DBSP \\ 
AT\,2020bot & 20200202 & Transient & $0$ & P200+DBSP \\ 
AT\,2020bot & 20200203 & Transient & $1$ & Gemini+GMOS \\ 
SN\,2020ikq & 20200429 & Transient & $-3$ & P60+SEDM \\ 
SN\,2020ikq & 20200503 & Transient & $0$ & P60+SEDM \\ 
SN\,2020ikq & 20200510 & Transient & $7$ & LT+SPRAT \\ 
SN\,2020ikq & 20200511 & Transient & $8$ & P60+SEDM \\ 
SN\,2020ikq & 20200515 & Transient & $12$ & NOT+ALFOSC \\ 
SN\,2020ikq & 20200517 & Transient & $14$ & P60+SEDM \\ 
SN\,2020jmb & 20200511 & Transient & $-1$ & P60+SEDM \\ 
SN\,2020jmb & 20200516 & Transient & $3$ & LT+SPRAT \\ 
SN\,2020jmb & 20200523 & Transient & $10$ & P60+SEDM \\ 
SN\,2020jmb & 20200528 & Transient & $15$ & P200+DBSP \\ 
SN\,2020jji & 20200511 & Transient & $0$ & P60+SEDM \\ 
SN\,2020jji & 20200527 & Transient & $16$ & P200+DBSP \\ 
AT\,2020kfw & 20200528 & Transient & $5$ & P200+DBSP \\ 
% SN\,2020ntt & 20200713 & Transient & $9$ & LT+SPRAT \\ 
% SN\,2020ntt & 20200715 & Transient & $11$ & P60+SEDM \\ 
AT\,2020aexw & 20200812 & Host & $21$ & P200+DBSP \\ 
AT\,2020yqt & 20200825 & Transient & $5$ & Gemini+GMOS \\ 
AT\,2020yqt & 20200829 & Transient & $9$ & P200+DBSP \\ 
AT\,2020yqt & 20200920 & Transient & $31$ & Keck1+LRIS \\ 
SN\,2020rsc & 20200822 & Transient & $-2$ & Gemini+GMOS \\ 
SN\,2020rsc & 20200824 & Transient & $0$ & MMT+Binospec \\ 
SN\,2020rsc & 20200825 & Transient & $0$ & GTC+OSIRIS \\ 
SN\,2020rsc & 20200915 & Transient & $21$ & Keck1+LRIS \\ 
SN\,2020vyv & 20201014 & Transient & $1$ & Keck1+LRIS [3] \\ 
SN\,2020xlt & 20201028 & Transient & $8$ & P60+SEDM \\ 
SN\,2020xlt & 20201030 & Transient & $10$ & GTC+OSIRIS \\ 
\enddata 
\tablereferences{[1] \citet{Gromadzki2018_ZTF18abwkrbl}, [2] \citet{Smith2020_ZTF20aaivtof}, [3] \citet{Siebert2020}} 
\end{deluxetable*} 

\newpage

\onecolumngrid

\section{Spectroscopic evolution of individual events}

\begin{figure*}[p]
    \includegraphics[width=0.45\textwidth]{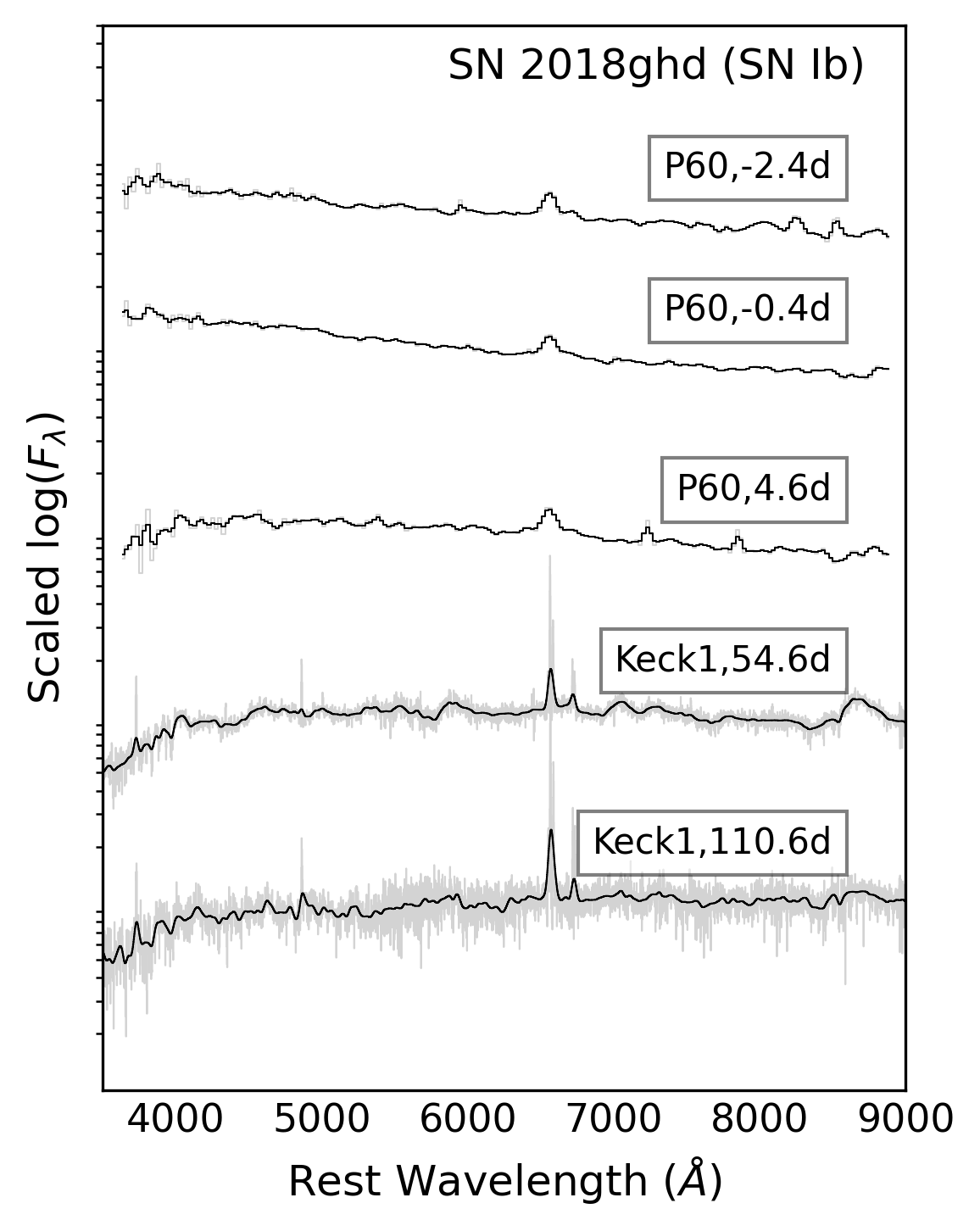}
    \includegraphics[width=0.45\textwidth]{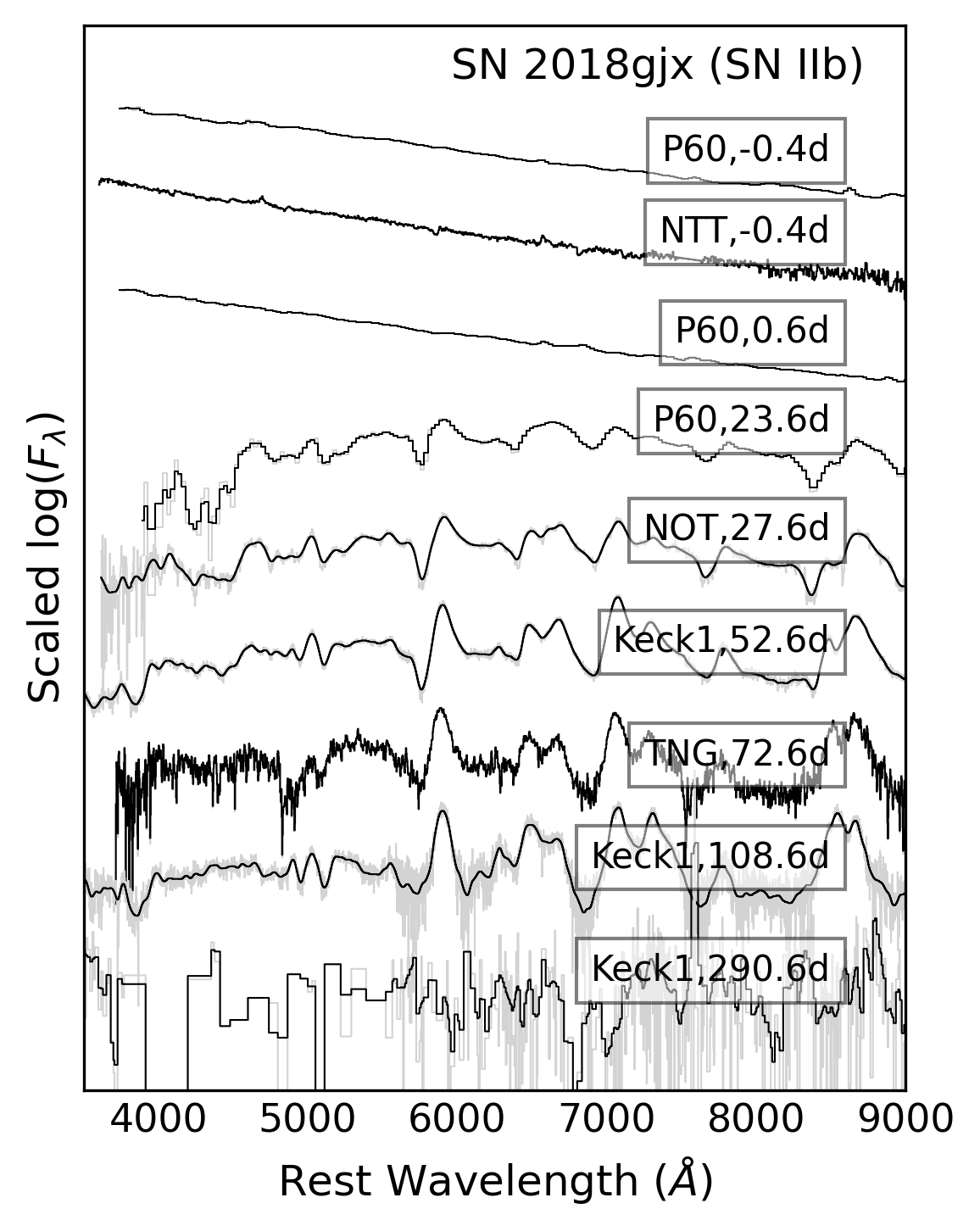}
    \includegraphics[width=0.45\textwidth]{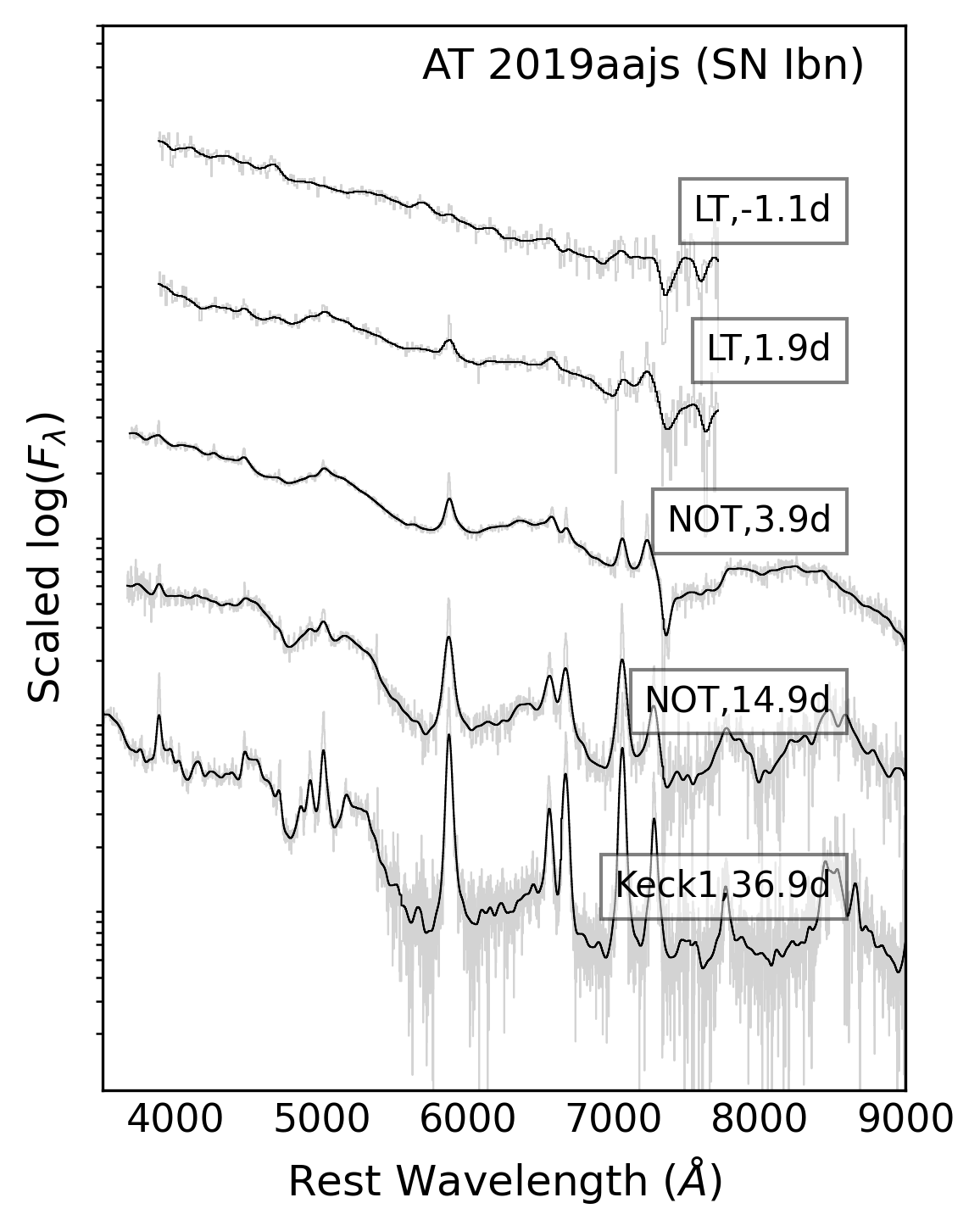}
    \includegraphics[width=0.45\textwidth]{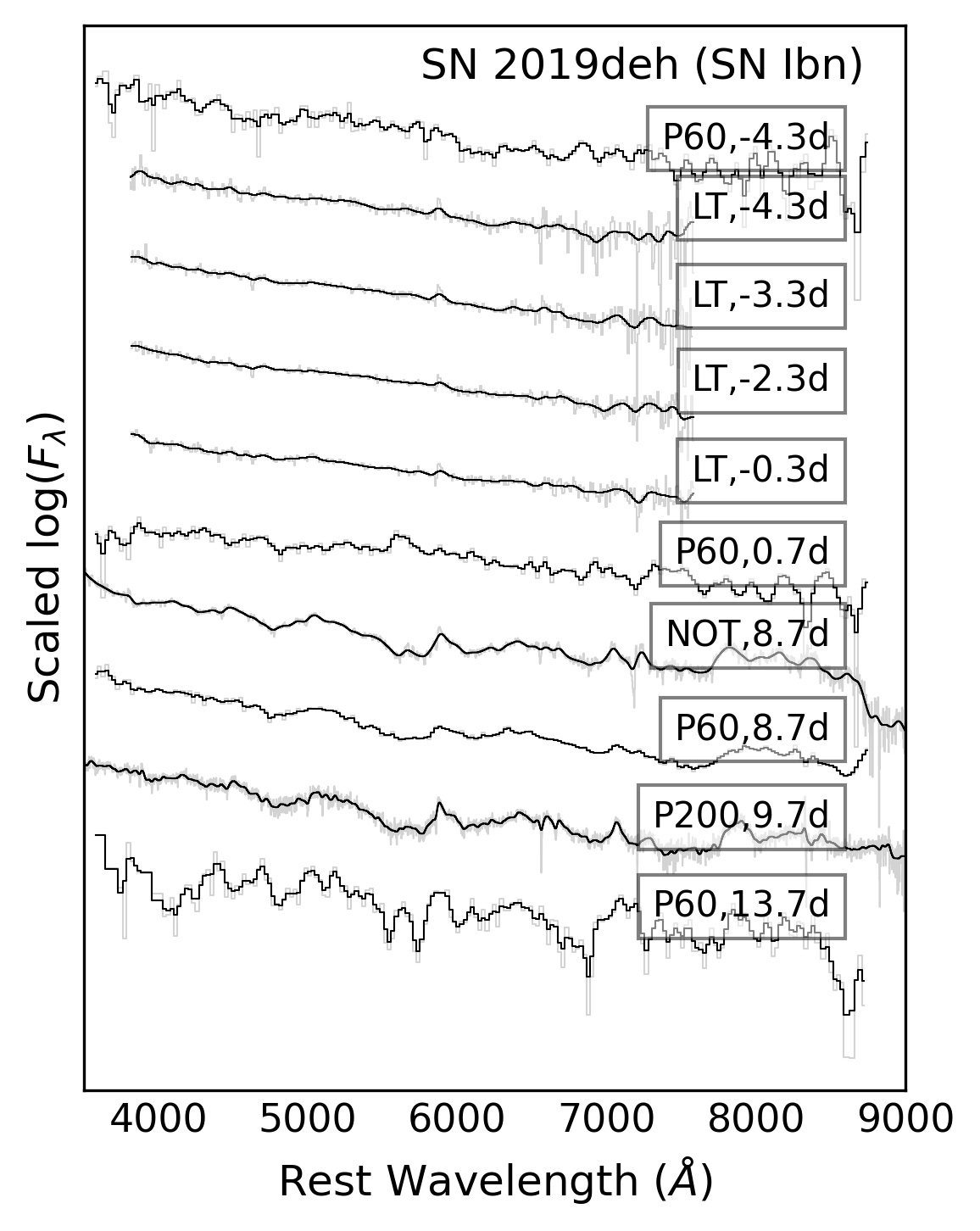}
    \caption{\edit2{Spectroscopic evolution for the ZTF transients in Table~\ref{tab:sources_all}.} Raw spectra are shown in light grey, and smoothed spectra are overlaid in black.}
    \label{fig:spec-events-1}
\end{figure*}

\begin{figure*}[p]
    \includegraphics[width=0.45\textwidth]{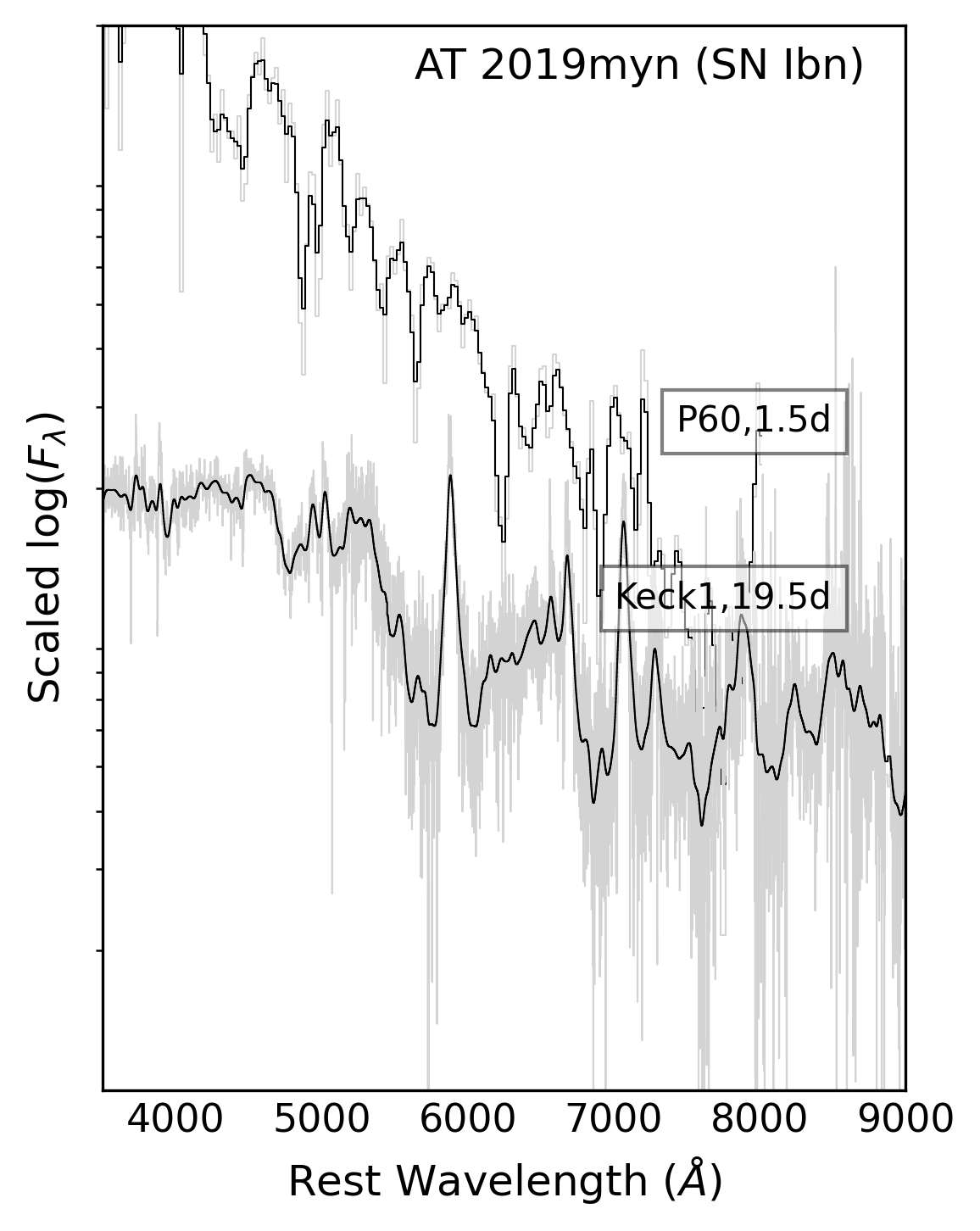}
    \includegraphics[width=0.45\textwidth]{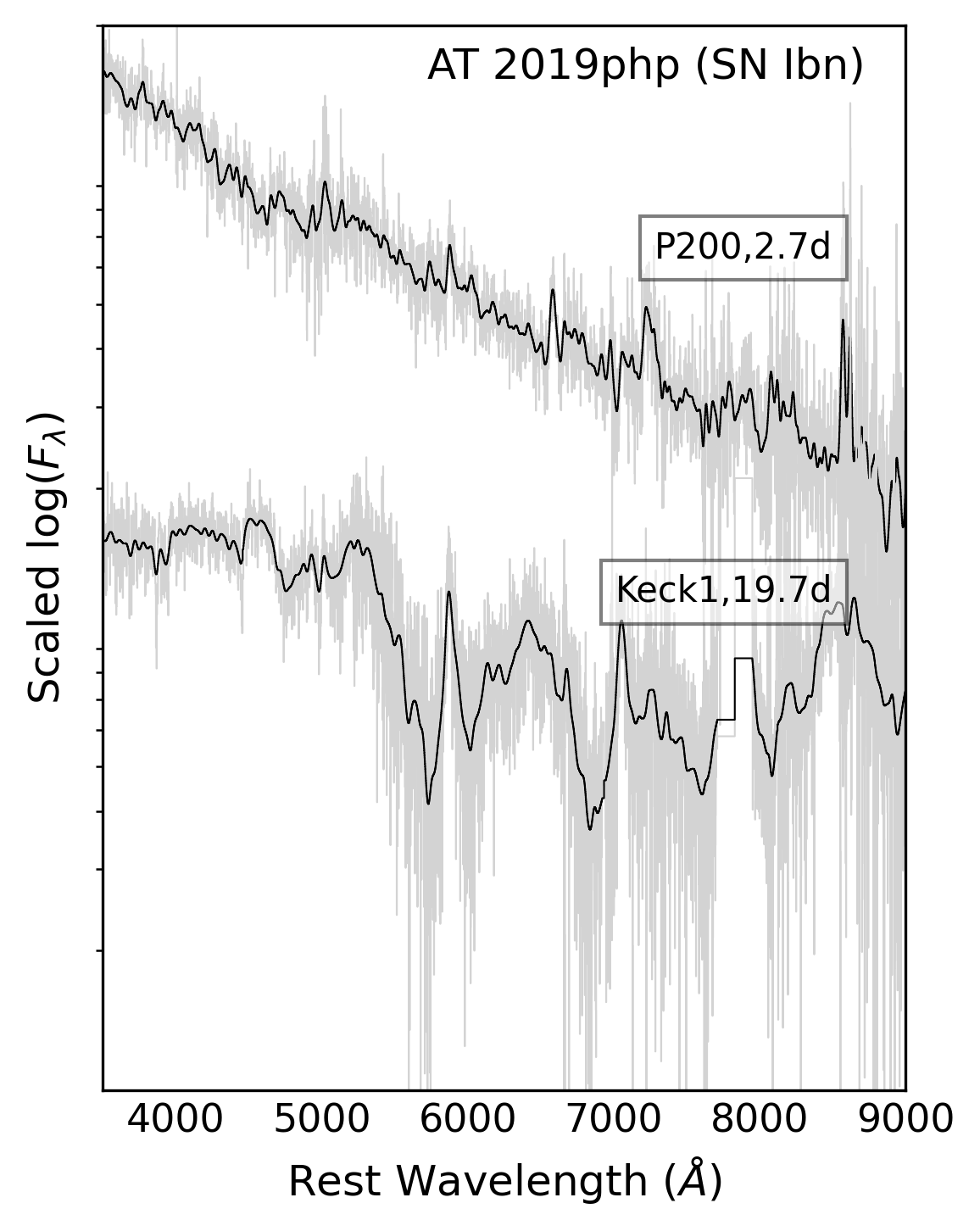}
    \includegraphics[width=0.45\textwidth]{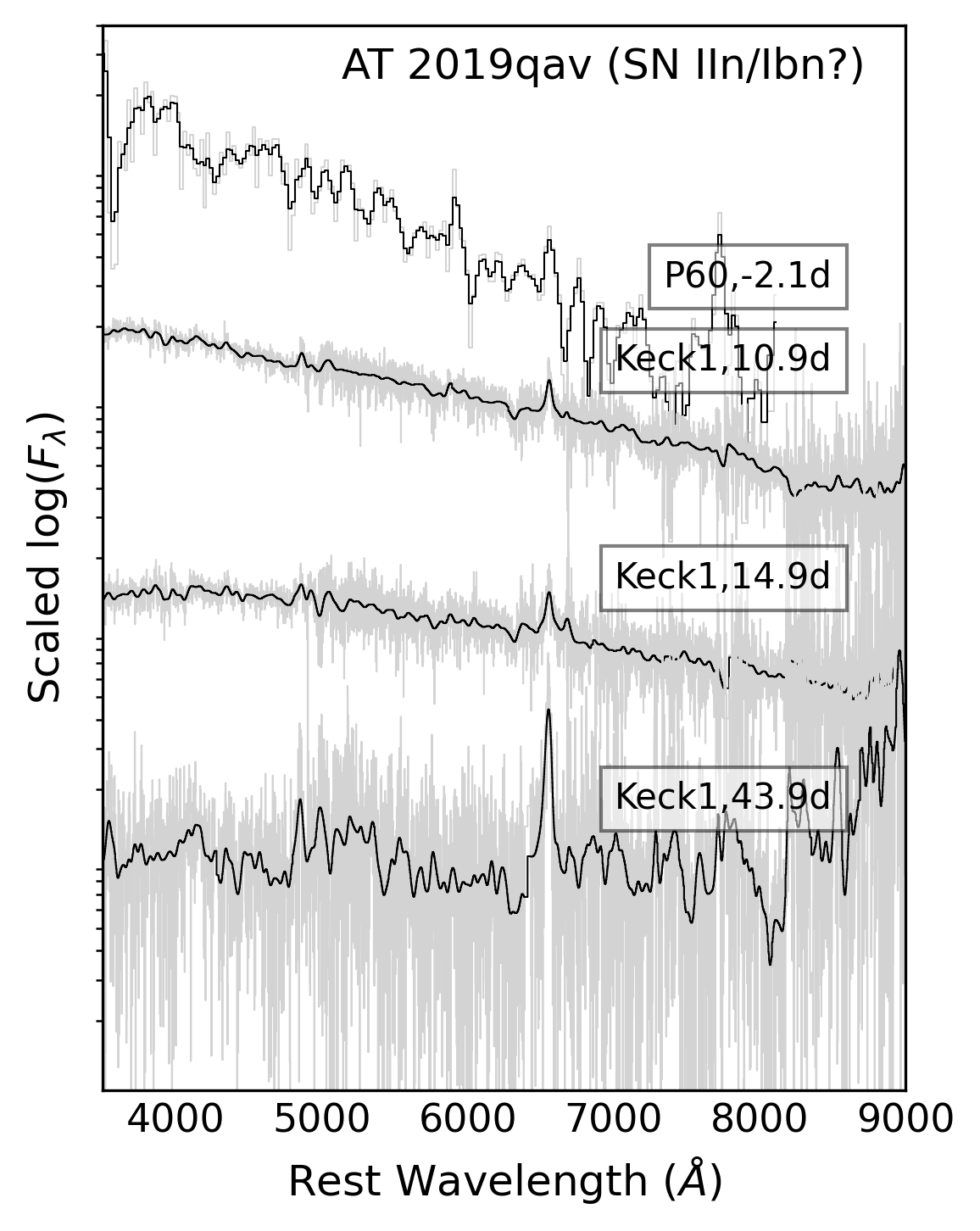}
    \includegraphics[width=0.45\textwidth]{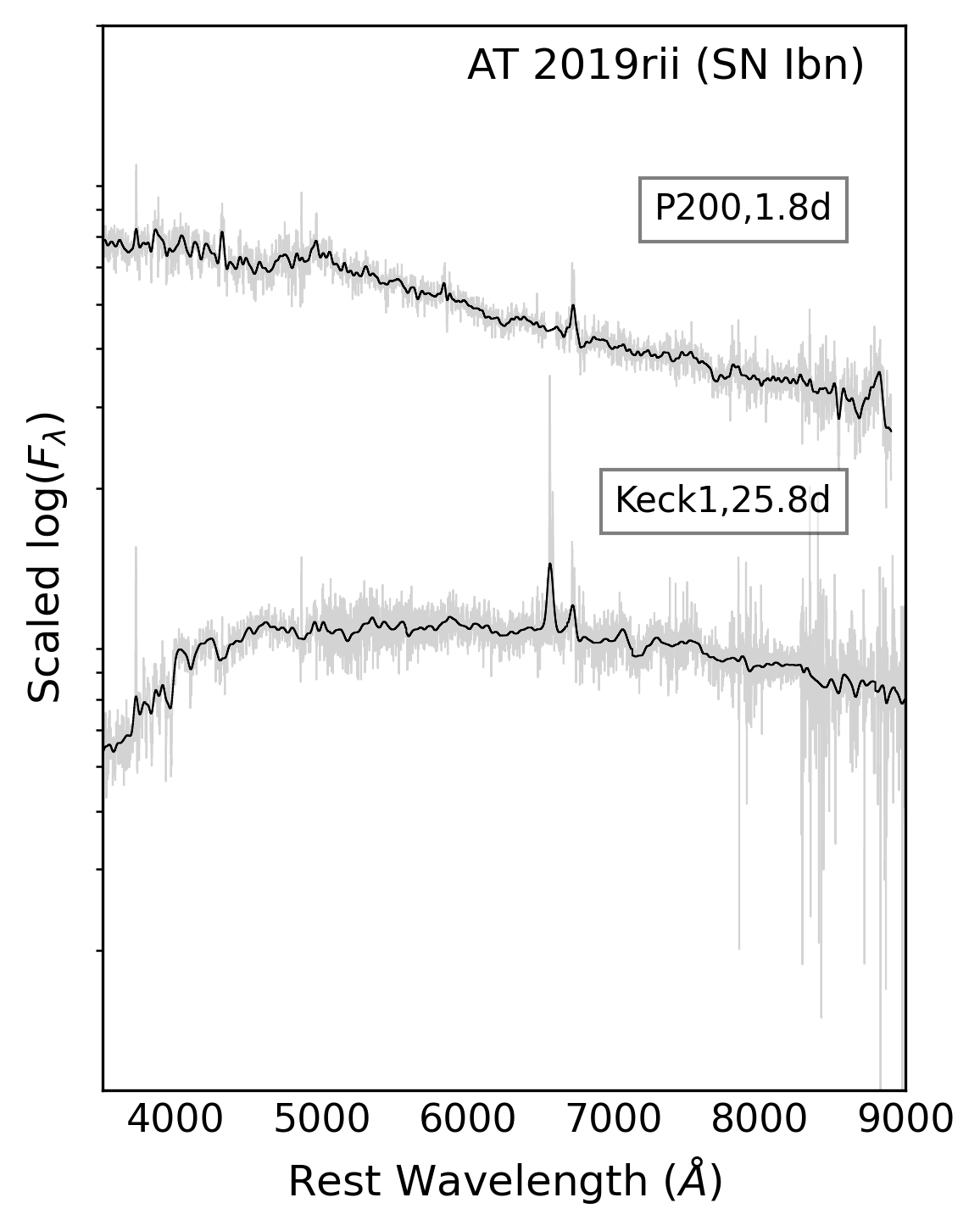}
    \caption{\edit2{Spectroscopic evolution for the ZTF transients in Table~\ref{tab:sources_all}.} Raw spectra are shown in light grey, and smoothed spectra are overlaid in black.}
    \label{fig:spec-events-2}
\end{figure*}

\begin{figure*}[p]
    \includegraphics[width=0.45\textwidth]{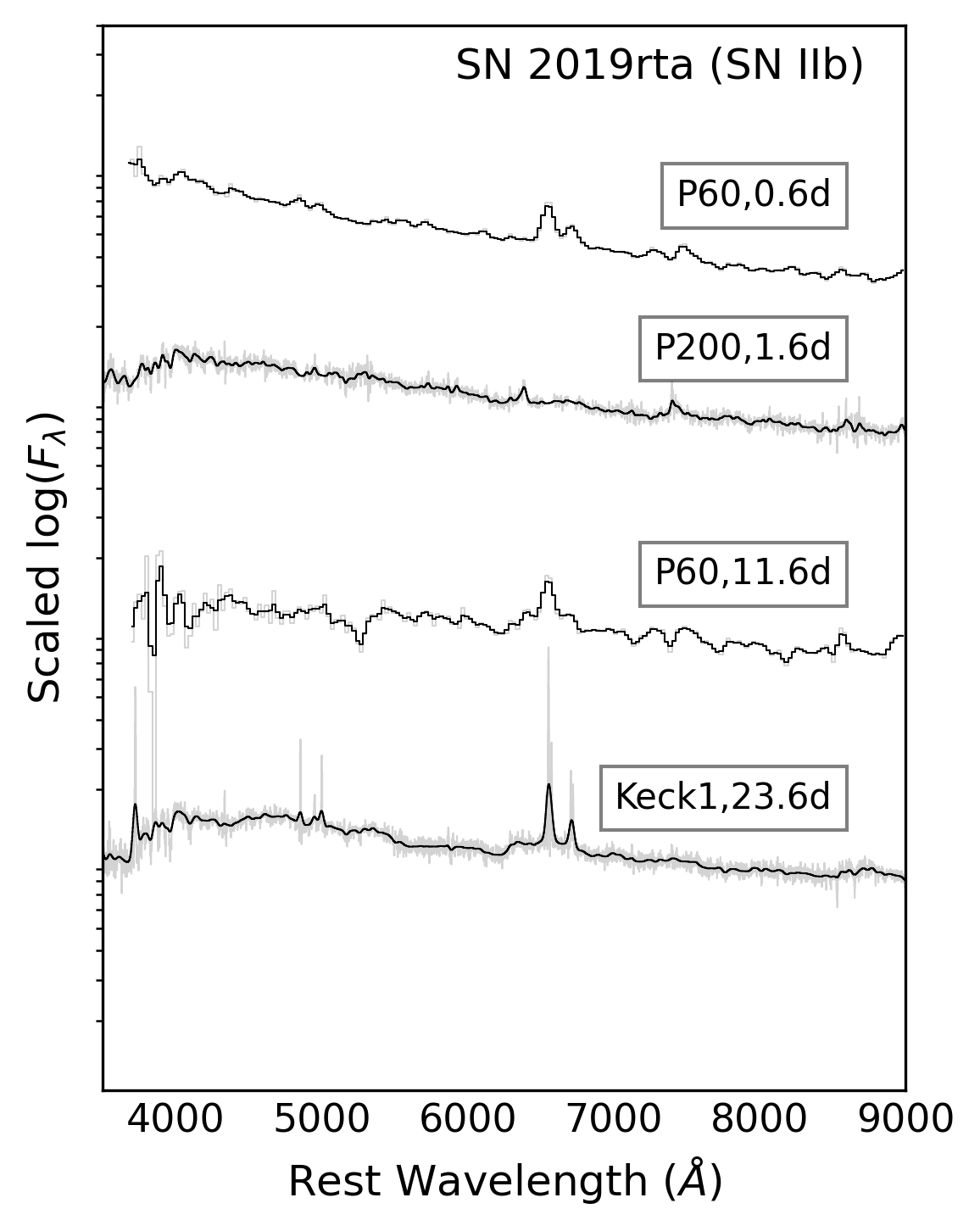}
    \includegraphics[width=0.45\textwidth]{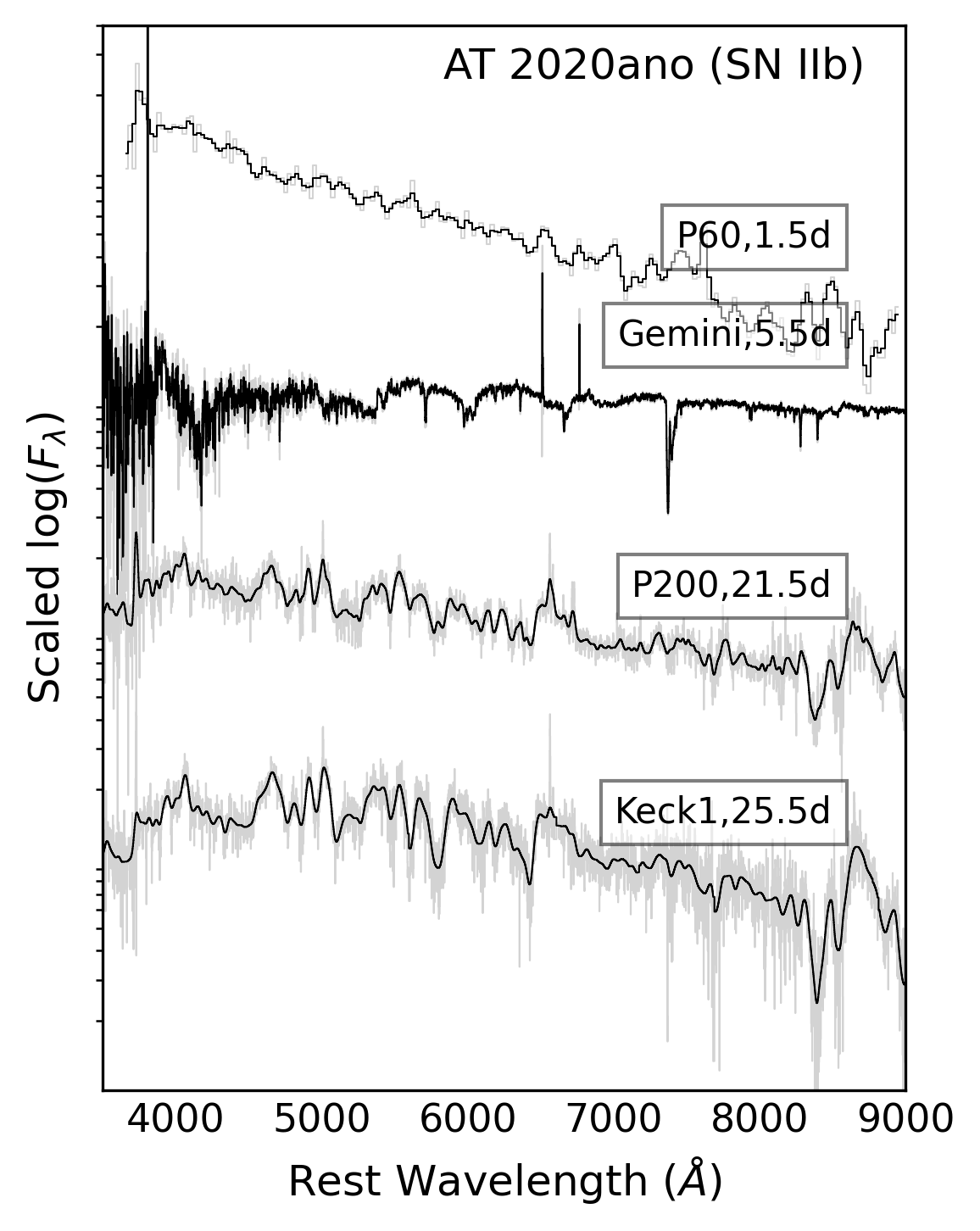}
    \includegraphics[width=0.45\textwidth]{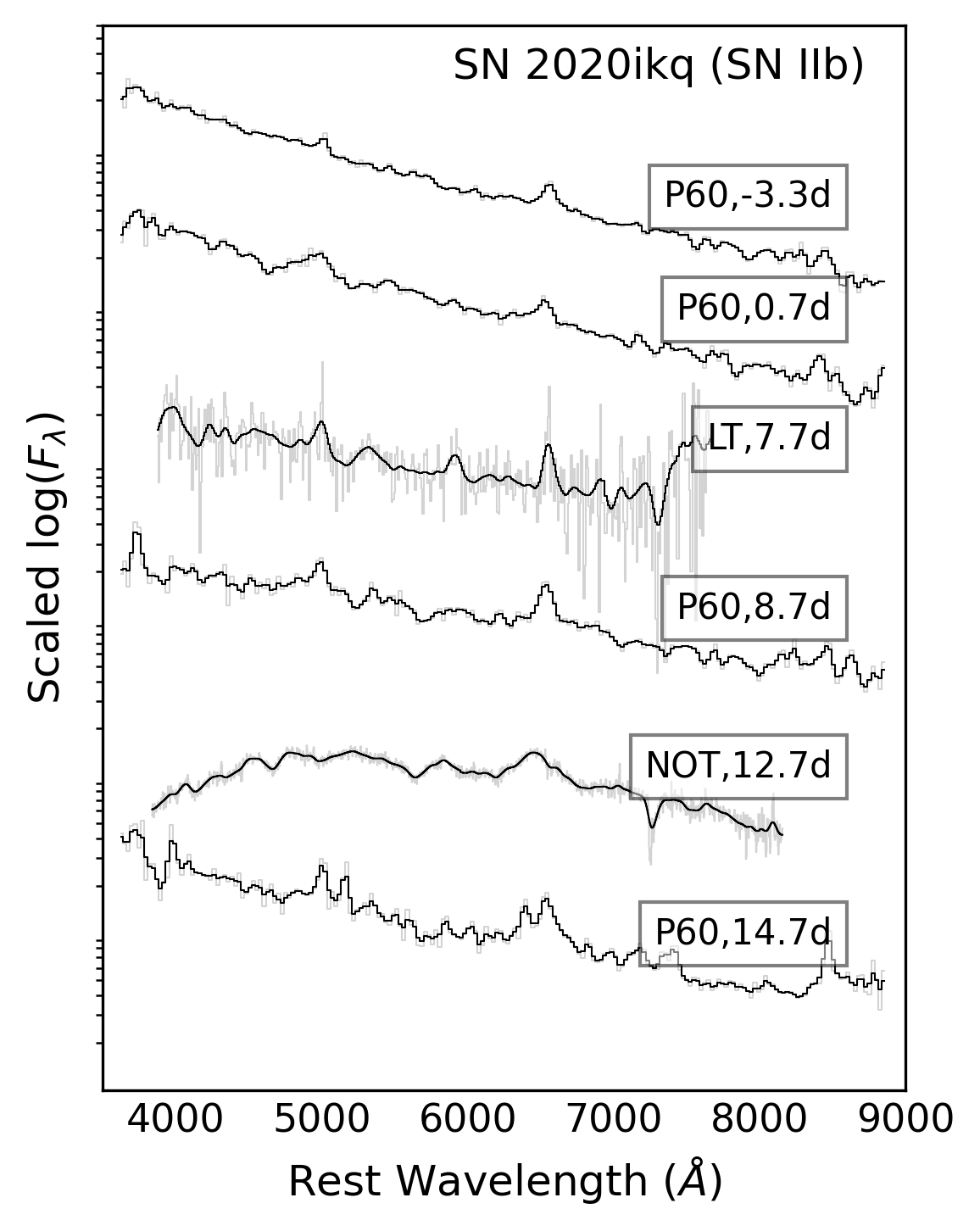}
    \includegraphics[width=0.45\textwidth]{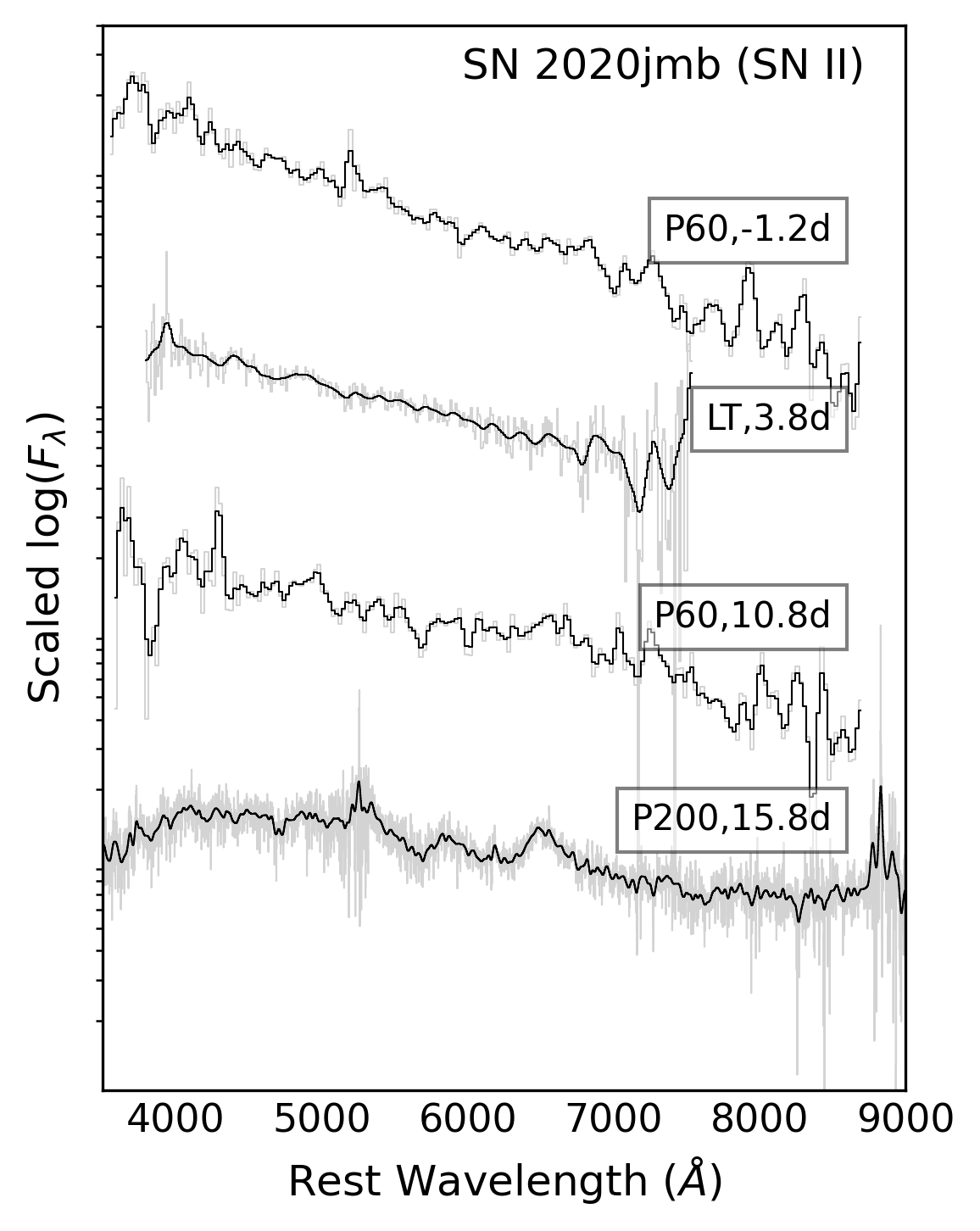}
    \caption{\edit2{Spectroscopic evolution for the ZTF transients in Table~\ref{tab:sources_all}.} Raw spectra are shown in light grey, and smoothed spectra are overlaid in black.}
    \label{fig:spec-events-3}
\end{figure*}

\begin{figure*}[p]
    \includegraphics[width=0.45\textwidth]{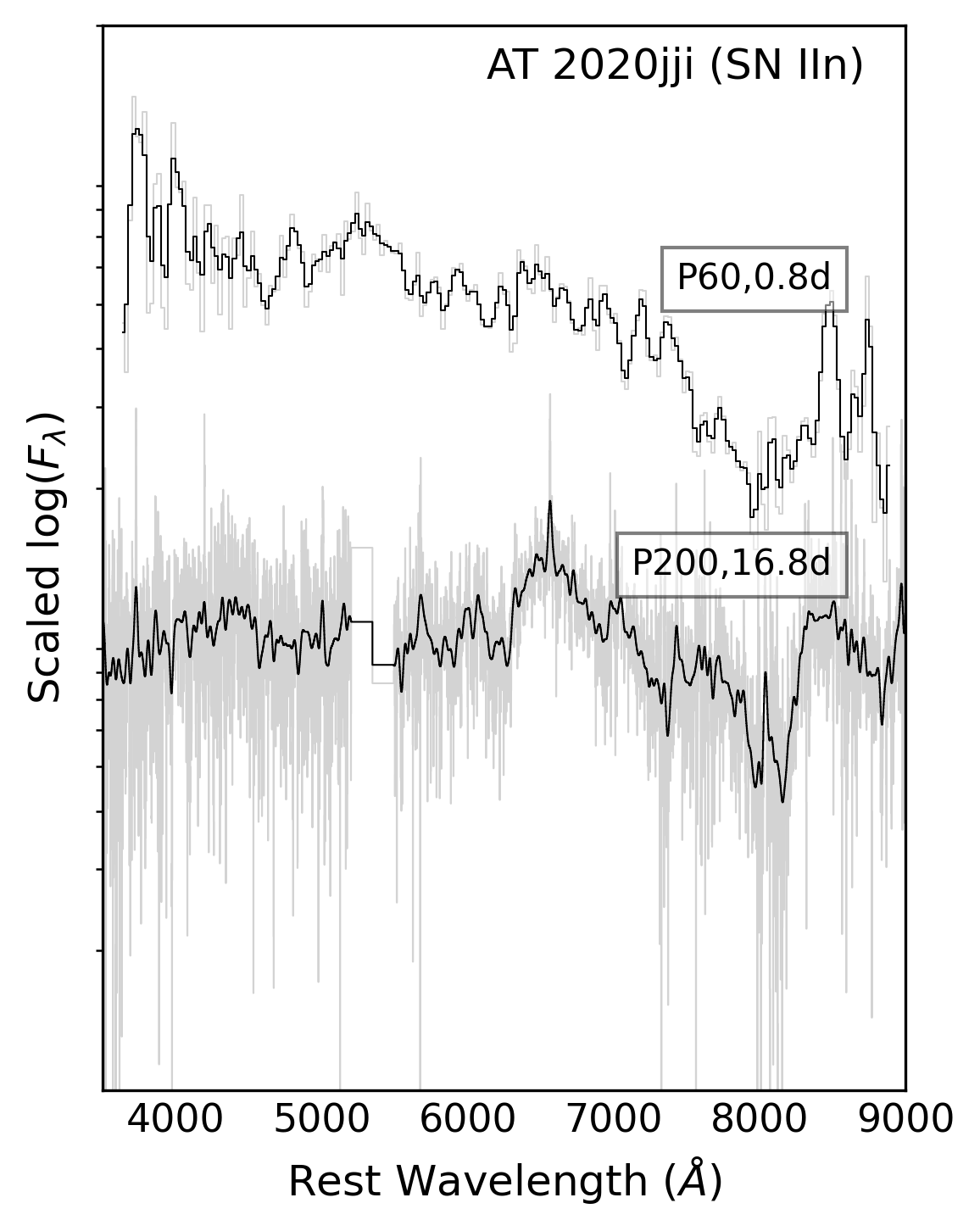}
    \includegraphics[width=0.45\textwidth]{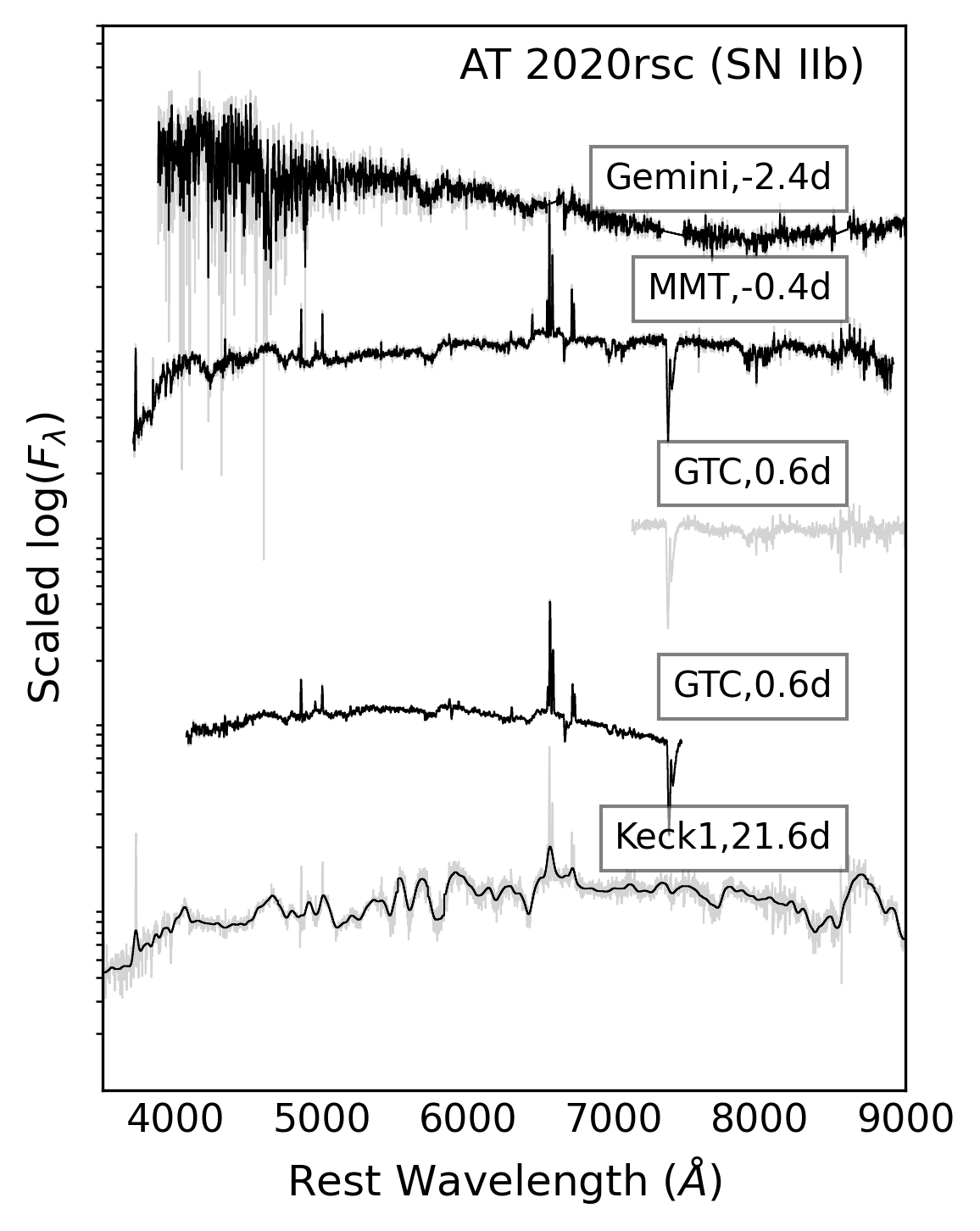}
    \caption{\edit2{Spectroscopic evolution for the ZTF transients in Table~\ref{tab:sources_all}.} Raw spectra are shown in light grey, and smoothed spectra are overlaid in black.}
    \label{fig:spec-events-4}
\end{figure*}

In Figure~\ref{fig:spec-events-1}, Figure~\ref{fig:spec-events-2}, Figure~\ref{fig:spec-events-3}, and Figure~\ref{fig:spec-events-4}, we plot the full set of spectra for each object.

\onecolumngrid

\section{Host Galaxy Properties}
\label{sec:appendix-gal}

\begin{figure*}
\centering
\includegraphics[width=1\textwidth]{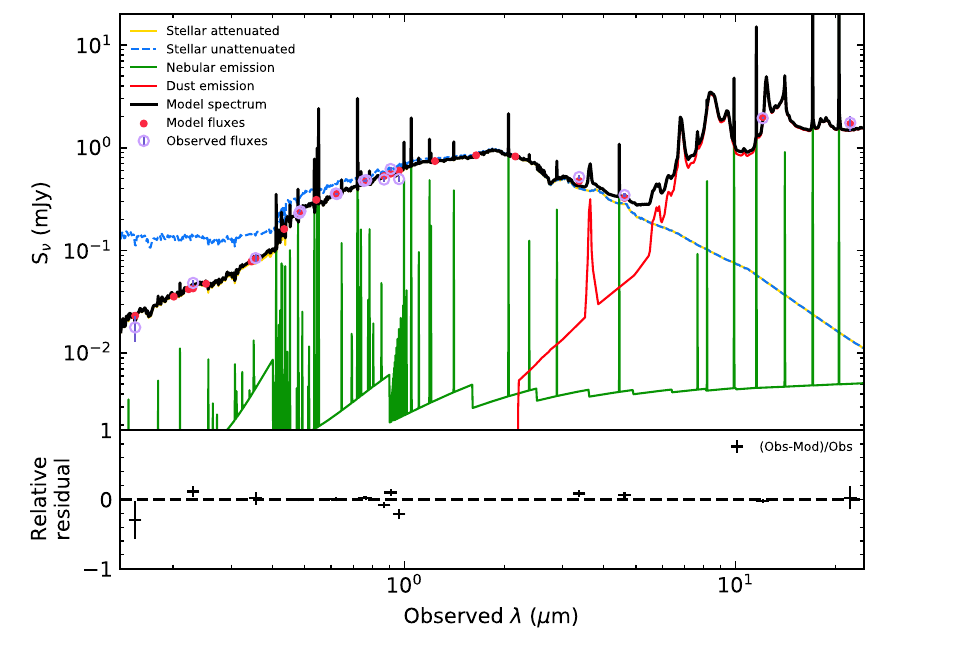}
\caption{The SED of AT\,2020yqt's host galaxy (empty circles) from the UV to mid-IR. The solid black line shows the best-fitting model of the SED. The SED model consists of a stellar component (yellow curve), a contribution from ionised gas, e.g., star-forming regions (green curve) and emission from heated dust (red curve) at $>30,000$~\AA. The photometry predicted by the best-fit model is shown by the red circles. The blue curve represents the stellar component corrected for host attenuation.
}
\label{fig:host_sed}
\end{figure*}

In this section we provide additional details and figures for the host-galaxy properties described in Section~\ref{sec:host-analysis}.
We model the spectral energy distributions (SEDs) with the software package \cigale{} \citep[Code Investigating GALaxy Evolution;][] {Burgarella2005, Noll2009,  Boquien2019}. The models and fitting parameters are summarized in Table~\ref{t:sed_input}. We adopt the \citet{Bruzual2003} simple stellar population model to compute the stellar emission and the \citet{Chabrier2003} initial mass function.
Furthermore, we assume a linear-exponential star-formation history [functional form $t \times \exp\left(-t/\tau\right)$, where $t$ is the age of the SFH episode and $\tau$ is the $e$-folding timescale]. To calculate the nebular emission from the ionised gas in \ion{H}{2} regions, we fix the \cigale{} ionisation parameter log $\text{U}_{\text{ion}}$ as $-2$. 
We use a modified \citet{Calzetti2000} starburst attenuation curve to model the dust attenuation. Dust emission was included via the \citet{Dale2014} dust templates. More details on the models used can be found in \citet{Boquien2019}. We generate 24385536 models and choose the best-fit SED using Bayesian inference. In Figure~\ref{fig:host_sed} we show the best-fit SED for the host galaxy of AT\,2020yqt as an example, and we list the derived host-galaxy properties in Table~\ref{tab:host_sed}.

\begin{table*}
\begin{center}
\caption{
\label{t:sed_input}
The models and fitting parameters used for \cigale{}
}
\begin{tabular}{p{3.5cm} p{9.5cm}}
\hline 
\hline 
Galaxy attributes & Brief description \\
\hline 
SFH & SFR $\propto$ $\frac{t}{\tau^2}$ $\exp(-t/\tau)$ \\
& $\tau= 250, 500, 1000, 2000, 4000, 6000, 8000$~Myr  \\
& Age $t = 250, 500, 1000, 2000, 4000, 8000,  12000$~Myr \\
\hline 
SSPs models & BC03 \citep{Bruzual2003} $+$ Chabrier IMF \citep{Chabrier2003} \\
& Stellar metallicity = 0.0004, 0.004, 0.008, 0.02 $Z_\odot$  \\
\hline 
Dust Attenuation & Modified power law curves \citep{Calzetti2000} + differential reddening of stars according to age \\
& $E(B-V)_{\rm young}$ = 0.0, 0.2, 0.3, 0.5, 0.8,  1.0, 1.5, 2.0, 3.0 \\
& $E(B-V)_{\rm old}$ = 0.3, 0.50, 1.0 \\
& ${\rm UV-bump~wavelength} = 217.5~{\rm nm}$\\
& ${\rm UV-bump~amplitude} = 0.0, 1.0, 2.0, 3.0 $\\
& ${\rm powerlaw~slope} = -0.13, -0.2 , -0.5$ \\
\hline 
Dust emission &  Dust templates of \citet{Dale2014} $+$ Energy Balance  \\
& AGN fraction = 0 \\
& alpha = 1.0, 1.5, 2.0, 2.5 \\
\hline 
Nebular & $\log U_{\rm ion} =$ -2.0 \\
& emission line width = 300.0 km/s  \\
\hline
\end{tabular}
\end{center}
\end{table*}

\begin{table*}
\caption{Summary of the host galaxy SED modelling}\label{tab:host_sed}
\begin{tabular}{ccccccc}
\hline
Object & $\chi^2/{\rm n.o.f.}$ & $M_B$~(mag) & $\log\,M/M_\odot$ & $\log\,{\rm SFR}/\left(M_\odot\,{\rm yr}^{-1}\right)$ & $\log {\rm Age}/{\rm yr}$ & $E(B-V)\,({\rm mag})$ \\
\hline
SN\,2018bcc & 2.10/9 & -18.57 & $9.47^{+0.15}_{-0.10}$ & $-0.48^{+0.09}_{-0.08}$ & $9.91^{+0.12}_{-0.19}$ & $0.25^{+0.06}_{-0.03}$ \\
AT2018cow & 56.74/24 & -18.46 & $9.59^{+0.06}_{-0.12}$ & $-0.19^{+0.10}_{-0.13}$ & $9.93^{+0.10}_{-0.15}$ & $0.25^{+0.07}_{-0.04}$ \\
SN\,2019dge & 0.61/14 & -16.19 & $8.65^{+0.12}_{-0.14}$ & $-1.22^{+0.28}_{-0.23}$ & $9.89^{+0.13}_{-0.21}$ & $0.31^{+0.15}_{-0.08}$ \\
SN\,2018gep & 9.85/19 & -16.59 & $8.45^{+0.09}_{-0.08}$ & $-1.08^{+0.04}_{-0.07}$ & $9.81^{+0.18}_{-0.19}$ & $0.25^{+0.06}_{-0.03}$ \\
SN\,2018ghd & 2.09/15 & -19.23 & $10.30^{+0.12}_{-0.13}$ & $0.37^{+0.32}_{-0.20}$ & $9.89^{+0.13}_{-0.23}$ & $0.58^{+0.43}_{-0.24}$ \\
SN\,2018gjx & 3.11/16 & -19.58 & $10.19^{+0.08}_{-0.13}$ & $0.42^{+0.15}_{-0.14}$ & $9.91^{+0.12}_{-0.21}$ & $0.28^{+0.12}_{-0.06}$ \\
SN\,2019aajs & 1.55/9 & -17.76 & $9.64^{+0.17}_{-0.14}$ & $-0.61^{+0.22}_{-0.49}$ & $9.94^{+0.09}_{-0.13}$ & $0.28^{+0.11}_{-0.06}$ \\
SN\,2019deh & 5.75/17 & -20.94 & $10.72^{+0.05}_{-0.07}$ & $0.89^{+0.07}_{-0.12}$ & $9.95^{+0.09}_{-0.12}$ & $0.25^{+0.06}_{-0.03}$ \\
SN\,2019myn & 5.43/14 & -18.32 & $9.42^{+0.13}_{-0.10}$ & $-0.54^{+0.06}_{-0.08}$ & $9.90^{+0.13}_{-0.19}$ & $0.25^{+0.06}_{-0.03}$ \\
SN\,2019php & 1.27/3 & -12.66 & $6.51^{+0.56}_{-0.39}$ & $-2.21^{+0.37}_{-0.34}$ & $8.94^{+0.80}_{-0.38}$ & $0.37^{+0.58}_{-0.13}$ \\
SN\,2019qav & 14.49/18 & -19.83 & $9.87^{+0.20}_{-0.16}$ & $0.60^{+0.21}_{-0.25}$ & $9.66^{+0.22}_{-0.50}$ & $0.25^{+0.08}_{-0.04}$ \\
SN\,2019rii & 4.23/10 & -19.81 & $10.27^{+0.18}_{-0.31}$ & $0.67^{+0.37}_{-0.20}$ & $9.84^{+0.17}_{-0.63}$ & $0.28^{+0.15}_{-0.06}$ \\
SN\,2019rta & 10.75/11 & -18.50 & $9.43^{+0.20}_{-0.18}$ & $0.01^{+0.29}_{-0.22}$ & $9.72^{+0.24}_{-0.55}$ & $0.33^{+0.19}_{-0.10}$ \\
SN\,2020oi & 1.22/10 & -21.82 & $10.90^{+0.23}_{-0.14}$ & $1.46^{+0.22}_{-0.41}$ & $9.75^{+0.23}_{-0.52}$ & $0.43^{+0.49}_{-0.15}$ \\
SN\,2020ano & 0.64/7 & -19.52 & $9.57^{+0.19}_{-0.21}$ & $0.54^{+0.45}_{-0.25}$ & $9.23^{+0.55}_{-0.49}$ & $0.32^{+0.26}_{-0.09}$ \\
SN\,2020ikq & 8.57/16 & -18.02 & $9.25^{+0.07}_{-0.31}$ & $-0.53^{+0.31}_{-0.07}$ & $9.91^{+0.12}_{-0.49}$ & $0.25^{+0.06}_{-0.03}$ \\
SN\,2020jmb & 5.93/13 & -17.38 & $8.77^{+0.16}_{-0.15}$ & $-0.70^{+0.11}_{-0.09}$ & $9.78^{+0.20}_{-0.41}$ & $0.26^{+0.08}_{-0.04}$ \\
SN\,2020jji & 1.00/14 & -18.53 & $9.49^{+0.20}_{-0.15}$ & $-0.24^{+0.19}_{-0.32}$ & $9.86^{+0.15}_{-0.21}$ & $0.25^{+0.08}_{-0.04}$ \\
%SN\,2020ntt & 4.73/13 & -17.98 & $9.16^{+0.24}_{-0.22}$ & $-0.19^{+0.37}_{-0.33}$ & $9.68^{+0.27}_{-0.62}$ & $0.34^{+0.26}_{-0.10}$ \\
SN\,2020rsc & 9.05/14 & -19.46 & $10.15^{+0.13}_{-0.14}$ & $0.41^{+0.18}_{-0.21}$ & $9.88^{+0.14}_{-0.25}$ & $0.32^{+0.15}_{-0.09}$ \\
SN\,2020vyv & 4.68/18 & -19.48 & $9.86^{+0.15}_{-0.15}$ & $0.46^{+0.20}_{-0.12}$ & $9.74^{+0.23}_{-0.50}$ & $0.26^{+0.11}_{-0.05}$ \\
SN\,2020xlt & 6.18/11 & -20.65 & $10.70^{+0.09}_{-0.12}$ & $0.70^{+0.16}_{-0.16}$ & $9.93^{+0.10}_{-0.15}$ & $0.26^{+0.09}_{-0.04}$ \\
AT2018lug & 19.49/14 & -19.34 & $9.32^{+0.16}_{-0.21}$ & $0.31^{+0.10}_{-0.10}$ & $9.29^{+0.22}_{-0.32}$ & $0.25^{+0.06}_{-0.03}$ \\
AT2019esf & 23.53/17 & -19.65 & $9.91^{+0.18}_{-0.15}$ & $0.42^{+0.20}_{-0.16}$ & $9.78^{+0.21}_{-0.50}$ & $0.29^{+0.11}_{-0.06}$ \\
AT2019kyw & 4.10/11 & -17.56 & $8.76^{+0.18}_{-0.18}$ & $-0.64^{+0.33}_{-0.18}$ & $9.70^{+0.25}_{-0.56}$ & $0.28^{+0.11}_{-0.06}$ \\
AT2020bdh & 2.95/11 & -18.38 & $8.93^{+0.20}_{-0.27}$ & $-0.08^{+0.14}_{-0.41}$ & $9.22^{+0.64}_{-0.32}$ & $0.25^{+0.06}_{-0.03}$ \\
AT2020bot & 10.77/17 & -21.28 & $11.67^{+0.04}_{-0.12}$ & $0.41^{+0.92}_{-0.10}$ & $9.78^{+0.09}_{-0.12}$ & $0.29^{+0.53}_{-0.07}$ \\
AT2020kfw & 2.96/15 & -19.66 & $10.31^{+0.17}_{-0.22}$ & $0.61^{+0.33}_{-0.25}$ & $9.84^{+0.17}_{-0.50}$ & $0.42^{+0.51}_{-0.16}$ \\
AT2020aexw & 10.15/16 & -19.48 & $9.86^{+0.12}_{-0.38}$ & $0.18^{+0.36}_{-0.05}$ & $9.90^{+0.12}_{-0.66}$ & $0.26^{+0.11}_{-0.05}$ \\
AT2020yqt & 7.74/16 & -20.01 & $10.04^{+0.17}_{-0.20}$ & $0.98^{+0.27}_{-0.26}$ & $9.26^{+0.55}_{-0.36}$ & $0.42^{+0.48}_{-0.14}$ \\
AT2020xnd & 4.62/3 & -15.62 & $7.23^{+0.25}_{-0.19}$ & $-1.12^{+0.16}_{-0.15}$ & $8.71^{+0.35}_{-0.22}$ & $0.27^{+0.11}_{-0.05}$ \\
\hline
\end{tabular}
\tablecomments{The absolute magnitudes are not corrected for host reddening. The SFRs are corrected for host reddening. The abbreviation `n.o.f.' stands for number of filters. The `age' in the last column refers to the age of the stellar population.}
\end{table*}

\bibliography{refs}
\bibliographystyle{aasjournal}

\end{document}